%% file: ms.tex
\documentclass[12pt,preprint]{aastex}

\usepackage{hyperref}

\newcommand{\dm}{\Delta{\rm m}_{15}(B)}

\newcommand{\kms}{${\rm km~s}^{-1}$}
\newcommand{\cmsq}{${\rm cm}^{-2}$}
\newcommand{\myemail}{mmp@lco.cl}

\slugcomment{Revised version}

\shorttitle{Dust Extinction and \ion{Na}{1} Absorption in SNe~Ia}
\shortauthors{Phillips et al.}

\begin{document}

%% LaTeX will automatically break titles if they run longer than
%% one line. However, you may use \\ to force a line break if
%% you desire.

\title{On the Source of the Dust Extinction in Type~Ia Supernovae and the Discovery of Anomalously Strong \ion{Na}{1} Absorption\altaffilmark{1}}
%\title{Anomalously-Strong \ion{Na}{1} Absorption in Type~Ia Supernovae\altaffilmark{1}}
%% \title{Interstellar Absorption Lines and Dust Extinction in Type~Ia Supernovae\altaffilmark{1}}

%% Use \author, \affil, and the \and command to format
%% author and affiliation information.
%% Note that \email has replaced the old \authoremail command
%% from AASTeX v4.0. You can use \email to mark an email address
%% anywhere in the paper, not just in the front matter.
%% As in the title, use \\ to force line breaks.

\author{
M.~M.~Phillips\altaffilmark{2},
Joshua~D.~Simon\altaffilmark{3},
Nidia~Morrell\altaffilmark{2},
Christopher R.~Burns\altaffilmark{3},
Nick~L.~J.~Cox\altaffilmark{4},
Ryan~J.~Foley\altaffilmark{5},
Amanda~I.~Karakas\altaffilmark{6},
F.~Patat\altaffilmark{7},
A.~Sternberg\altaffilmark{8},
R.~E.~Williams\altaffilmark{9},
A.~Gal-Yam\altaffilmark{10},
E.~Y.~Hsiao\altaffilmark{2},
D.~C.~Leonard\altaffilmark{11},
Sven~E.~Persson\altaffilmark{3},
Maximilian Stritzinger\altaffilmark{12},
I.~B.~Thompson\altaffilmark{3},
Abdo~Campillay\altaffilmark{2},
Carlos~Contreras\altaffilmark{2},
Gast\'{o}n~Folatelli\altaffilmark{13},
Wendy~L.~Freedman\altaffilmark{3},
Mario~Hamuy\altaffilmark{14},
Miguel~Roth\altaffilmark{2},
Gregory~A.~Shields\altaffilmark{15},
Nicholas~B.~Suntzeff\altaffilmark{16},
%Joshua~S.~Bloom\altaffilmark{15},
Laura~Chomiuk\altaffilmark{5},
Inese~I.~Ivans\altaffilmark{17},
Barry~F.~Madore\altaffilmark{3,18},
%John~E.~Norris\altaffilmark{6},
B.~E.~Penprase\altaffilmark{19},
Daniel~Perley\altaffilmark{20},
G.~Pignata\altaffilmark{21}
G.~Preston\altaffilmark{3},
and
Alicia~M.~Soderberg\altaffilmark{5}
%James~F.~Steiner\altaffilmark{22}
%and
%David~Yong\altaffilmark{5}
}

\altaffiltext{1}{
 This paper includes data gathered with the 6.5 meter Magellan telescopes 
at Las Campanas Observatory, Chile.}

\altaffiltext{2}{Carnegie Observatories, Las Campanas Observatory, 
  Casilla 601, La Serena, Chile; \myemail}

\altaffiltext{3}{Observatories of the Carnegie Institution for
 Science, 813 Santa Barbara St., Pasadena, CA 91101, USA}
 
 \altaffiltext{4}{Instituut voor Sterrenkunde, KU Leuven, Celestijnenlaan 200D bus 2401, 3001 Leuven, Belgium}
 
 \altaffiltext{5}{Harvard-Smithsonian Center for Astrophysics, 60 Garden Street, Cambridge, MA 02138, USA}
 
\altaffiltext{6}{Research School of Astronomy and Astrophysics, The Australian National University, Weston, ACT 2611, Australia}
 
\altaffiltext{7}{European Southern Observatory (ESO), Karl Schwarschild Strasse 2, 85748, 
Garching bei M\"{u}nchen, Germany}

\altaffiltext{8}{Minerva Fellow, Max Planck Institute for Astrophysics, 
Karl Schwarzschild Strasse 1, D-85741 Garching bei M\"{u}nchen, Germany}

\altaffiltext{9}{Space Telescope Science Institute, 3700 San Martin Drive, Baltimore, MD 21218, USA} 

\altaffiltext{10}{Benoziyo Center for Astrophysics, Faculty of Physics, Weizmann Institute of Science, Rehovot 76100, Israel}

\altaffiltext{11}{Department of Astronomy, San Diego State University, San Diego, CA 92182, USA}
  
\altaffiltext{12}{Department of Physics and Astronomy, Aarhus University, Ny Munkegade 120, DK-8000 Aarhus C, Denmark}
 
\altaffiltext{13}{Kavli Institute for the Physics and Mathematics of the Universe, Todai Institutes for Advanced Study, 
the University of Tokyo, Kashiwa, Japan 277-8583}

\altaffiltext{14}{Universidad de Chile, Departamento de Astronom\'{\i}a,
  Casilla 36-D, Santiago, Chile.}
  
\altaffiltext{15}{Department of Astronomy, University of Texas, Austin, TX 78712, USA}

\altaffiltext{16}{
George P. and Cynthia Woods Mitchell Institute for
Fundamental Physics and Astronomy, Texas A\&M University,
Department of Physics and Astronomy,  College Station, TX 77843, USA}
  
%\altaffiltext{15}{Department of Astronomy, University of California, Berkeley, CA 94720-3411, USA}

\altaffiltext{17}{Department of Physics and Astronomy, University of Utah, Salt Lake City, UT 84112, USA}
  
\altaffiltext{18}{Infrared Processing and Analysis Center, Caltech/Jet
  Propulsion Laboratory, Pasadena, CA 91125, USA}
  
\altaffiltext{19}{Department of Physics and Astronomy, Pomona College, 610 N. College Ave., Claremont, CA 91711, USA}

\altaffiltext{20}{Cahill Center for Astrophysics, California Institute of Technology, Pasadena, CA 91125, USA}

\altaffiltext{21}{Departamento de Ciencias Fisicas, Universidad Andres Bello, Avda. Republica 252, Santiago, Chile}

%\altaffiltext{22}{Department of Astronomy, Cambridge University, Madingley Road, Cambridge, CB3 0HA, UK}

%% Mark off your abstract in the ``abstract'' environment. In the manuscript
%% style, abstract will output a Received/Accepted line after the
%% title and affiliation information. No date will appear since the author
%% does not have this information. The dates will be filled in by the
%% editorial office after submission.

\begin{abstract}
High-dispersion observations of the \ion{Na}{1}~D $\lambda\lambda$5890,~5896 
and \ion{K}{1} $\lambda\lambda$7665,~7699 interstellar lines, and the diffuse 
interstellar band at 5780~\AA\ in the spectra of 32 Type~Ia supernovae are used 
as an independent means of probing dust extinction.  We show that the dust 
extinction of the objects where the diffuse interstellar band at 5780~\AA\ is detected 
is consistent with the visual extinction derived from the supernova colors. This 
strongly suggests that the dust producing 
the extinction is predominantly located in the interstellar medium 
of the host galaxies and not in circumstellar material associated with the progenitor 
system.  One quarter of the supernovae display anomalously large 
\ion{Na}{1} column densities in comparison to the amount of dust extinction derived 
from their colors.  Remarkably, all of the cases of unusually  strong 
\ion{Na}{1}~D absorption correspond to ``Blueshifted'' profiles in the classification
scheme of \citet{sternberg11}.  This coincidence suggests that outflowing 
circumstellar gas
%with a low dust-to-gas ratio 
is responsible for at least some of 
the cases of anomalously large \ion{Na}{1} column densities. 
%although for one of
%the more heavily reddened supernovae, the observations appear more to 
%favor an origin in the host galaxy interstellar medium.  
Two supernovae with unusually strong \ion{Na}{1}~D 
absorption showed essentially normal \ion{K}{1} column densities for the dust 
extinction implied by their colors, but this does not appear to be a universal 
characteristic.  
Overall, we find the most accurate predictor of individual supernova 
extinction to be the equivalent width of the diffuse interstellar band 
at 5780~\AA, and provide an empirical relation for its use.
Finally, we identify ways of producing significant 
enhancements of the Na abundance of circumstellar material in both the 
single-degenerate and double-degenerate scenarios for the progenitor system.
%% We briefly speculate on the possiblity that the progenitors
%% of these strong-sodium events are single-degenerate systems with a 
%% massive ($> 3 M_\sun$) AGB star as the donor star, in analogy to the 2002ic-like events.
\end{abstract}

%% Keywords should appear after the \end{abstract} command. The uncommented
%% example has been keyed in ApJ style. See the instructions to authors
%% for the journal to which you are submitting your paper to determine
%% what keyword punctuation is appropriate.

\keywords{circumstellar matter --- dust, extinction --- galaxies: ISM --- supernovae: general}

%% From the front matter, we move on to the body of the paper.
%% In the first two sections, notice the use of the natbib \citep
%% and \citet commands to identify citations.  The citations are
%% tied to the reference list via symbolic KEYs. The KEY corresponds
%% to the KEY in the \bibitem in the reference list below. We have
%% chosen the first three characters of the first author's name plus
%% the last two numeral of the year of publication as our KEY for
%% each reference.

%% Authors who wish to have the most important objects in their paper
%% linked in the electronic edition to a data center may do so by tagging
%% their objects with \objectname{} or \object{}.  Each macro takes the
%% object name as its required argument. The optional, square-bracket 
%% argument should be used in cases where the data center identification
%% differs from what is to be printed in the paper.  The text appearing 
%% in curly braces is what will appear in print in the published paper. 
%% If the object name is recognized by the data centers, it will be linked
%% in the electronic edition to the object data available at the data centers  
%%
%% Note that for sources with brackets in their names, e.g. [WEG2004] 14h-090,
%% the brackets must be escaped with backslashes when used in the first
%% square-bracket argument, for instance, \object[\[WEG2004\] 14h-090]{90}).
%%  Otherwise, LaTeX will issue an error. 

\section{INTRODUCTION}
\label{sec:intro}

Type~Ia supernovae (SNe~Ia) are one of the most effective observational
tools for measuring the expansion history of the Universe.  Their successful use in cosmology
is due to the discovery of empirical relations that dramatically decrease the dispersion in peak 
luminosities at optical wavelengths.  The first of these is the well-known correlation with 
light curve shape: intrinsically brighter SNe~Ia have broader light curves that 
decline more slowly from maximum than do the light curves of less luminous SNe~Ia 
\citep{phillips93}. The second is a strong dependence of peak luminosity on color 
that is in the same sense as dust reddening, but with an average value of 
the ratio of total-to-selective extinction, $R_V$, that is  significantly less than 
would be produced by normal interstellar dust in the Milky Way 
\citep{tripp98}.  The latter result has variously been interpreted as possible evidence that the
extinction arises in circumstellar dust \citep{wang05,goobar08}, as the consequence
of intrinsic differences in color between SNe~Ia with ``normal'' and ``high''
\ion{Si}{2} expansion velocities \citep{foley_kasen11}, or as a bias due to a
misidentification of the dispersion in the luminosity/color-corrected 
Hubble diagram with an intrinsic scatter in luminosity rather than color \citep{scolnic13}.

Our understanding of the progenitors and explosion mechanism(s) that produce SNe~Ia is 
still quite limited.  Although there is widespread agreement that these objects correspond to the
thermonuclear disruption of a white dwarf in a binary system, it is not yet clear if the companion
to the white dwarf is a main sequence or giant star (``single-degenerate''  or ``SD'' model) or another
white dwarf (``double-degenerate'' or ``DD'' model).  In recent years, observational evidence favoring both
scenarios has been put forward \citep[e.g., see][]{howell11,maoz12,patat13a}.  
In both the SD  and DD scenarios, material ejected from the 
system prior to the explosion may remain as circumstellar material (CSM) 
\citep{moore12,raskin13,shen13}.  
\citet{sternberg11} found a strong statistical preference for blueshifted structures in the 
narrow \ion{Na}{1}~D absorption observed in the line-of-sight to many SNe~Ia, suggestive of gas 
outflows from the progenitor systems.  In a few such cases, temporal variations of blueshifted
components of the \ion{Na}{1}~D lines apparently due to changing ionization conditions in the CSM
have also been observed \citep[e.g.,][]{patat07,simon09,dilday12}.  On the other hand, radio and X-ray
observations of the prototypical type~Ia SN~2011fe place tight upper limits on the amount of
CSM in the progenitor system before explosion \citep[e.g.,][]{horesh12}, and early-time
photometry of this event apparently rules out either a red giant or main sequence 
companion \citep{bloom12}.

In the Milky Way, the strengths of certain interstellar absorption features such as the
\ion{Na}{1}~D lines and the diffuse interstellar bands (DIBs) have been known for many
years to correlate with dust extinction \citep[e.g.,][]{merrill38,hobbs74}.  In this paper, we employ 
high-dispersion spectroscopy to use these features as an independent probe of the dust
affecting the colors of SNe~Ia.  As we will show,
the data indicate that the dust extinction for the objects where DIBs are observed is
generally consistent with the extinction derived from the SN colors, and therefore most
likely arises in the interstellar medium of the host galaxy.  However, one-fourth of the 
SNe~Ia, all with blueshifted structures as per \citet{sternberg11}, display anomalously large 
\ion{Na}{1} column densities that, in the interstellar medium (ISM) of the Milky Way, would 
correspond to an order of magnitude or more greater dust extinction than that implied by
the SN colors.

\section{OBSERVATIONS AND ANALYSIS}

%% In a manner similar to \objectname authors can provide links to dataset
%% hosted at participating data centers via the \dataset{} command.  The
%% second curly bracket argument is printed in the text while the first
%% parentheses argument serves as the valid data set identifier.  Large
%% lists of data set are best provided in a table (see Table 3 for an example).
%% Valid data set identifiers should be obtained from the data center that
%% is currently hosting the data.
%%
%% Note that AASTeX interprets everything between the curly braces in the 
%% macro as regular text, so any special characters, e.g. "#" or "_," must be 
%% preceded by a backslash. Otherwise, you will get a LaTeX error when you 
%% compile your manuscript.  Special characters do not 
%% need to be escaped in the optional, square-bracket argument.

\subsection{Column Densities and Equivalent Widths}
\label{sec:coldensities}

Our approach is to first examine the relationship between dust extinction 
and interstellar absorption lines in the Milky Way.  These results will then 
be contrasted with a similar comparison between the dust extinction in the 
line-of-sight to the SN Ia as derived from their optical and near-infrared (NIR) light
curves, and the narrow absorption lines produced by the host galaxy ISM 
and/or a pre-existing CSM (hereafter referred to collectively as
``host absorption'').

In the first case, we employ a sample of 46 SNe and AGNs as external
beacons to study the absorption lines produced by the ISM
of the Milky Way.  Echelle spectra of 22 of these objects are 
drawn from the observations of thermonuclear and core-collapse SNe 
published by \citet{sternberg11} and available through WISeREP 
\citep[][\url{http://www1.weizmann.ac.il/astrophysics/wiserep/}]{yaron12}.  
The remaining 24 spectra in our data 
set correspond to unpublished observations of SNe and AGNs 
obtained with the Magellan Inamori Kyocera Echelle (MIKE) 
\citep{bernstein03} on the 6.5~m Clay 
telescope.  Table~\ref{tab:tab1} lists the objects in this sample.  Henceforth,
we refer to these objects as the ``Milky Way'' sample.

To study the relationship between the SN dust extinction and the narrow host
absorption lines, we have put together a sample of 32 SNe~Ia with both 
high-dispersion spectra {\em and} well-observed light curves.  
Spectra for 21 of these SNe~Ia were drawn from the \citet{sternberg11} 
study (also available through WISeREP), and an additional 6 are taken 
from \citet{foley12b}.  Results for the remaining 5 SNe are taken from the 
literature.  Table~\ref{tab:tab2} lists the full sample of 32 SNe~Ia along 
with host galaxy names, morphologies, and references to the 
SN photometry.  Table~\ref{tab:tab3} gives the sources and wavelength 
resolutions of the high-dispersion spectral observations.
These SNe are referred to as the ``host absorption'' sample in the remainder
of this paper.

Column densities of neutral sodium and potassium were measured for 
both the Milky Way and host absorption components of the \ion{Na}{1}~D 
$\lambda\lambda$5890,~5896 and \ion{K}{1} $\lambda\lambda$7665,~7699 
doublets using the Voigt profile fitting program, 
VPFIT\footnote[22]{\url{http://www.ast.cam.ac.uk/~rfc/vpfit.html}},
developed by R.~F.~Carswell, J.~K.~Webb, and others, in combination with the 
VPGUESS\footnote[23]{\url{http://www.eso.org/~jliske/vpguess/}} interface of 
J.~Liske.   Upper limits for non-detections of both
\ion{Na}{1} and \ion{K}{1} were calculated by first estimating an upper limit
to the equivalent width, and then converting this to a column density using
empirical relations between equivalent width and column density derived
from weak, unsaturated lines in other objects.
In cases where the \ion{Na}{1}~D lines were significantly
saturated ($\log{N_{Na~I}} \gtrsim 13$~\cmsq), the much weaker \ion{K}{1} lines 
were used to determine the velocity and Doppler parameter, $b$, of each visible
component, and this information was employed in fitting the saturated
portion of the \ion{Na}{1}~D profiles.  This procedure was possible for most
of the objects observed from 2006 onward.  Four of the SNe in the host absorption sample
had extremely strong D~lines ($\log{N_{Na~I}} > 13.5$~\cmsq).  For one of 
these --- SN~2002bo ---the spectral coverage did not include the \ion{K}{1} 
lines, and so the uncertainty of our measurement of $\log{N_{Na~I}}$ 
is large.  A model with two absorption components provided a 
significantly better fit to the severely saturated profiles of the D~lines
in this SN than did one with a single component, and so we 
have adopted the results of the two-component model in this paper. 
However, without the additional information provided by the \ion{K}{1}
lines, the error associated with the measurement of 
$\log{N_{Na~I}}$ is large.

\citet{kemp02} found that \ion{Na}{1} column densities measured
from fitting profiles to the D~lines  for values
$\log{N_{Na~I}} > 12.5$~\cmsq\ were systematically underestimated by 
0.40--0.70~dex compared to column densities measured from the much weaker 
 \ion{Na}{1} UV $\lambda\lambda$3302, 3303 doublet.  The UV lines are not 
 covered by our echelle spectra so we cannot confirm this, although as 
 mentioned in  \S\ref{sec:ki}, a comparison of the $N_{Na~I} / N_{K~I}$ 
 ratio derived for our Milky Way sample with the measurements of 
 \citet{kemp02} suggests that our $N_{Na~I}$ values may be similarly 
 affected.  Nevertheless, since our approach is to compare the SN host
 absorption measurements {\em relative} to our same measurements for
 the Milky Way, this problem should not affect our conclusions.
 
Although the \ion{Ca}{2} H~\&~K lines
were also present in many of the SNe spectra, except for a few
specific objects, they are not included
in this study since the column density of \ion{Ca}{2} is poorly correlated with dust
extinction in the Milky Way, presumably due to variations in the large
depletion factor of calcium \citep{hobbs74}.

Equivalent widths of the DIB at 5780~\AA\ were calculated using the 
IRAF\footnote[24]{IRAF is distributed by the National Optical Astronomy Observatory, 
which is operated by the Association of Universities for Research in Astronomy 
(AURA) under cooperative agreement with the National Science Foundation.} task
{\tt fitprofs} assuming a Gaussian profile of 2.1~\AA~FWHM, a typical
value in the Milky Way \citep{tuairisg00,welty06,hobbs08}\footnote[25]{Although
very high dispersion spectroscopy has shown that the profile of the 5780~\AA\ feature
is not Gaussian \citep{galazutdinov08}, the wavelength resolution and 
signal-to-noise ratio of our observations
do not warrant a more sophisticated method of determining the equivalent width.}.
This feature was selected since, among the DIBs,  it offers an optimal 
combination of strength and narrow line width.  Upper limits (3-$\sigma$) were 
calculated from signal-to-noise ratio measurements using an empirical relationship 
derived from the error estimates returned by {\tt fitprofs} for those objects where 
the 5780~\AA\ feature was detected.  In the case of SN~2009ig, a high signal-to-noise ratio
spectrum obtained with UVES on the ESO VLT was used to calculate a tighter upper limit
on the equivalent width of the host 5780~\AA\ absorption than allowed by our MIKE
data.

\begin{figure}
\epsscale{.7}
\plotone{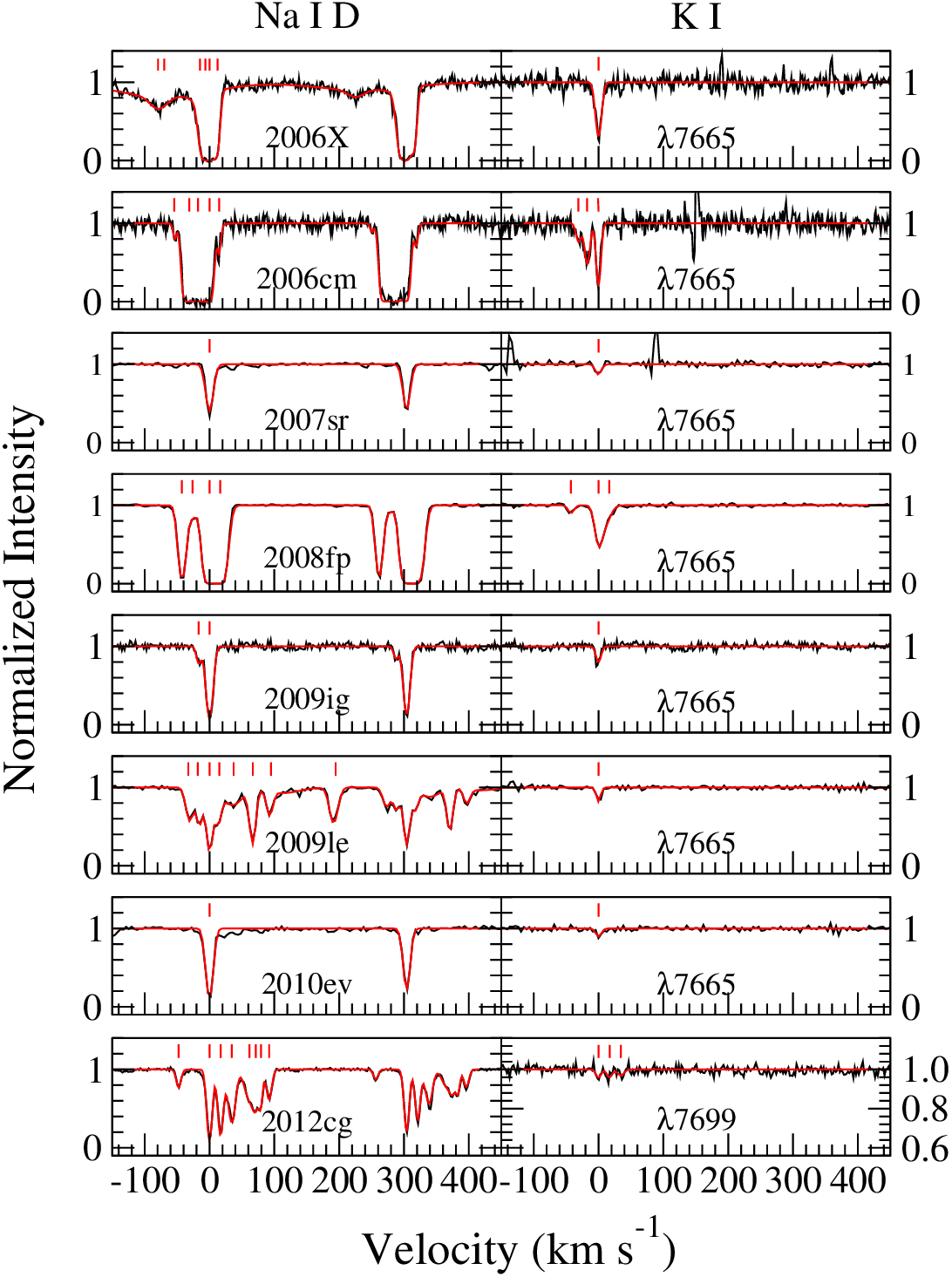}
\caption{VPFIT models for the \ion{Na}{1}~D and \ion{K}{1} absorption 
lines for those SNe~Ia where both features were detected. The \ion{Na}{1}~D
lines are shown in the left side of the figure.  In the right half the
\ion{K}{1} absorption is displayed.  For all but SN~2009le, the stronger
component of the \ion{K}{1} doublet, $\lambda7665$, is plotted.  In
the case of SN~2009le, the $\lambda7665$ line is blended with strong
telluric absorption, and so for this object the $\lambda7699$ transition
is shown. The red lines in each spectrum indicate the velocities of the 
components used to fit the profiles.  The most prominent (deepest) 
component is used to set the zero point of the velocity scale.}\label{fig:fig1}
\end{figure}

\begin{figure}
\epsscale{.7}
\plotone{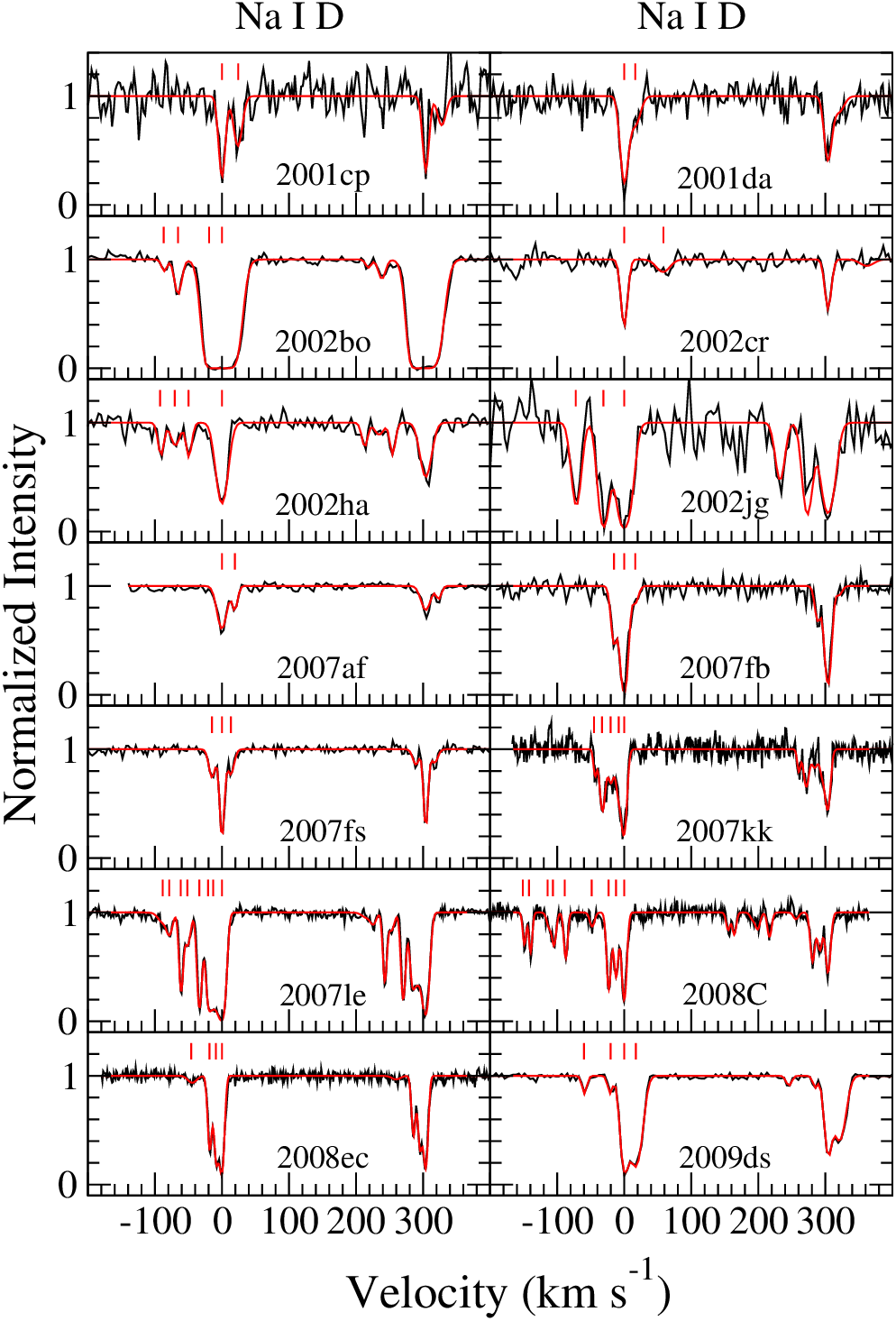}
\caption{VPFIT models for the \ion{Na}{1}~D absorption lines for SNe~Ia
where the \ion{K}{1} lines were not observed or detected. The red lines
above the D2 line in each spectrum indicate the velocities of the 
components used to fit the profiles.  The most prominent (deepest) 
component is used to set the zero point of the velocity scale.}\label{fig:fig2}
\end{figure}

Total column densities for the 46 objects in our Milky Way sample are 
given in Table~\ref{tab:tab1}.  Total column density and equivalent width measurements 
for the 32 SNe~Ia in the host absorption sample are found in Table~\ref{tab:tab3}.
Figure~\ref{fig:fig1} shows the VPFIT host absorption profile fits for the SNe where
both the  \ion{Na}{1}~D and \ion{K}{1} lines were detected, while Figure~\ref{fig:fig2} 
shows the fits of the host  \ion{Na}{1}~D lines for those
objects for which the \ion{K}{1} lines were not observed or detected.

\subsection{Dust Extinction}
\label{sec:extinction}

We adopt the \citet{schlafly11} re-calibration of the \citet{schlegel98} extinction 
map to estimate the Milky Way component of the visual extinction, $A_V$, for 
the objects in both our Milky Way and host absorption samples.  
Measuring the dust extinction produced in the SN 
host galaxy is a more complicated matter.  In general, SNe~Ia in 
low-reddening environments display intrinsic colors at maximum light that vary uniformly 
with the decline rate parameter, $\dm$, and these relations can be used to 
derive color excesses such as $E(B-V)$.  Fundamentally, however, it is $A_V$ 
that is proportional to the column density of the dust, and converting from 
$E(B-V)$ to $A_V$  requires knowledge of the ratio of total-to-selective extinction,
$R_V$.  In the Milky Way, the distribution of $R_V$ values is strongly peaked at 
$\sim$3 \citep[e.g., see][]{fitzpatrick07}, whereas the situation is not so clear for 
SNe~Ia in other galaxies.  Cosmological studies solving for an average value of  $R_V$ by 
minimizing the Hubble diagram dispersion inevitably yield values $\la 2$
\citep[e.g.,][]{astier06,conley07,kessler09}, 
whereas ratios of color excesses at optical and NIR wavelengths 
for individual SNe~Ia indicate that while most heavily-reddened events have 
$R_V < 2$, the color excesses of many SNe~Ia with $E(B-V) < 0.3$ 
are consistent with the standard Galactic value of $R_V \sim 3$ 
\citep{folatelli10,phillips12,burns13}.

In this study, we employ a Markov-Chain Monte Carlo (MCMC) code to 
simultaneously estimate most-likely values of $A_V$ and $R_V$ from 
optical and (when available) NIR light curves of the SNe~Ia listed in 
Table~\ref{tab:tab2}.  Sources for these data are given in 
columns 4 and 5 of the table.
Briefly, the code models the observed pseudo-colors\footnote{The pseudo-colors at
maximum are computed by taking the difference of the maxima of each filter's
light-curve.} of each SN as a combination
of an intrinsic color dependent on the decline rate parameter $\dm$ 
and a color excess due to dust extinction along the line-of-sight. The dust 
extinction, $A_{X}$, in any band, $X$, is modeled using the \citet{cardelli89} 
extinction law, which has two free parameters:  $A_V$ and $R_V$.  
Given these two parameters 
and the observed value of $\dm$, the observed pseudo-color at maximum,
$m_X-m_Y$, can be computed as:
\begin{equation}
   m_X - m_Y = P^{2}_{XY}\left(\dm - 1.1\right) + A_X\left(A_V,R_V\right) - A_Y\left(A_V,R_V\right)
\end{equation}
where $P^{2}_{XY}$ is a second-degree polynomial that describes the intrinsic 
$m_X-m_Y$ color at maximum of  SNe~Ia \citep[e.g., see Figure 2 of][]{phillips12}.  
Note that nowhere do we need to know the distance 
to the SN Ia as we are dealing with colors only. We determine the pseudo-colors
at maximum for both the training sample and the objects in this paper
using SNooPy \citep{Burns12} fits to the observed light-curves. For $N$
filters, there are only $N-1$ independent colors and we model the
set $B-X$, where $X=ugriVYJH$. Any other color can be derived as the sum
or difference of two of these colors. However, the errors in the colors
for a single object
are correlated as they all contain the same $B$ magnitude. These correlated
errors are accounted for in the MCMC code.

For objects that are significantly reddened ($E(B-V) \gtrsim 0.2$) and have full optical 
and NIR coverage, the value of $R_V$ can be constrained quite well. However, as the reddening 
decreases, the 
value of $R_V$ cannot be constrained so readily, and if we do not use a more 
restrictive prior, we cannot get an upper bound for $A_V$ (a lower bound of zero is 
implicitly assumed). This is also the situation when we do not have NIR coverage. 
We therefore adopt a flat prior for  $A_V$, while for $R_V$  a two-Gaussian prior
is employed consisting of a sharp component centered at $R_V = 2.23$ that carries 
most of the distribution, plus a smaller ($\simeq20$\%) contribution from a Gaussian centered
at $R_V = 3.27$ with very large spread.  This prior is expressed mathematically as:
\begin{equation}
p\left(R_V | A, \mu_1, \sigma_1, \mu_2, \sigma_2\right) \sim A \times e^{\left(-0.5 \times \left(R_V - \mu_1\right)^2/\sigma_1^2\right)} + \left(1-A\right) \times e^{\left(-0.5 \times \left(R_V - \mu_2\right)^2/\sigma_2^2\right)}
\end{equation}
where $A$, $\mu_1$, $\mu_2$, $\sigma_1$, and $\sigma_2$ are fixed at the following values 
derived from a training set of 74 SNe~Ia observed in the optical ($ugriBV$) by the
CSP, 54 of which also had NIR ($YJH$) coverage 
\citep[see][for further details]{burns13}:
\begin{equation}
A = 0.97; \mu_1 = 2.23; \sigma_1 = 0.4; \mu_2 = 3.27; \sigma_2 = 8
\end{equation}
The two-Gaussian prior was motivated by the observed distribution of values 
of $R_V$ derived from the Carnegie Supernova Project (CSP) \citep{hamuy06} 
sample of SNe~Ia, which is strongly peaked at $R_V \sim 2$, 
but has long tails to larger and smaller values of $R_V$.  One additional constraint 
on $R_V$ is that it be strictly positive.

For each color $m_X-m_Y$, we assume 
$P^{2}_{XY}$ is the same for all SNe~Ia, but that each SN has its own $A_V$ and 
$R_V$. Of course, $A_V$ and $R_V$ are highly covariant (indeed are 
completely degenerate for a single color).  We therefore require several 
{\em independent} colors to properly constrain $A_V$ and $R_V$. Having NIR 
colors is of great help because the relative extinction becomes quite insensitive to $R_V$ 
at longer wavelengths, allowing us to break the $A_V$/$R_V$ degeneracy.  In these
cases, the two-Gaussian prior on $R_V$ has little effect on the derived value of $R_V$.  
However, when there is no NIR photometry or $E(B-V)$ is very small, the prior on $R_V$ limits the possible
values of $A_V$.

Using MCMC techniques, we simultaneously solve for $R_V$ 
and $A_V$ for each SN using all independent colors available. 
We construct histograms for $R_V$ and $A_V$ by binning the Markov chains.
We then compute the mode and 
1-$\sigma$ errors by bracketing 34\% of the area to each side.  The final calculated 
values are given in columns 6 and 7 of Table~\ref{tab:tab2}. 
Where the posterior probability distribution is
significantly non-symmetric, we report upper and lower bounds.

\section{RESULTS}
\label{sec:results}

\subsection{\ion{Na}{1} and \ion{K}{1}}
\label{sec:NaK}

\begin{figure}
\epsscale{1.}
\plotone{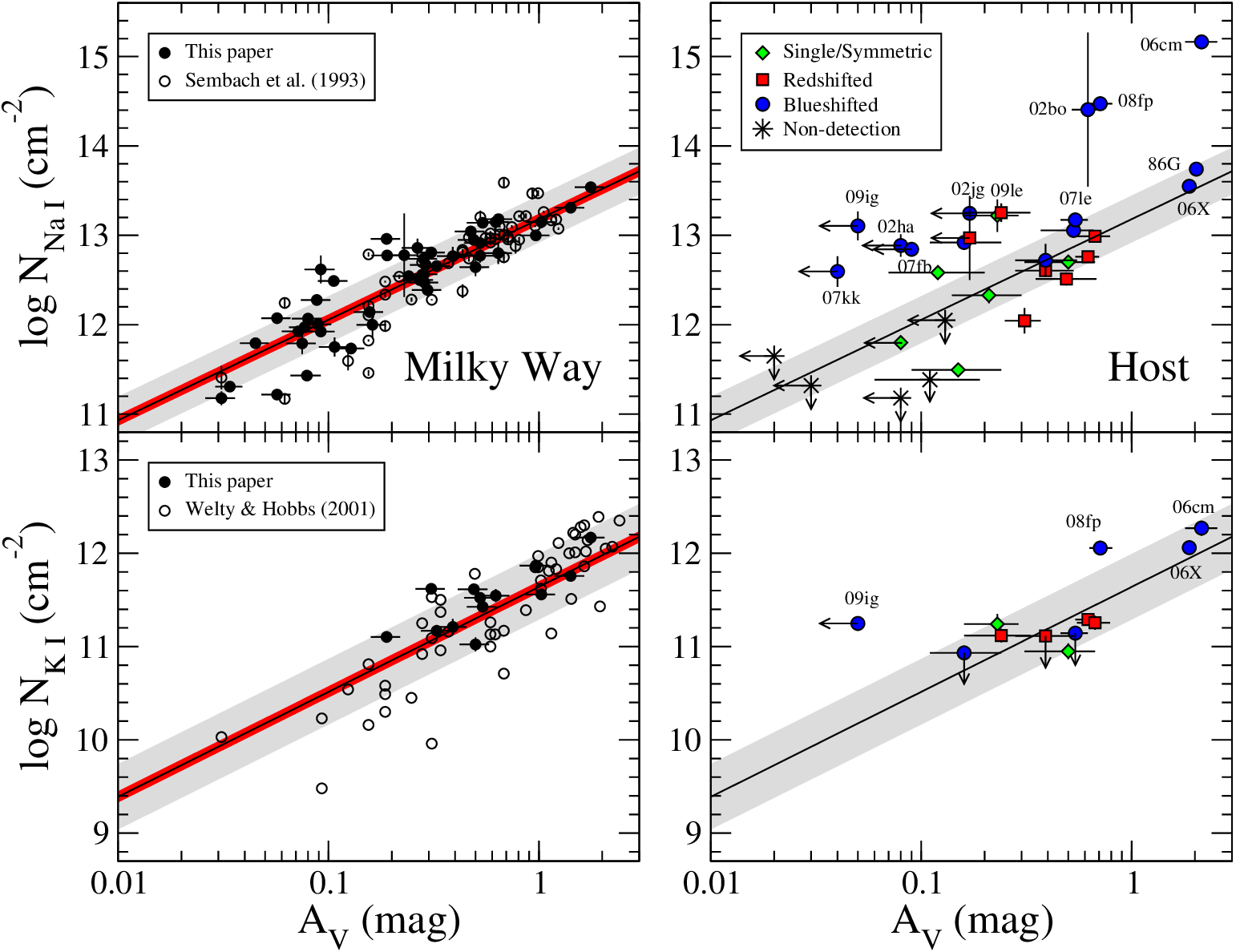}
\caption{(Left) Column densities of \ion{Na}{1} and \ion{K}{1} in the ISM
of the Milky Way plotted versus $A_V$.  Open circles are 
the \ion{Na}{1}~D measurements of \citet{sembach93} and the \ion{K}{1} data
of \citet{welty01}; solid circles in both plots show column densities 
for our Milky Way sample 
with the $A_V$ values taken from \citet{schlegel98} 
as rescaled by \citet{schlafly11}.  
The solid lines are fits to these combined data
(see text), with the gray shading corresponding to the 1-$\sigma$  dispersion.
%The dashed line is a fit to the data where the slope
%of the relation has been forced to a value of one.
The shaded red area illustrates the
uncertainty introduced by the 1-$\sigma$ dispersion in $R_V$ values
observed in the Milky Way \citep{fitzpatrick07};
(Right) Column densities of \ion{Na}{1} and \ion{K}{1} for
the host absorption sample are
plotted versus the $A_V$ values derived
from the SN colors.
The fits and 1-$\sigma$ dispersions observed
in the Milky Way are reproduced from the left half of the Figure.  
The different symbols used to plot the SNe~Ia
correspond to the \ion{Na}{1}~D profile classification scheme of
\citet{sternberg11}.
%The two open blue circles in both plots 
%correspond to the Na I and K I column densities of the absorption
%components closest in velocity to the DIB features at 5780~\AA\
%in SNe~2006cm and 2008fp.
}\label{fig:fig3}
\end{figure}

The upper-left panel of Figure~\ref{fig:fig3} displays total Galactic \ion{Na}{1} column 
densities for the 46 objects in our Milky Way sample plotted versus the $A_V$ values
inferred from the \citet{schlafly11} Galactic reddenings.  In calculating $A_V$, a value
of $R_V = 3.1$ is assumed.  The red shaded area in Figure~\ref{fig:fig3} illustrates the 
uncertainty in these $A_V$ values introduced by the 1-$\sigma$ dispersion of $\pm0.27$
in $R_V$ values observed in the Milky Way \citep{fitzpatrick07}.
Shown for comparison are column density measurements 
obtained from profile fits to the \ion{Na}{1}~D~lines by \citet{sembach93} for a 
sample of 50 distant (d~$>$~1~kpc) late-O and early-B stars in low-density 
regions of the Milky Way disk and halo.  A fit to the combined (Milky Way + \citet{sembach93})
sample of 96 measurements gives
\begin{equation}
\label{eq:mwnai}
\log{N_{Na~I}} = 13.180 (0.003) + 1.125 (0.005) \times \log{A_V}{\rm ,}
\end{equation}
where the uncertainties in the slope and intercept are given in parentheses.
This fit is plotted in the upper-left panel of Figure~\ref{fig:fig3},
with the gray shading indicating the 1-$\sigma$ dispersion in $\log{N_{Na~I}}$
of 0.26~dex.

In the lower-left panel of Figure~\ref{fig:fig3}, a similar plot of \ion{K}{1} column densities
versus $A_V$ is shown for the objects in our Milky Way sample for which a 
Galactic component 
of the \ion{K}{1} $\lambda\lambda$7665,~7699 doublet was detected.
The open circles are \ion{K}{1} column densities derived from the 
\ion{K}{1} $\lambda$7699 line for 52 stars in the Milky Way by \citet{welty01}.  The solid
line is a fit to the combined data, with the gray shading indicating the
1-$\sigma$ dispersion of 0.35~dex.  Since an essentially linear relationship 
exists in the Milky Way between $\log{N_{Na~I}}$ and $\log{N_{K~I}}$
 \citep{welty01}, the slope of 
the fit was set equal to the slope of the $\log{N_{Na~I}}$ vs. $\log{A_V}$ relation in
the upper-left panel of Figure~\ref{fig:fig3}, yielding
\begin{equation}
\log{N_{K~I}} = 11.639 (0.005) + 1.125 \times \log{A_V}{\rm .}
\end{equation}

As the left half of Figure~\ref{fig:fig3} shows, the measurements of the
Galactic $N_{Na~I}$ and $N_{K~I}$ values derived from our Milky Way sample 
are consistent with the fits to the stellar data.  Note that the 
dispersions in the stellar relations are large, and if one
were to use the fits to estimate $A_V$ from a measurement of the \ion{Na}{1} or
 \ion{K}{1} column density, the error  would be 54\% or 72\% of $A_V$ itself,
 respectively.
 
 The right half of Figure~\ref{fig:fig3} shows the same plots, but this time 
 for the host absorption components of the \ion{Na}{1} and \ion{K}{1} lines.  The fits and
 1-$\sigma$ dispersions corresponding to the Milky Way relations are 
 reproduced from the left half of Figure~\ref{fig:fig3} for comparison. The symbols used to 
 plot the SNe~Ia measurements follow the classification scheme based on the structure 
 of the \ion{Na}{1}~D profile employed by \citet{sternberg11}, with the three
 categories defined as follows:
 \begin{itemize}
 \item Blueshifted: One prominent absorption feature with weaker features at shorter 
 wavelengths with respect to it.
\item Redshifted: One prominent absorption feature with weaker features at longer 
 wavelengths with respect to it.
\item Single/Symmetric: A single absorption feature, or several features with both 
 blue and redshifted structures of similar magnitude.
 \end{itemize}

Figure~\ref{fig:fig3} shows that only 1 of 96 objects in the Milky Way + \citet{sembach93}
sample has a \ion{Na}{1} column density that lies $> 3$-$\sigma$ from the fit
defined by equation~\ref{eq:mwnai}, consistent with the 
statistical expectation.  However, 8 of 27 (30\%) of the SNe~Ia host galaxy
\ion{Na}{1} column densities 
fall $> 3$-$\sigma$ above the Galactic fit.  
Even more striking is the fact that all eight of these deviant SNe~Ia have
``Blueshifted''  host \ion{Na}{1}~D absorption profiles.  
Some of these (e.g., 
SNe~2007kk and 2009ig) are objects with significant \ion{Na}{1} column 
densities, but for which the optical and NIR colors imply quite small 
amounts of dust extinction.  Note that one of the ``Blueshifted'' SNe, 2006cm, was 
originally classified by \citet{sternberg11} as ``Single/Symmetric''.  The 
\ion{Na}{1}~D1 and \ion{K}{1} $\lambda$7665 profiles for 
this SN are displayed in Figure~\ref{fig:fig4}.  The  D1 line is 
highly saturated making it difficult to categorize, but the much weaker \ion{K}{1} 
$\lambda$7665 absorption reveals that SN~2006cm should actually be classified 
as ``Blueshifted''.

In our sample, there are 14 ``Blueshifted'', 7 ``Redshifted'', and 6 ``Single/Symmetric''
SNe.  The \citet{sternberg11} sample had very similar numbers --- 15, 6, and 6 SNe, 
respectively, in these three categories --- and so our sample should be representative
of theirs.
The probability in a random draw that all eight of the SNe~Ia in Figure~\ref{fig:fig3} 
lying more than 3-$\sigma$ from the Milky Way relationship would belong to the 
``Blueshifted'' class is 0.14\% and, therefore, highly unlikely\footnote[26]
{It might be argued that the ``Single/Symmetric'' SN~1994D and the
``Redshifted'' SN~2001cp, for which we are
only able to place an upper limit on $A_V$, should also be included in this calculation.
Doing so increases the probability of a chance occurrence to $\sim$1.5\%}.

%Of the 12 SNe~Ia in Figure~\ref{fig:fig3} that fall more than 3-$\sigma$ above the
%Milky Way relationship, 9 have host \ion{Na}{1}~D absorption profiles belonging to 
%the ``Blueshifted'' class, 2 correspond to the ``Redshifted'' morphology, and 1 to
%the ``Single/Symmetric''.  The
%probability of this occurring by chance in a random draw is 
%1.4\%\footnote[26]{It might be argued that the ``Single/Symmetric'' SN~1994D, which
%lies on the Galactic $N_{Na~I}$ versus $A_V$ relation but for which we are
%only able to place an upper limit for $A_V$, should also be included in this calculation.
%Doing so increases the probability of a chance occurrence to 3.1\%}. 
%Eight
%SNe~Ia lie more than 4-$\sigma$ above the Milky Way relationship and all belong to 
%the ``Blueshifted'' class, an occurrence that has only a 0.14\% likelihood of being
%due to chance.  If the ``Redshifted'' SN~2001cp is included since our measurement
%of $A_V$ is only an upper limit, and hence it could have host \ion{Na}{1}~D absorption
%that is 4-$\sigma$ above the Milky Way relationship, the probability of this
%being a chance event is still very low (0.43\%).

Five of the SNe~Ia in our sample (2007hj, 2007on, 2008hv, 2008ia, and 
SNF20080514-002) did not show detectable host \ion{Na}{1}~D absorption 
in their spectra.  The optical and NIR colors of these objects, all of which 
occurred in E or S0 host galaxies, identify them as having suffered little or no
dust reddening, consistent with the absence of detectable \ion{Na}{1}~D lines.
Upper limits (3-$\sigma$) on $\log{N_{Na~I}}$ for these non-detections are 
given in Table~\ref{tab:tab3} and plotted in Figure~\ref{fig:fig3}. 

Due to the weakness of the \ion{K}{1} $\lambda\lambda$7665,~7699 doublet, 
we were able to measure host \ion{K}{1} column densities for
only eight of the SNe~Ia in our host absorption sample.  Upper limits (3-$\sigma$) were measured
for an additional three SNe with high signal-to-noise ratio spectra.  
These measurements, along with the host  \ion{K}{1} 
column density for SN~2001el published by \citet{sollerman05}, are given in 
Table~\ref{tab:tab3} and plotted in the lower-right panel 
of Figure~\ref{fig:fig3}.  Interestingly, of the three ``Blueshifted'' objects that lie more than
5-$\sigma$ above the Galactic $N_{Na~I}$ vs. $A_V$ relationship ---
SNe~2006cm, 2008fp and 2009ig --- only SN~2009ig falls 
significantly above the corresponding $N_{K~I}$ vs. $A_V$ relationship.

\subsection{DIB at 5780~\AA}

Although the carriers of DIBs remain unknown \citep{sarre06}, 
they are ubiquitously present in the ISM of the Milky Way and beyond --- from the 
Magellanic Clouds, M31, and M33, to starburst galaxies, SNe host galaxies and 
even damped Ly$\alpha$ systems \citep[e.g.,][]{cox08a,heckman00,sollerman05,york06}.
The upper-left panel of Figure~\ref{fig:fig5} displays the equivalent width of the
DIB at 5780~\AA\ plotted as a function of $A_V$ for 131 
late-O and early-B stars in the Milky Way \citep{friedman11} (open circles), and
for the 12 objects in our Milky Way sample for which we were able to detect a Galactic
component of the 5780~\AA\ DIB absorption feature (solid circles).  Previous studies
\citep[e.g.,][]{herbig93,welty06,friedman11,vos11,yuan12} have found that the strength of the 5780~\AA\ 
feature is essentially linearly proportional to
the amount of reddening, and so we assume this in deriving the following 
fit to these combined data:
\begin{equation}
\label{eq:mwdib}
\log{EW\left(5780\right)} = 2.283 (0.001) + \log{A_V}{\rm .}
\end{equation}
The black line in the upper-left panel of Figure~\ref{fig:fig5}
shows this fit, which is in excellent agreement with that recently obtained 
by \citet{yuan12} for a much larger sample of stars.  The gray shading indicates 
the 1-$\sigma$ dispersion of 0.22 dex that
%, which also is consistent with the scatter in the \citet{yuan12} data. 
translates to a 50\% error in $A_V$ 
if the 5780~\AA\ feature is used to estimate the dust extinction for any single
object.
%The solid circles in Figure~\ref{fig:fig5} correspond to the objects in
%our Milky Way sample for which we were able to detect a Galactic
%component of the 5780~\AA\ DIB absorption feature.  These equivalent widths
%fit nicely within the dispersion of the stellar measurements.

As shown in the lower-left panel of Figure~\ref{fig:fig5}, 
the strength of the 5780~\AA\ feature also correlates reasonably well with
the column density of \ion{Na}{1} in the Milky Way \citep{herbig93,welty06}.  
The open circles are
DIB 5780~\AA\ equivalent widths for 55 early-type stars from \citet{friedman11}
combined with \ion{Na}{1} column densities measured from the D~lines 
for the same stars by \citet{welsh10}; the solid circles in the figure correspond
to the measurements of the Galactic 5780~\AA\ feature in our Milky Way sample. 
Equations~\ref{eq:mwnai} and \ref{eq:mwdib}
yield
\begin{equation}
\log{EW\left(5780\right)} = -9.433 (0.004) + 0.889 (0.004)\times \log{N_{Na~I}}{\rm ,}
\end{equation}
which provides a reasonable fit to the observations, albeit with a large dispersion of 0.44 dex.   

The right half of Figure~\ref{fig:fig5} shows the same plots  for 
the host absorption component of the DIB at 5780~\AA.  This feature
was detected in 34\% (11 of 32) of the SNe~Ia in our sample.  Upper limits
(3-$\sigma$) are plotted for an additional eight objects.  
Figure~\ref{fig:fig6} illustrates three examples of detections (SNe~2006cm,
2008fp, and 2007sr), and one where only an upper limit could be
measured (SN~2009ig).
The upper-right panel of Figure~\ref{fig:fig5}
indicates that in all cases where the 5780~\AA\ feature was detected, 
its strength is compatible with the dust extinction
implied by the SN colors. In general, the upper limits are also consistent with 
the Milky Way relation, with the exception of SN~2006X for which 
\citet{cox08} found the DIBs at 6196~\AA\ and 6283~\AA\ to be 2.5--3.5 times
weaker than expected.  Unfortunately, the echelle spectra obtained by these
authors did not cover the 5780~\AA\ feature, and the 
signal-to-noise ratio of our own observation does not allow a tight upper 
limit to be measured.  Based on the relations given by \citet{friedman11}, the
\citet{cox08} measurements of the equivalent widths of the 6196~\AA\ and 6283~\AA\ 
features imply $EW$~5780~\AA\ $=$ 50--160~m\AA, consistent with
our upper limit.

The lower-right panel in Figure~\ref{fig:fig5} shows that the correlation between the strength of the 
5780~\AA\ feature and the \ion{Na}{1} column density is particularly poor for the ``Blueshifted'' 
SNe~Ia. 
As in Figure~\ref{fig:fig3}, SNe~2002bo, 2006cm, and 2008fp, and 2009ig have values of $N_{Na~I}$ 
that are an order of magnitude or more greater than predicted by the Milky Way relationship.
While the discrepancy of these four SNe~Ia in Figure~\ref{fig:fig3} could perhaps be attributed to 
large underestimates of $A_V$ from the SN colors, the fact that errors in $A_V$ have no 
effect on the points in the lower-right panel of  Figure~\ref{fig:fig5} argues against this explanation.  We 
conclude, therefore, that the anomalously large \ion{Na}{1} column densities in these SNe~Ia
is a real effect.

\section{DISCUSSION}
\label{sec:discussion}

The observations presented in \S\ref{sec:results} can be summarized as follows:

 \begin{itemize}
 
 \item One-fourth of the SNe~Ia in our sample (8 of 32) exhibit 
 host \ion{Na}{1}~D absorption that is more than 3-$\sigma$
 stronger than expected based on the amount of dust extinction
 implied by the SN colors and the Galactic relationship between 
 $N_{Na~I}$ and $A_V$.  Nearly all of these SNe~Ia 
 with anomalously large \ion{Na}{1} column densities belong to the ``Blueshifted'' class of 
 \citet{sternberg11}.
 
 \item The relationship between the host \ion{K}{1} absorption column densities and $A_V$ 
 for the SNe~Ia appear to be more consistent with 
 the Milky Way relationship.  While the number of detections
 is small and this result therefore requires verification, it is interesting to 
 note that two of the ``Blueshifted'' objects with highly-discrepant \ion{Na}{1} 
 column densities, SNe~2006cm and 2008fp, showed essentially normal 
 \ion{K}{1} absorption for the dust extinction implied by their colors.
 
 \item When detected in SNe~Ia, the strength 
 of the host absorption DIB 5780~\AA\ feature is consistent with the amount of dust extinction implied by 
 the SN colors.  Notably, this statement
 applies to the ``Blueshifted'' SNe~2002bo, 2006cm, 2008fp, and 2009ig, all of which
 displayed anomalously large \ion{Na}{1} column densities.  
%% The fact that these same three objects fall significantly off the Galactic 
%% $EW$~5780~\AA\ versus $N_{Na~I}$ relation, implies that the exceptionally strong
%% \ion{Na}{1}~D absorption is not due to an erroneous estimate of the host dust extinction 
%% based on the SN colors. 
 
\end{itemize}

Figure~\ref{fig:fig7} illustrates just how unusually strong the
 \ion{Na}{1}~D absorption is for two SNe with ``Blueshifted'' host 
absorption profiles.  In the upper-left panel, the Milky Way 
component of  the  \ion{Na}{1}~D lines in the spectrum of SN~2009le is
plotted.  The Galactic extinction for this object 
%on the scale of  \citet{schlafly11}
is $A_V = 0.05 \pm 0.01$~mag.  In the upper-right panel,
the host absorption component of the D~lines in the spectrum of the
``Blueshifted''  SN~2009ig is shown for comparison.  Analysis of the 
light curves of SN~2009ig  yields an upper limit for the host extinction of 
$A_V < 0.05$~mag, yet the D~lines are clearly much stronger than the 
Milky Way absorption in SN~2009le.  The lower half of Figure~\ref{fig:fig7} 
shows a similar comparison of the Milky Way component of the \ion{Na}{1} 
absorption in the spectrum of SN~2011ek [$A_V \left({\rm Galactic}\right) = 
0.97 \pm 0.15$~mag], which is contrasted with the host absorption component of 
the D~lines in the spectrum of the ``Blueshifted'' SN~2008fp 
[$A_V \left({\rm Host}\right) = 0.71^{+0.10}_{-0.08}$~mag].  Note the strongly
saturated line profiles in the spectrum of the latter object, whereas the \ion{Na}{1} lines 
produced by the interstellar medium of the Milky Way in SN~2011ek
are considerably weaker despite the extinction being nominally somewhat
greater.  These two comparisons demonstrate visually that the anomalous strength of
the host  \ion{Na}{1}~D absorption in these two SNe is very real, and not
due to inaccuracies in measuring column densities of saturated lines.

\subsection{\ion{Na}{1}, \ion{K}{1}, and DIBs in the Local Group and Beyond}
\label{sec:mc}

In order to understand if systematic differences in the sodium and potassium 
abundances might occur due to potential metallicity effects in supernova host 
galaxies, we can examine similar trends for the Large and Small Magellanic Clouds 
(LMC and SMC) --- two nearby galaxies which have significantly lower metallicities, 
by factors of 2 and 5, respectively, compared to the Milky Way
\citep{dufour84,rolleston02,peimbert76,bouret03}. 
The observed \ion{Na}{1} and \ion{K}{1} column densities of the diffuse ISM in both the LMC
and SMC \citep{cox06,cox07,welty06} are plotted as a function of $A_V$ in the left half
of Figure~\ref{fig:fig8}.  When only color excesses are provided by these authors, visual extinctions were 
computed using the average $R_V$ values given by \citet{gordon03}.  The measurements
clearly reproduce the Galactic trends shown in Figure~\ref{fig:fig3}. 
Despite marked environmental differences such as the higher gas-to-dust ratio and 
stronger UV radiation field of the LMC, both $N_{Na~I}$ and $N_{K~I}$ follow the Galactic trend 
although the scatter in $N_{Na~I}$ is somewhat larger.  The \ion{Na}{1} and \ion{K}{1}
column densities are also highly correlated in the Magellanic Clouds, with the ratio being 
similar to that found in the ISM of the Milky Way (see \S\ref{sec:ki}).

In this comparison it is important to understand that, due to the lower metallicity and 
stronger radiation fields, both $N_{Na~I}$ and $N_{K~I}$ are lower with respect to $N_H$ 
in the Magellanic Clouds.  However, the gas-to-dust ratio $N_H / E(B-V)$, which scales
approximately with metallicity, is also higher by a similar factor in the Magellanic Clouds 
\citep{welty06}.  Both effects essentially cancel, resulting in similar correlations between 
$N_{Na~I}$, $N_{K~I}$, and $A_V$ for both the Magellanic Clouds and the Milky Way. 
We conclude that it is unlikely that the enhanced \ion{Na}{1} 
absorption observed for some SNe~Ia is due to global
host galaxy metallicity differences.

The panels on the right in Figure~\ref{fig:fig8} display the observed 5780~\AA\ DIB 
equivalent width versus $A_V$ (top) and the \ion{Na}{1} column density (bottom), 
respectively. In both cases, the equivalent width measurements of the 5780~\AA\ feature
fall, on average, below the Galactic relation by a factor of $\sim$2, 
consistent with the findings of \citet{welty06}. 
As discussed by these authors, this is likely a consequence of lower metallicity, lower 
dust-to-gas ratios and molecular fractions, and the generally stronger radiation 
fields in the Magellanic Clouds.
The similar behaviour
of the 5780~\AA\ DIB strength with respect to both $N_{Na~I}$ and $A_V$ is consistent 
with the good correlation between the latter two quantities (see the upper left panel of 
Figure~\ref{fig:fig8}). This is notably dissimilar to the situation for the SNe~Ia, 
where the 5780~\AA\ DIB strengths follow the Galactic EW(5780) vs $A_V$ relation, 
but do not follow the Galactic EW(5780) vs $N_{Na~I}$ relation.

Unfortunately, recent determinations of \ion{Na}{1} column densities for 
extra-Galactic sight-lines towards M31 and M33 suffer from saturation due to 
lower spectral resolution observations \citep{cordiner11}; accurate comparisons 
for these galaxies will require more sensitive high-resolution spectroscopic data. 
The measured 5780~\AA\ DIB strengths versus $E(B-V)$ in M31 are slightly above 
the Milky Way trend, but the sample is limited and prone to observational 
bias in preferentially detecting the strongest bands.

Beyond the Local Group, moderate-resolution observations of the 
\ion{Na}{1}~D lines and several DIBs, including the 5780~\AA\ feature, in 
starburst galaxies indicate that the equivalent widths follow the same 
dependence as Galactic DIBs on $E(B-V )$ and the \ion{Na}{1} column 
density \citep{heckman00}.  The 5780~\AA\ feature was detected by
\citet{york06} in the $z = 0.5$ damped Ly$\alpha$ system of the 
QSO AO~0235+164  and also found to be consistent with the Milky Way
EW(5780) vs. $E(B-V)$ relation.  These authors also detected the
\ion{Na}{1}~D2 line in this same system at an equivalent width of 
$\sim$0.8~\AA\ .  As discussed in the next section, equivalent widths 
are not as reliable as column densities for inferring dust extinction, but 
this value is not inconsistent with the reddening of $E(B-V) = 0.23$ derived 
for the QSO \citep{junkkarinen04} assuming the correlation observed in
the Milky Way.

\citet{sparks97} detected extended interstellar \ion{Na}{1}~D absorption lines
from gas associated with the compact emission filament system and dust lane 
in the central 20~arcsec of the dominant elliptical galaxy in the Centaurus 
cluster, NGC~4696.  Due to the low resolution of their spectral observations,
only total equivalent widths of the \ion{Na}{1} absorption could be measured.
Interestingly, the correlation between equivalent width and $E(B-V)$ was
found to be significantly steeper than the standard Milky Way relation, 
suggesting a low dust-to-gas ratio.  This is one of the few cases we are
aware of in the literature where enhanced \ion{Na}{1} absorption is observed in the ISM
of a galaxy, although it should be emphasized that the environment is quite 
different from the disks of the spiral galaxy hosts of the majority of the SNe~Ia 
in our sample.

\subsection{Equivalent Width of Na~I~D as an Indicator of Extinction}
\label{sec:ew}

The equivalent  width of the \ion{Na}{1}~D lines, EW(Na~I~D) is commonly 
measured in low-dispersion spectra to estimate the dust reddening 
for SNe~Ia with poor photometric coverage, or peculiar SNe whose intrinsic colors 
are unknown \citep[see][and references therein]{turatto03}.  It is interesting to
compare the results for the \ion{Na}{1}~D host  absorption column densities 
presented in \S\ref{sec:results} with what equivalent widths give.  This is shown 
in Figure~\ref{fig:fig9} where the equivalent widths of the D~lines measured 
from our echelle spectra are plotted versus $A_V$.
Values for the
Milky Way sample, augmented by high dispersion measurements of 82 stars 
\citep{sembach93,munari97} and 30 QSOs \citep{poznanski12},
are plotted as asterisks.
The dashed line corresponds to the relation derived by 
\citet{poznanski12} based on 117 moderate-to-high-resolution observations
of QSOs and nearly a million low-resolution Sloan Digital Sky Survey spectra
of galaxies and QSOs.  This line provides a reasonably adequate
representation of the Milky Way EW(Na~I~D) measurements, but the rms dispersion 
of 0.08~dex in $\log{A_V}$ quoted by these authors and shown in gray in
Figure~\ref{fig:fig9} is clearly underestimated.  Presumably this is because 
the points used by Poznanski et al. to derive the final relations and dispersion 
(see Figure~9 of their paper) were obtained by averaging individual
measurements of very large numbers of objects.  
Also plotted in
Figure~\ref{fig:fig9} is the curve given by \citet{munari97} for 
single-absorption components of the D1 line.  As
noted by \citet{poznanski12}, this relation provides an equally good
fit when scaled to match the observations.  The dispersion in $A_V$ 
with respect to this latter relation is 0.30~dex, corresponding to a 68\% 
error in $A_V$ for any single object rather than the $\sim$20\% error 
implied by \citet{poznanski12}.  It is also clear from Figure~\ref{fig:fig9} 
that the equivalent  width of the \ion{Na}{1}~D lines becomes 
insensitive to estimating the dust extinction for 
EW(Na~I~D)~$\gtrsim~1$~\AA\
\citep[see also][]{munari97}.

Turning to the host absorption EW(Na~I~D) measurements, 
Figure~\ref{fig:fig9} reveals an effect similar to that seen in the column densities
plotted in Figure~\ref{fig:fig3}, with a significant number of the points lying 
more than 3-$\sigma$ above the Milky Way relationship and the majority
of these belonging to the ``Blueshifted'' class.  As more SNe~Ia with 
good photometric and spectroscopic coverage have been observed, it has 
become clear that  the EW(Na~I~D) is {\em not} a reliable 
indicator of the host dust extinction of SNe~Ia \citep{blondin09,folatelli10,poznanski11}.  
Figure~\ref{fig:fig3} shows that the reason for this 
poor correlation is the relatively large fraction of SNe~Ia with strong 
\ion{Na}{1}~D absorption that is an order of magnitude greater in column 
density than expected from the SN~Ia colors and the Milky Way relation.  
Indeed, based on our 
high-dispersion observations, it would seem that the only reliable deduction 
that can be made using either column densities or equivalent widths is
that the {\em absence} of detectable  \ion{Na}{1} absorption in a high
signal-to-noise ratio spectrum is consistent with 
low dust extinction.

Note that in some cases equivalent widths and column densities give 
quite different answers.  An example is the case of SN~1986G 
and SN~2006cm.  While both of these events suffered a similar amount of 
dust extinction, the host absorption \ion{Na}{1} column density for SN~1986G 
was fully consistent with the Milky Way $\log{N_{Na~I}}$ vs. $A_V$ relation, 
whereas for SN~2006cm it was greater by $\sim$1.2~dex (see 
Figure~\ref{fig:fig3}).  Nevertheless, Figure~\ref{fig:fig9} shows that the 
EW(Na~I~D) for SN~2006cm was significantly {\em smaller} than that 
of SN~1986G.  The explanation for this apparent inconsistency is found in 
Figure~\ref{fig:fig10}, where the \ion{Na}{1} column density is plotted versus 
the equivalent width of the D~lines for both the Milky Way and host 
absorption samples.  Curves of growth for three different values 
of the Doppler parameter, $b$, are also displayed in this figure.  These are 
shown for both single and multiple absorption components with the
same Doppler parameter, with the 
single-component cases corresponding to the left-most curve for each value 
of $b$.  Note that the curve for a single component with $b = 8$~\kms\ is 
nearly identical to the curves for two components with $b = 4$~\kms\ or 
four components with $b = 2$~\kms.  This ambiguity is inherent to equivalent 
widths, and explains why column densities are to be preferred whenever possible.

Returning to SN~1986G, the profile of the host absorption D~lines consisted 
of five major components covering a wide swath ($\sim$200~\kms) of the 
spectrum \citep{dodorico89}.  The two strongest components had Doppler 
parameters of 10~\kms\ and column densities $\log{N_{Na~I}}$ of 13.4 and 
13.1, respectively.  The third-strongest component also had
$\log{N_{Na~I}} = 13.1$, but with $b = 7$~\kms\, and the two weakest 
both had $b \sim 5$~\kms, with column densities of 12.6 and 11.7.
Reading off the equivalent widths corresponding to these individual 
components from Figure~\ref{fig:fig10} and summing them yields a
total equivalent width EW(Na~I~D) very close to the
observed value of 3.8~\AA.  In the case of SN~2006cm,
the \ion{Na}{1}~D profile was dominated by three major components, but
concentrated into a considerably narrower ($\sim$60~\kms) portion of the 
spectrum.  The two strongest components had $b = 5$~\kms\ and 
$\log{N_{Na~I}}$ of 15.0 and 14.5, respectively, while the third component
was slightly narrower ($b \sim 4$~\kms) with a column density of 13.7. 
Summing the corresponding equivalent widths using Figure~\ref{fig:fig10}
gives a total equivalent width close to, but slightly larger than, the observed 
value, with the discrepancy accounted for by the significant overlapping 
in velocity space of the absorption components.

\subsection{Relation to Previous Studies of Na~I~D in SNe~Ia}
\label{sec:relation}

The fact that nearly all of the SNe~Ia showing
anomalously large \ion{Na}{1} column densities belong to the \citet{sternberg11} 
 ``Blueshifted'' class suggests an association with CSM.
 \citet{sternberg11} estimated that a quarter
 to a third of local SNe~Ia with detectable \ion{Na}{1}~D absorption show
 evidence of outflowing CSM, which is remarkably similar to the percentage
 of objects with anomalously large \ion{Na}{1} column densities that we find in this study.
 It should be kept in mind, however,  that the \citet{sternberg11} classifications are valid
 in a statistical sense only. When considering single objects, it is 
 impossible to ascertain the systemic velocity of the progenitor and thus
 ascribe any particular absorption component to either inflow or outflow.
 Rather, it is the unexplained surplus of SNe
 with blueshifted absorption features
 that implies an association with 
 some kind of outflow.
 Note that the strong preference observed for blueshifted features compared
 to non-blueshifted ones does not depend on whether the velocity zero point
 is taken to be that of the strongest absorption component or if, alternatively, 
 it is defined with
 respect to emission or absorption lines in the SN spectrum due to the host
 galaxy or using the radial velocity of the host \citep{sternberg11,maguire13}.
 %Hence, the unusually strong D~line absorption observed for the "Redshifted" 
 %SN~2009le, for example, could possibly be produced by an outflowing CSM, depending
 %on the actual velocity of the progenitor system.
 
 \citet{foley12b} found that SNe~Ia of
 the ``Blueshifted'' class have systematically higher ejecta velocities 
 at maximum brightness relative to the rest of the SN~Ia population,
 suggesting a link between the SN progenitor system and the explosion properties.
 \citet{wang13} recently presented evidence that high-velocity SNe~Ia, defined as
 having $v({\rm Si~II}~\lambda6355) \geq 12,000$~\kms\ at $B$ maximum, are more
 concentrated in the inner and brighter regions of their host galaxies, and that
 they tend to occur in more-luminous hosts than do normal-velocity events.  They
 suggest that high-velocity SNe represent a subtype of the Type~Ia class originating 
 from younger, more metal-rich progenitor systems.  The left half of 
 Figure~\ref{fig:fig11} shows 
 plots of $\log{N_{Na~I}}$ vs. $A_V$ for the SNe~Ia in our host absorption 
 sample split into the high-velocity and normal-velocity subsamples.
 Taken as a group, the high-velocity events deviate significantly from 
 the Milky Way relation, and 70\% of them belong to the ``Blueshifted'' class.  
 The median of the absolute deviations of the
 high-velocity SNe~Ia with respect to the Milky Way relation 
 is 0.88~dex in column density, whereas it
 is 0.29~dex for the normal-velocity subsample.  Nevertheless, the 
 correlation is imperfect, with four of the most deviant ``Blueshifted''
 SNe (2002ha, 2006cm, 2007fb, and 2008fp) belonging to the 
 normal-velocity subsample.
 
 Our observations are also relevant to a recent paper by \citet{forster12} who 
 found an apparent correlation between the late-phase nebular velocity 
 shift, $v_{neb}$, and the equivalent width of the narrow host \ion{Na}{1}~D 
 absorption as measured in low-dispersion spectra.  The sense of this 
 dependence was that SNe with $v_{neb} > 0$ have generally stronger 
 \ion{Na}{1}~D lines than those with $v_{neb} < 0$.  \citet{maeda10,maeda11}
 had previously found a correlation of both ejection velocity and peak $B-V$ 
 color with $v_{neb}$, which they interpreted as a viewing angle effect of 
 off-center explosions.  \citet{forster12} argued that the color effect might 
 alternatively be due to dust mixed with previously-existing CSM, thus
 accounting for not only the stronger \ion{Na}{1}~D lines in SNe
 with $v_{neb} > 0$, but also the redder colors of these objects.  Nevertheless,
 our discovery that SNe with ``Blueshifted'' \ion{Na}{1}~D profiles 
 often show anomalously large \ion{Na}{1} column densities taken together with the 
 \citet{foley12b} finding that ``Blueshifted'' SNe~Ia have systematically 
 higher ejecta velocities and redder colors at  maximum provides an alternative
 explanation to the EW(Na~I~D) vs. $v_{neb}$ correlation that does 
 not require a dusty CSM.
 
 Recently, \citet{maguire13} presented single-epoch, intermediate-resolution 
 ($\lambda/\Delta\lambda \sim 18,200$) spectral observations for a sample of 17 SNe~Ia,
 which they then combined with the sample of high-resolution spectra published by
 \citet{sternberg11}.  These authors confirmed the \citet{sternberg11} finding that 
 an excess of events showed blueshifted structure in the \ion{Na}{1}~D 
lines, and also pointed out that the strength of the \ion{Na}{1}~D 
absorption was stronger in SNe~Ia displaying blueshifted profiles than in those 
without such structure, in accordance with our own findings.  In addition,
\citet{maguire13} found that the strength of the blueshifted subcomponent of the 
D~lines for those objects with blueshifted profiles was 
correlated with the $(B-V)$ colors of the SNe at maximum light.
These findings were interpreted as being strongly suggestive of absorbing material 
(CSM) near the SN. \citet{maguire13} went on to speculate that SNe~Ia with
strong, blueshifted absorption may represent a distinct population of events 
produced preferentially in late-type galaxies.

In the upper right half of Figure~\ref{fig:fig11}, the column densities of the most 
prominent (deepest) component of the host \ion{Na}{1}~D absorption for
each ``Blueshifted'' SN in our sample are plotted versus the values of $A_V$ 
inferred from the SN colors.  The column densities were derived from the VPFIT 
results shown in Figures~\ref{fig:fig1} and \ref{fig:fig2}, and correspond to the 
zero velocity components.  Following \citet{maguire13}, we also plot in the lower
right half of Figure~\ref{fig:fig11} the sum of the column densities of the absorption
components lying blueward of the most prominent component, again for the
 ``Blueshifted'' SNe in our sample.\footnote[27]{Note that rather than using
 the most prominent absorption component to define zero velocity, \citet{maguire13} 
 employed galaxy emission or absorption lines observed in the SN Ia spectrum or, 
 lacking this, the recessional velocity of the host galaxy.
 While both approaches have their merits, they do not result in
 significant differences in defining the blueshifted absorption components.}
 In both cases, a general correlation is 
 observed, but with considerable scatter.  To quantify the goodness of fit to 
 the Galactic $N_{Na~I}$ vs. $A_V$ relationship (equation~\ref{eq:mwnai}), 
 we calculate the chi-squared per degree of freedom, 
 $\chi_\nu^2$, for several different samples in 
 Table~\ref{tab:tab4}.  In the case of the SNe, the calculation is done both 
 including the objects with upper limits for $A_V$, and excluding them.  Note 
 that all of the $\chi_\nu^2$ values are large --- even for the 
 \citet{sembach93} + Milky Way 
 sample that was used to derive equation~\ref{eq:mwnai} --- reflecting 
 the large intrinsic dispersion in the $N_{Na~I}$ vs. $A_V$ relation, 
 which is considerably greater than the individual errors in the column 
 densities.
% The $\chi_\nu^2$ for the Milky Way absorption sample 
% considered on its own is 
% somewhat smaller than that of the \citet{sembach93} sample.  
 Interestingly,
 the value for the combined sample of ``Redshifted'' and ``Single/Symmetric''
 SNe is similar to that of the \citet{sembach93} + Milky Way sample, implying that 
 their column densities and dust extinctions are fully consistent with the 
 Galactic relation.  On the other hand, $\chi_\nu^2$ is very large for 
 ``Blueshifted'' SNe, consistent with the significant number of objects
 with anomalously large values of $N_{Na~I}$.  Table~\ref{tab:tab4} shows
 that using either the column density of the most prominent (zero velocity) 
 absorption component  or the sum of the column densities of the absorption 
 components blueward of zero velocity for the ``Blueshifted'' SNe gives $\chi_\nu^2$
 values that, while lower than that obtained for the total column densities,
 are still relatively large.  Our conclusion is that splitting the ``Blueshifted'' SNe 
 into these two subsamples of the absorption components does not provide 
 any particular insight into the location of the
 absorbing gas and dust.
 
 If the strong \ion{Na}{1} absorption is an indication of outflowing CSM, it is
 somewhat puzzling that  the three ``Blueshifted'' SNe~Ia in our sample that
 almost certainly had CSM --- SNe~2006X, 2007le, and 2012cg --- did 
 {\em not} show unusually  large \ion{Na}{1} column densities for the host extinction 
 implied by their colors (see Figure~\ref{fig:fig3}).  Echelle spectroscopy of
 these three SNe revealed variations in the strengths of certain components 
of the \ion{Na}{1}~D lines \citep{patat07,simon09,raskutti13}.  The evidence strongly 
suggests that the variations were due to changing ionization conditions in
CSM, particularly since similar variations were not observed for the
 \ion{Ca}{2} H~\&~K lines.  Multiple echelle spectra were not obtained for 
 the majority of the 
 SNe~Ia in our sample, and so we do not know if more objects might
 have shown variable \ion{Na}{1} absorption.  An exception is 
 SN~2008fp, one of the SNe with the strongest \ion{Na}{1} lines.   
 \citet{cox13} did not find evidence for variability of any of the absorption 
 components in this SN from echelle spectra obtained at +6, +11, +17, and +39
 days after maximum light.  Of course, whether or not variability is observed
 in any particular object will be a function of the density, distribution, 
 and geometry of the CSM, and also the timing of the observations.
 %However, if the anomalously strong \ion{Na}{1}~D lines observed in SN~2008fp
 %are an indication of CSM, we must explain why variability was not observed.
 Perhaps in the case of SN~2008fp, the variations were too small to detect
 or occurred before the first spectrum was obtained.  High signal-to-noise ratio
 observations beginning at the earliest epochs will be necessary
 to make more progress on this question.
 
 The rare class of 2002ic-like events that display a strong CSM
 interaction, for which a white dwarf with a massive AGB star companion  
 \citep{hamuy03} or a symbiotic nova \citep{dilday12} have been proposed 
 as progenitor systems, may be key objects for unraveling the CSM-progenitor
 connection.  Echelle spectroscopy of the best-observed member 
 of this class, PTF11kx \citep{dilday12}, obtained on four epochs 
 ranging from $-1$ to $+44$~days with respect to optical maximum
 revealed strong variations in the strength of the most 
 blueshifted component of the \ion{Na}{1}~D lines.
 Our analysis of these spectra gives a total column density of 
 $\log{N_{Na~I}} = 12.93 \pm 0.19$~\cmsq\ and a 3-$\sigma$ upper
 limit of EW(5780)~$< 40$~m\AA.  \citet{dilday12} estimated a dust
 extinction of $A_V \sim 0.5$~mag from the SN colors before the strong CSM
 interaction set in.  These numbers place PTF11kx squarely
 on the Galactic $N_{Na~I}$ vs. $A_V$ trend, and only slightly
 below the EW(5780) vs. $A_V$ relation.
 %(see Figures~\ref{fig:fig1}
 %and \ref{fig:fig5}). 

\subsection{SNe~2006cm, 2008fp, 2009ig, and 2009le}
\label{sec:indiv}
 
 Three of the best-observed examples of ``Blueshifted'' SNe~Ia in our sample with 
 anomalously large \ion{Na}{1} column densities  are 2006cm, 2008fp, and 2009ig, 
 In this section, we examine some of their properties 
 in more detail along with those of the ``Redshifted'' SN~2009le.  
 
\begin{figure}
\epsscale{1.}
\plotone{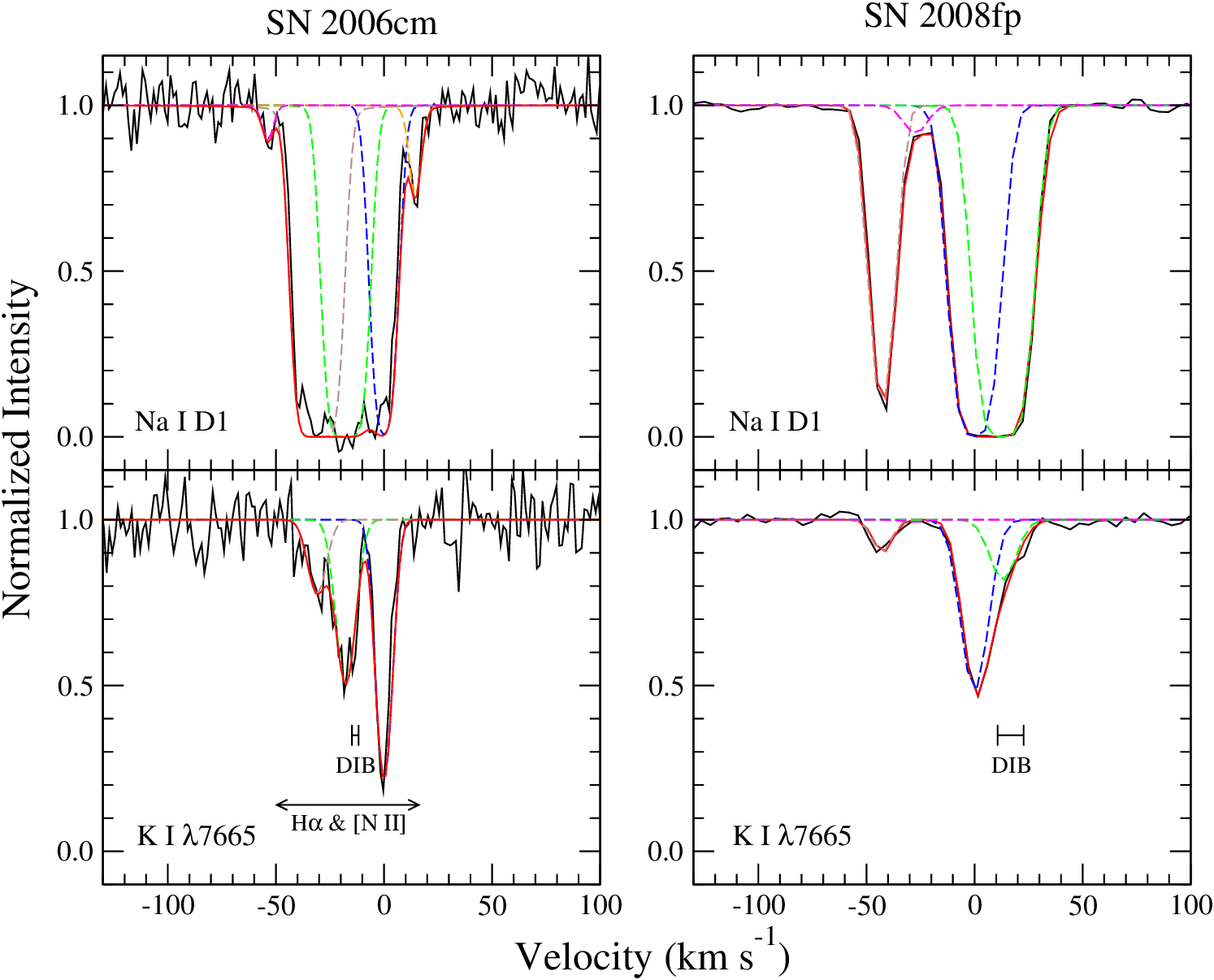}
\caption{Host \ion{Na}{1}~D1 and \ion{K}{1} $\lambda$7665 absorption 
in SNe~2006cm and 2008fp.  The observations correspond to the black line.  
The best-fitting VPFIT model is shown as the red line, with the individual 
Voigt profile components plotted as dashed lines.  Zero velocity has
been arbitrarily set to correspond to the strongest component of the 
\ion{K}{1} $\lambda$7665 profile. The velocities of the 5780~\AA\ DIB  in each 
SN are shown as $\pm$1-$\sigma$ error bars, and were derived assuming
a Gaussian profile of 2.1~\AA~FWHM and a rest wavelength of 5780.55~\AA\
\citep{tuairisg00}.
The FWHM velocity range covered by the H$\alpha$ and [N~II]
emission in the extracted spectrum of SN~2006cm is also indicated.
}\label{fig:fig4}
\end{figure}

\begin{figure}
\epsscale{1.}
\plotone{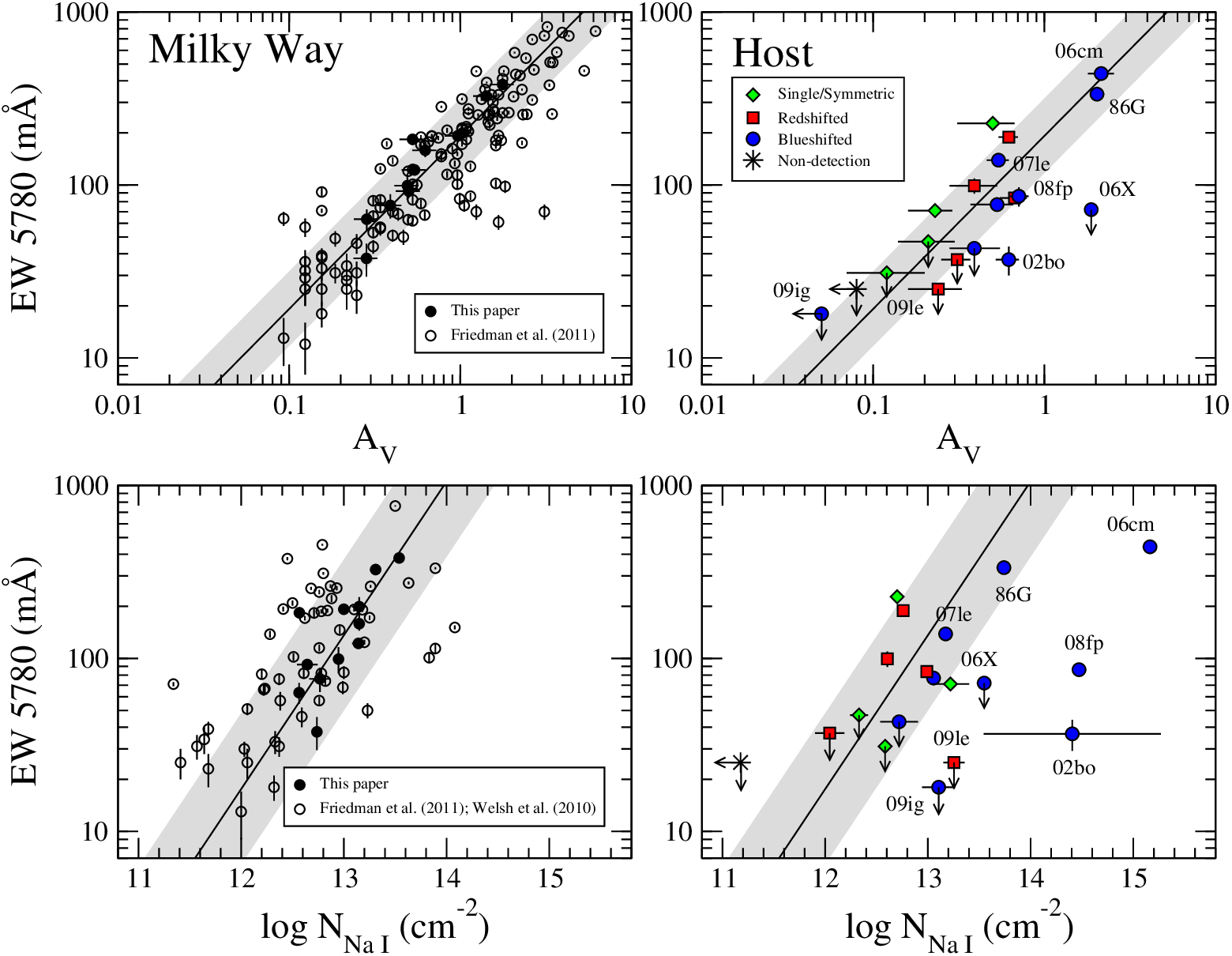}
\caption{(Left) Equivalent width of the DIB feature at 5780~\AA\ in the ISM
of the Milky Way plotted versus $A_V$ and the column 
density of \ion{Na}{1}.  The open circles in the upper plot correspond to the  
measurements of \citet{friedman11}; those in the lower plot are a combination
of the EW~5780~\AA\ values of \citet{friedman11} with \ion{Na}{1} column densities
from \citet{welsh10}.  The solid circles in both 
plots correspond to the objects in our Milky Way sample for which
a Galactic component of the 5780~\AA\ feature was detected.  
The $A_V$ values for the latter objects are taken from \citet{schlegel98} 
as rescaled by \citet{schlafly11}. The solid lines are fits to
these combined data, with the gray shading corresponding to 
the 1-$\sigma$ dispersions (see text);
(Right) Equivalent width of the host DIB absorption at 5780~\AA~plotted versus 
$A_V$ values derived from the SN colors (upper) and the column density 
of \ion{Na}{1} (lower).  The fits and 1-$\sigma$ dispersions observed
in the Milky Way are reproduced from the left half of the Figure.  
The different symbols used to plot the SNe~Ia
correspond to the \ion{Na}{1}~D profile classification scheme of
\citet{sternberg11}.}\label{fig:fig5}
\end{figure}

\begin{figure}
\epsscale{.75}
\plotone{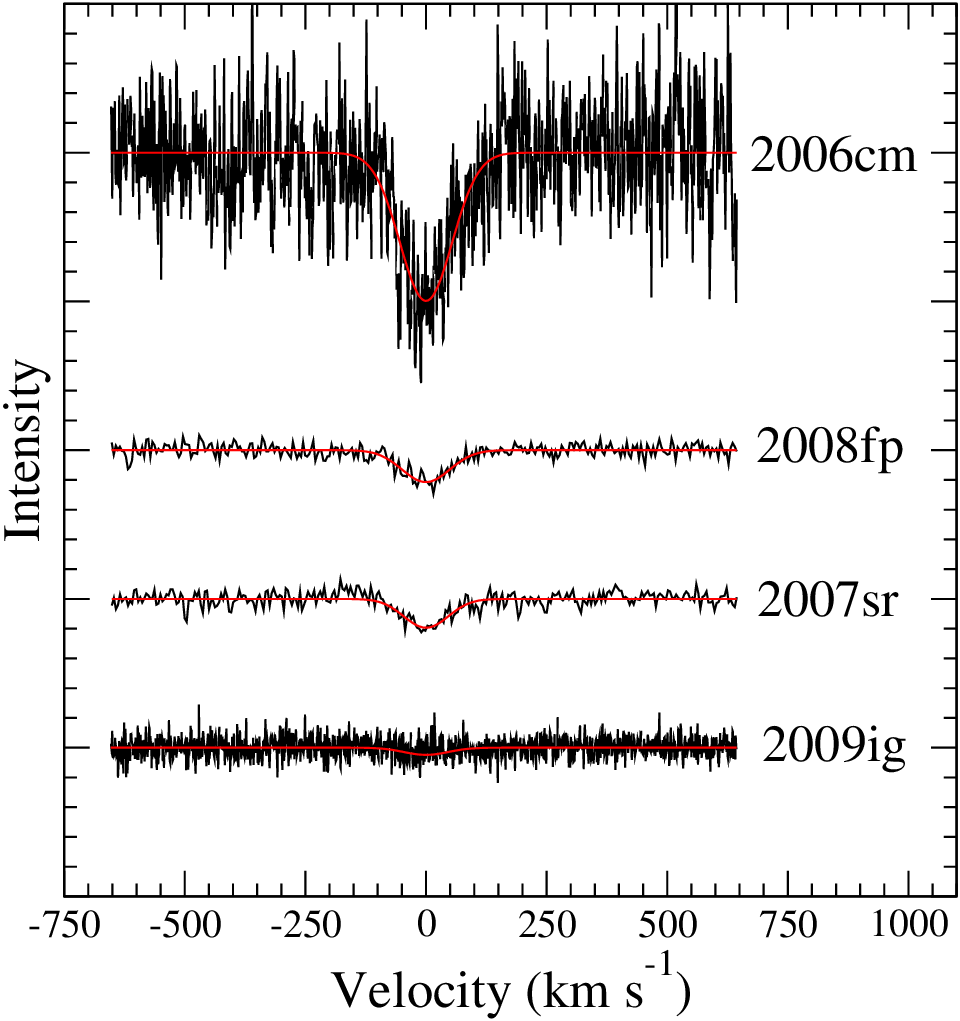}
\caption{Examples of three detections of host DIB absorption at 
5780~\AA\ (SNe~2006cm, 2008fp, and 2007sr).   Spectral observations are
plotted in black, and fits assuming a Gaussian profile of 2.1~\AA\ FWHM are
in red.  Also shown is the spectrum of SN~2009ig where the 5780~\AA\ is
not clearly detected, but where the signal-to-noise ratio of the spectrum allowed
 a meaningful upper limit to be measured.  The fit in this case corresponds
 to the derived 3-$\sigma$ upper limit on the equivalent width.  The scaling
 of the intensity axis is the same for each SN in order to preserve
 the relative differences in strength of the 5780~\AA\ feature; the spectra
 have been offset for illustration purposes.}
\label{fig:fig6}
\end{figure}

\begin{figure}
\epsscale{1.}
\plotone{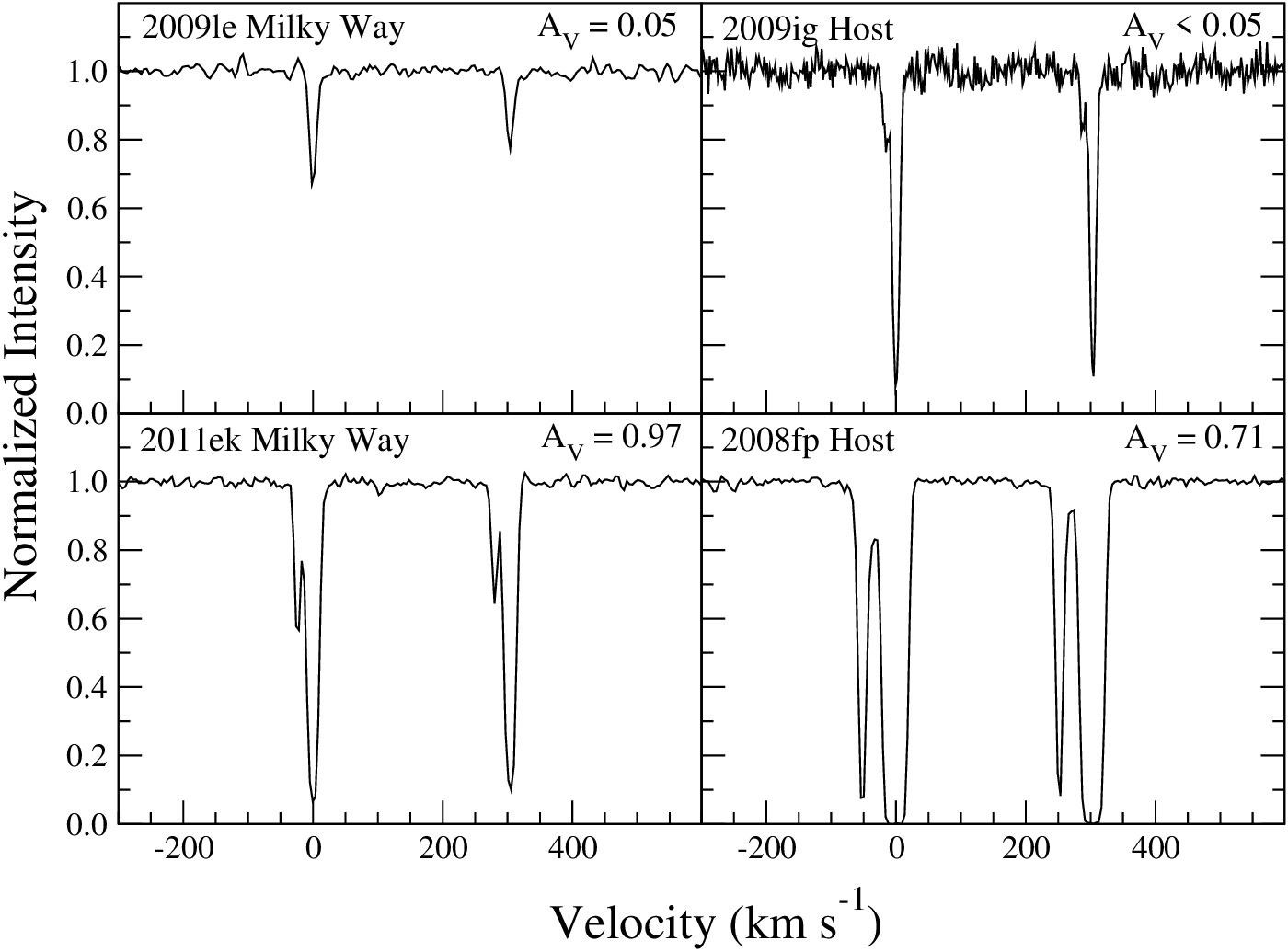}
\caption{\ion{Na}{1}~D profiles of two of the objects in the Milky Way sample (left),
and two SNe~Ia in the host absorption sample with similar dust extinctions (right).
Zero velocity has been arbitrarily set to correspond to the minimum of the D2 line 
absorption.  The anomalous strength of the host \ion{Na}{1}~D absorption in 
SNe~2009ig and 2008fp is readily apparent.}
\label{fig:fig7}
\end{figure}

\begin{figure}
\epsscale{1.}
\plotone{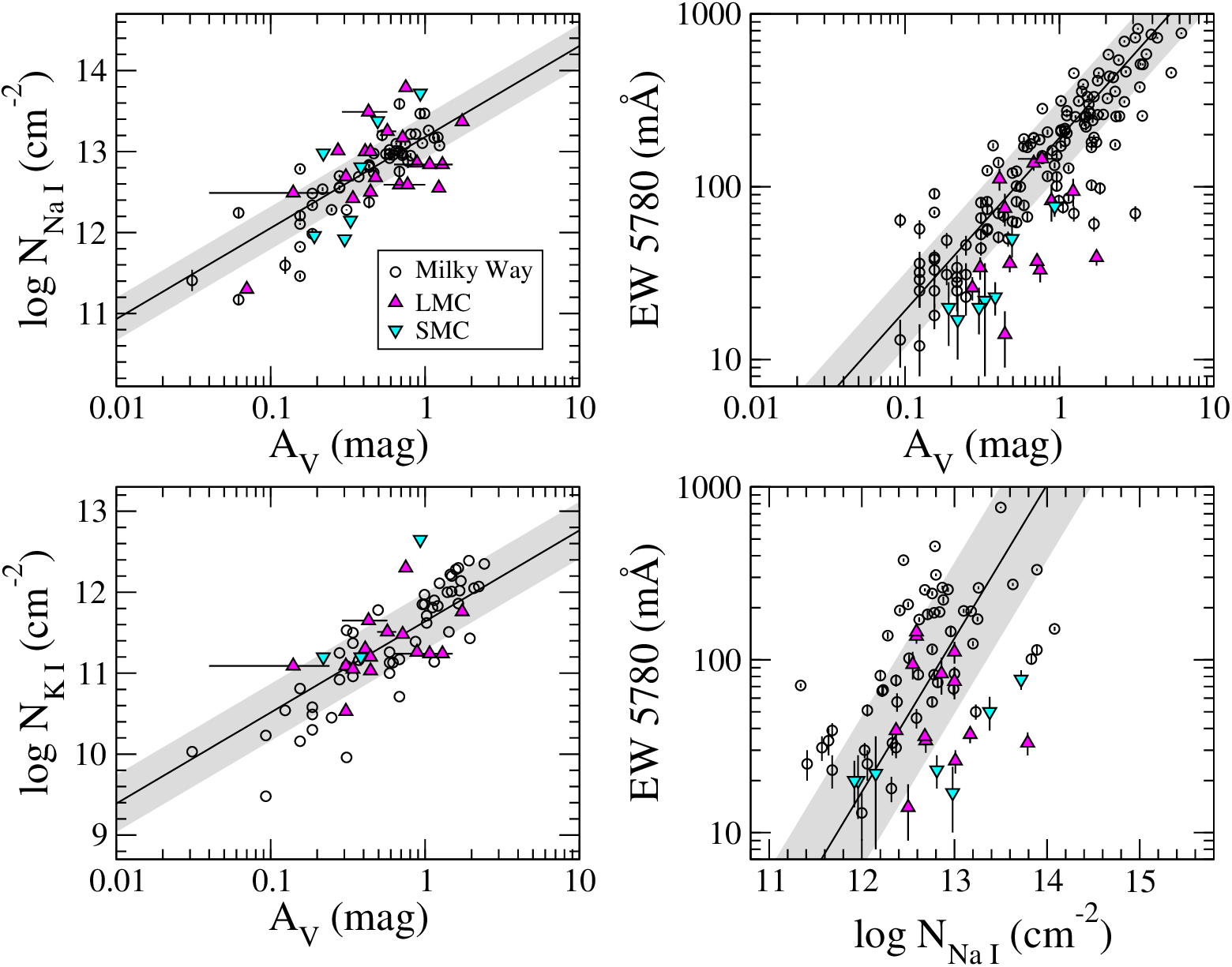}
\caption{(Left) Column densities of \ion{Na}{1} and \ion{K}{1} in the ISM
of the Milky Way, LMC, and SMC plotted versus $A_V$.  Open circles are 
the \ion{Na}{1}~D measurements of \citet{sembach93} and the \ion{K}{1} data
of \citet{welty01} shown in Figure~\ref{fig:fig3}. The fits and 1-$\sigma$ 
dispersions observed in the Milky Way are also reproduced from Figure~\ref{fig:fig3}. 
The triangles (magenta) are measurements for the LMC \citep{cox06,welty06} and
the inverted triangles (cyan) correspond to the SMC \citep{welty06}; (Right) Equivalent 
width of the DIB feature at 5780~\AA\ in the ISM
of the Milky Way, LMC, and SMC plotted versus $A_V$ and the column 
density of \ion{Na}{1}.  The open circles in the upper plot are the  
measurements of \citet{friedman11} reproduced from Figure~\ref{fig:fig5}.
Likewise, the EW~5780~\AA\ values of \citet{friedman11} are combined with 
the \ion{Na}{1} column densities from \citet{welsh10} as in Figure~\ref{fig:fig5}.  
The fits and 1-$\sigma$ dispersions observed in the Milky Way are 
reproduced from Figure~\ref{fig:fig5}.  Measurements 
for the LMC \citep{cox06,welty06} and SMC \citep{welty06} are plotted
with the same symbols and colors as in the left-hand side of the figure.}
\label{fig:fig8}
\end{figure}

\begin{figure}
\epsscale{.85}
\plotone{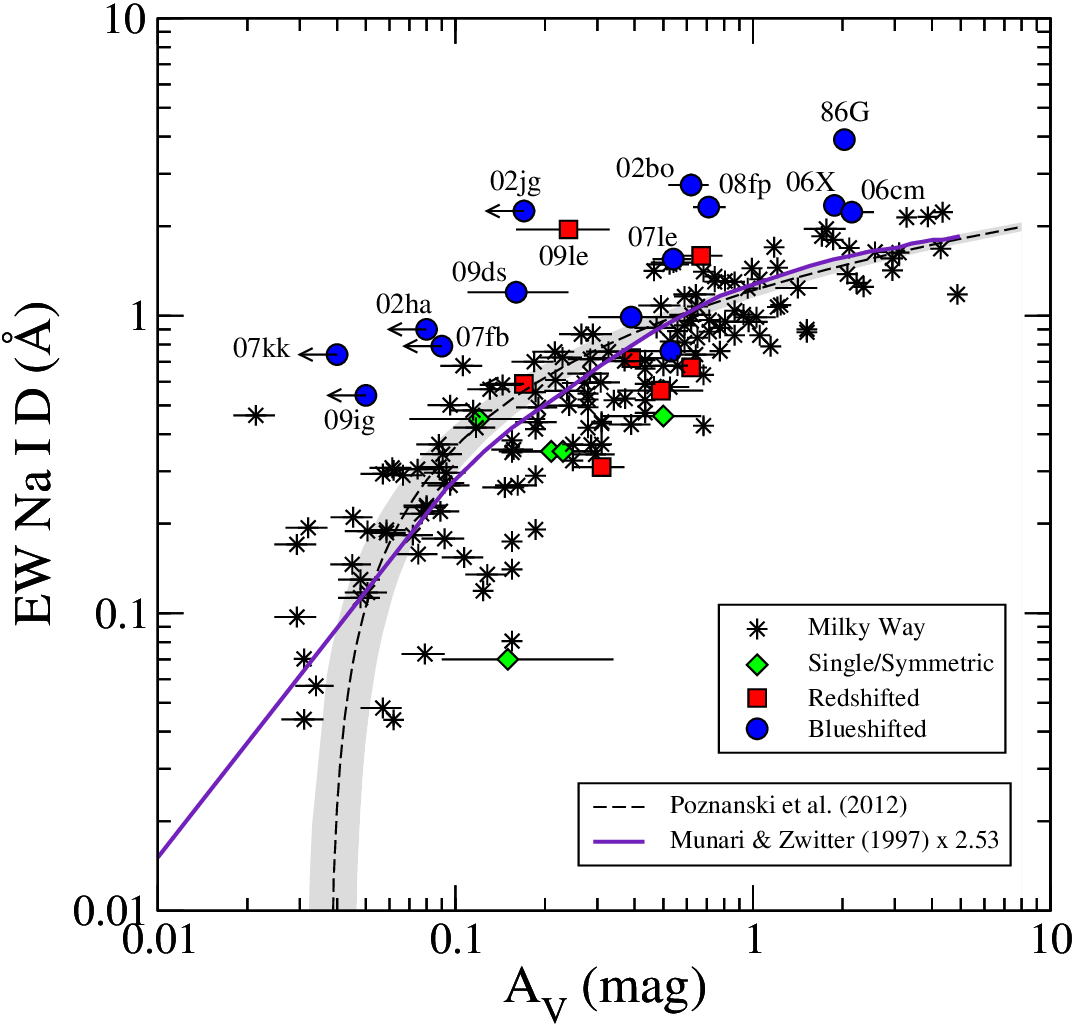}
\caption{\ion{Na}{1}~D equivalent widths of both the Milky Way and host absorption
samples are plotted versus the dust extinction, $A_V$.  The Milky Way sample has been
augmented by high dispersion measurements of 82 stars \citep{sembach93,munari97} 
and 30 QSOs \citep{poznanski12}. The dashed line corresponds to the Galactic 
relation derived by \citet{poznanski12}, with the 1-$\sigma$ dispersion found
by these authors indicated by the gray shading.  The solid purple line shows the
relation for the D1 line found by \citet{munari97}, multiplied by a factor of 2.53 
to match the total equivalent widths plotted here.}\label{fig:fig9}
\end{figure}

\begin{figure}
\epsscale{.85}
\plotone{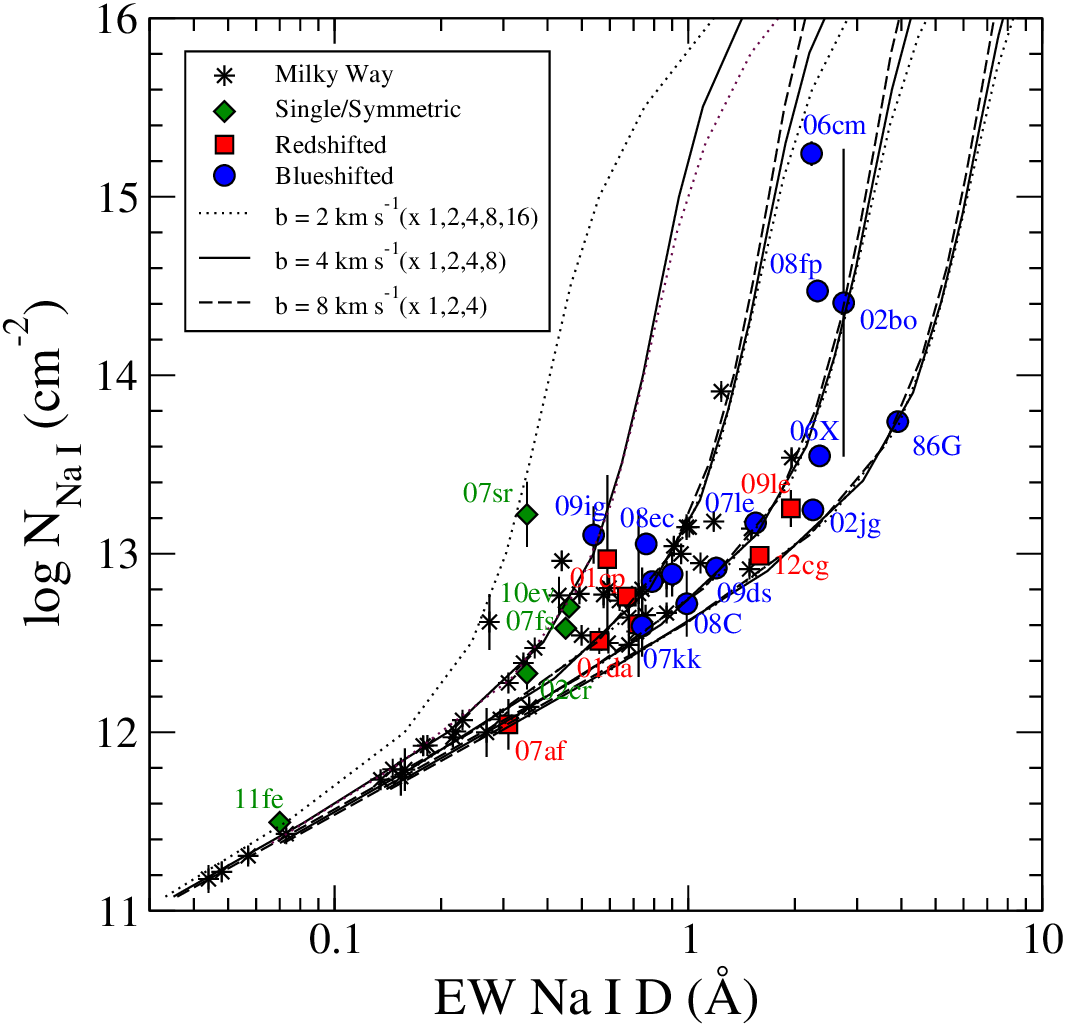}
\caption{\ion{Na}{1} column densities are plotted versus \ion{Na}{1}~D equivalent 
widths for both the Milky Way and host absorption samples.  Theoretical curve of
growth relations for three different values of the Doppler parameter, $b$, are shown 
for reference.  For each Doppler parameter, we show predictions for a single absorption 
component as well as for up to 16 components with the same value of
$b$ (as indicated in the legend.) In all cases, the single component predictions are 
the left-most curves with increasing number of components lying progressively to the right.}
%Curves for both single and multiple absorption components 
%are plotted, with the single-component cases corresponding to the left-most curve
%for each value of $b$.}
\label{fig:fig10}
\end{figure}

\begin{figure}
\epsscale{1.}
\plotone{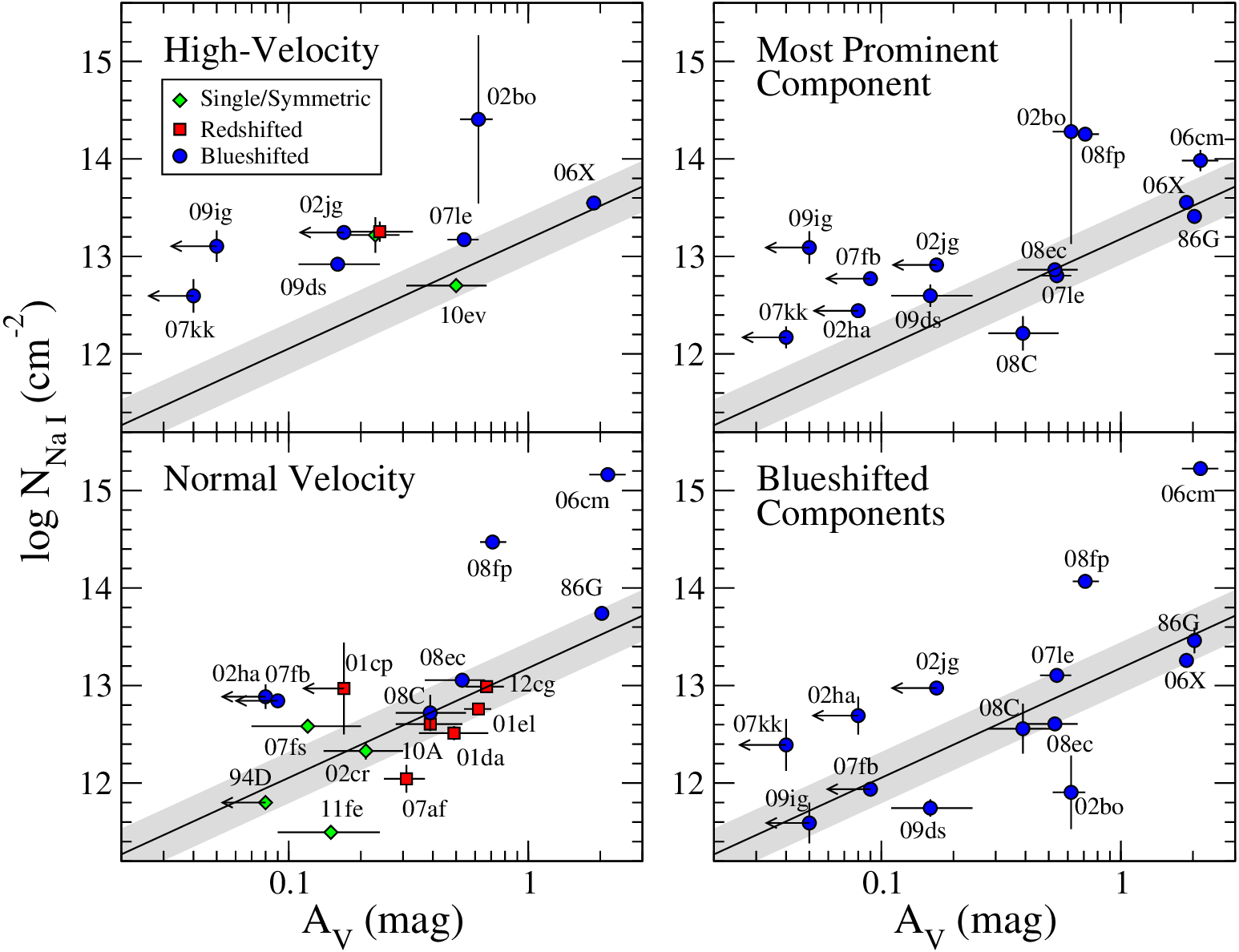}
\caption{(Left) Total host absorption column densities of \ion{Na}{1} plotted versus 
the $A_V$ values derived from the SN~Ia light curves.  The objects are separated into
the high-velocity (above) and normal-velocity (below) subclasses of \citet{wang13},
where high velocity is defined as $v(Si~II~\lambda6355) \geq$~12,000~\kms\
at $B$ maximum.  The fits and 1-$\sigma$ dispersions observed
in the Milky Way are reproduced from Figure~\ref{fig:fig3}.  
The different symbols used to plot the SNe~Ia
correspond to the \ion{Na}{1}~D profile classification scheme of
\citet{sternberg11};
 (Right) Similar plots, but in this case examining subcomponents
of the \ion{Na}{1}~D absorption profiles  for the ``Blueshifted''
sample only.  In the top diagram, the column density of the most prominent
\ion{Na}{1}~D profile is plotted, while in the lower diagram the sum of the
column densities of the components blueward of the most prominent 
component are shown.
}\label{fig:fig11}
\end{figure}

 SNe~2006cm and 2008fp were quite similar in their spectral and
 photometric characteristics.  Both 
 were substantially reddened by dust --- $E(B-V)_{host} = 1.08$ and 0.58, 
 respectively --- and both showed strong, saturated Na~I~D host absorption.
 Figure~\ref{fig:fig4} shows the \ion{Na}{1}~D1 and \ion{K}{1} $\lambda$7665 
 line profiles of these two SNe. \citet{cox13} recently presented an
 in-depth study of the host interstellar absorption features in 
 SN~2008fp.  They found a rich spectrum of narrow
 atomic and molecular lines characteristic of a cold, translucent cloud,  
 and argued that the host 
 dust extinction for this SN is produced in this cloud.
 This system is responsible for the \ion{K}{1} absorption at zero velocity in 
 Figure~\ref{fig:fig4} as well as the blue half of the strongly saturated component of
 the \ion{Na}{1}~D lines observed observed between $-20$ and 
 $+40$~\kms.  In modeling the latter absorption ``trough'' with
 VPFIT, we have used not only the information provided by the 
 \ion{K}{1} lines, but also the results of \citet{cox13} --- in
 particular, the velocity and column density of the \ion{Na}{1} UV 
$\lambda\lambda$3302, 3303 doublet and the velocities of the two
components of the \ion{Fe}{1} absorption observed by these authors. 
The fits yield  $\log{N_{Na~I}} = 14.43 \pm 0.04$~\cmsq\ and
 $\log{(N_{Na~I} / N_{K~I})} \sim 2.4$ for the ``trough''. 
 Measurement with VPFIT of the host \ion{Ca}{2} H~\&~K 
 absorption over this same velocity range yields 
 $N_{Na~I}/N_{Ca~II} \sim 13$,  typical of cold gas in the disk of 
 the Milky Way \citep{siluk74,vallerga93}.
 
 The evidence thus points to the ISM of the host galaxy of SN~2008fp
 as having produced at least a significant fraction of 
 the strongly saturated \ion{Na}{1}~D lines.  The origin
 of the blueshifted host absorption component at $-44$~\kms\ is less
 clear.  We measure 
 $\log{N_{Na~I}} = 13.43 \pm 0.14$~\cmsq, 
 $\log{(N_{Na~I} / N_{K~I})} \sim 2.5$, and 
 $N_{Na~I}/N_{Ca~II} \sim 110$.  The latter value
 suggests that this absorption is not produced in a high velocity
 cloud since, in the Milky Way, these typically have 
 $N_{Na~I}/N_{Ca~II} \lesssim 1$ \citep{siluk74,vallerga93}.
 
  \citet{cox13} also studied the host DIB lines in the spectrum
 of SN~2008fp, which they found to be slightly redshifted
 by $\sim$20~\kms\ with respect to the strong atomic and molecular
 lines.  This is consistent with our own measurement of the velocity 
 of the host 5780~\AA\ DIB in SN~2008fp which is very 
 similar to the velocity of the most redshifted component
 that we infer for the \ion{Na}{1}~D lines (see Figure~\ref{fig:fig4}).
 However, the 5780~\AA\ feature
 is broad (see Figure~\ref{fig:fig6}) and radial velocity measurements
 are likely to be sensitive to the exact profile used 
 to determine the observed wavelength, which is not a simple Gaussian as we assume
 in this paper \citep{galazutdinov08}.  The rest wavelength of the 5780~\AA\ DIB is also not very 
 accurately known, with the uncertainty corresponding to 10--15~\kms\ in radial velocity
 \citep[e.g.,][]{galazutdinov00,tuairisg00,hobbs08,hobbs09}.  Finally,
 the measured velocity of the DIB absorption will depend on the
 contributions of the various clouds in the line-of-sight, but this
 is impossible to decipher due to the significant width of the DIBs.  
 Hence, the fact that the velocity of the 5780~\AA\ DIB in SN~2008fp
 does not not coincide in redshift with the cloud responsible for
 the strongest component of the \ion{K}{1} lines should probably not 
 be given too much weight. 
 %Nevertheless, it is interesting that the
 %\ion{Na}{1} and \ion{K}{1} column densities of the cloud that 
 %coincides most closely in velocity with the 5780~\AA\ DIB feature
 %are more consistent with the Galactic relations (see Figure~\ref{fig:fig3}).
 
  In the case of SN~2006cm, three absorption systems are observed for the 
 \ion{K}{1}~$\lambda$7665 line (see Figure~\ref{fig:fig4}).  Using the 
 velocities and $b$ values of these as a guide to fitting the 
 \ion{Na}{1}~D~lines, we find 
 $\log{N_{Na~I}} = 15.23 \pm 0.07$~\cmsq\ and
 $\log{(N_{Na~I} / N_{K~I})} \sim 2.9$ for the strong absorption
 between $-50$ and $+10$~\kms.  As shown in Figure~\ref{fig:fig4},
 the velocity measured for the host 5780~\AA\ DIB falls near the 
 center of the D~line absorption.  SN~2006cm was located only
 2~arcsec from the center of its host, the edge-on Sb galaxy
 UGC~11723, and our spectrum contains
 weak, tilted emission lines of H$\alpha$ and
 [N~II]~$\lambda\lambda$6548,6584 along the slit,
 presumably due to diffuse ionized gas in the host.  The FWHM velocity
 of this emission in the extracted spectrum of the 
 SN is indicated in Figure~\ref{fig:fig4}, and
 closely coincides with the ``trough'' of \ion{Na}{1} absorption.
 %Hence, the conclusion again is that this absorption is most
 %likely produced in the host galaxy ISM.
 
 SN~2009ig is especially interesting since its colors are consistent with little or no host 
 reddening, yet its spectrum displays strong Na~I~D host absorption (see
 Figure~\ref{fig:fig7}).  Three other SNe in our sample --- 2002ha, 2007fb, 
 and 2007kk --- share this characteristic, and low-dispersion spectra
 have identified similar cases --- e.g., SNe~2006dd and 2006mr
 in NGC~1316 \citep{stritzinger10}.  The observed column density of
 \ion{Na}{1} for SN~2009ig implies a dust extinction in the range
 $A_V = $~0.5--1.2~mag using the Milky Way relation given in \S\ref{sec:results},
 yet both the SN colors and the upper limit on the strength of the host absorption
 DIB at 5780~\AA\ argue for $A_V < 0.1$~mag.  Host \ion{Ca}{2} H~\&~K 
 absorption is also present in our spectrum at 
 a ratio $N_{Na~I}/N_{Ca~II} \sim 3$, typical of cold gas in the disk of 
 the Milky Way \citep{siluk74}.  Whatever the source (CSM or ISM) of the gas 
producing the strong \ion{Na}{1} and \ion{Ca}{2} absorption in this SN, the
dust-to-gas ratio must be exceptionally low. 
 
 % \citet{garnavich13} recently announced that the optical light curve of 
 %SN~2009ig stopped fading approximately two years after explosion, and
 %suggested that the luminosity is now dominated by a light echo.  The
 %late appearance and faintness of the light echo implies that the dust is 
 %tens-to-hundreds of parsecs from the SN \citep{patat05}, and therefore 
 %associated with the ISM rather than the CSM.
 
\begin{figure}
\epsscale{.8}
\plotone{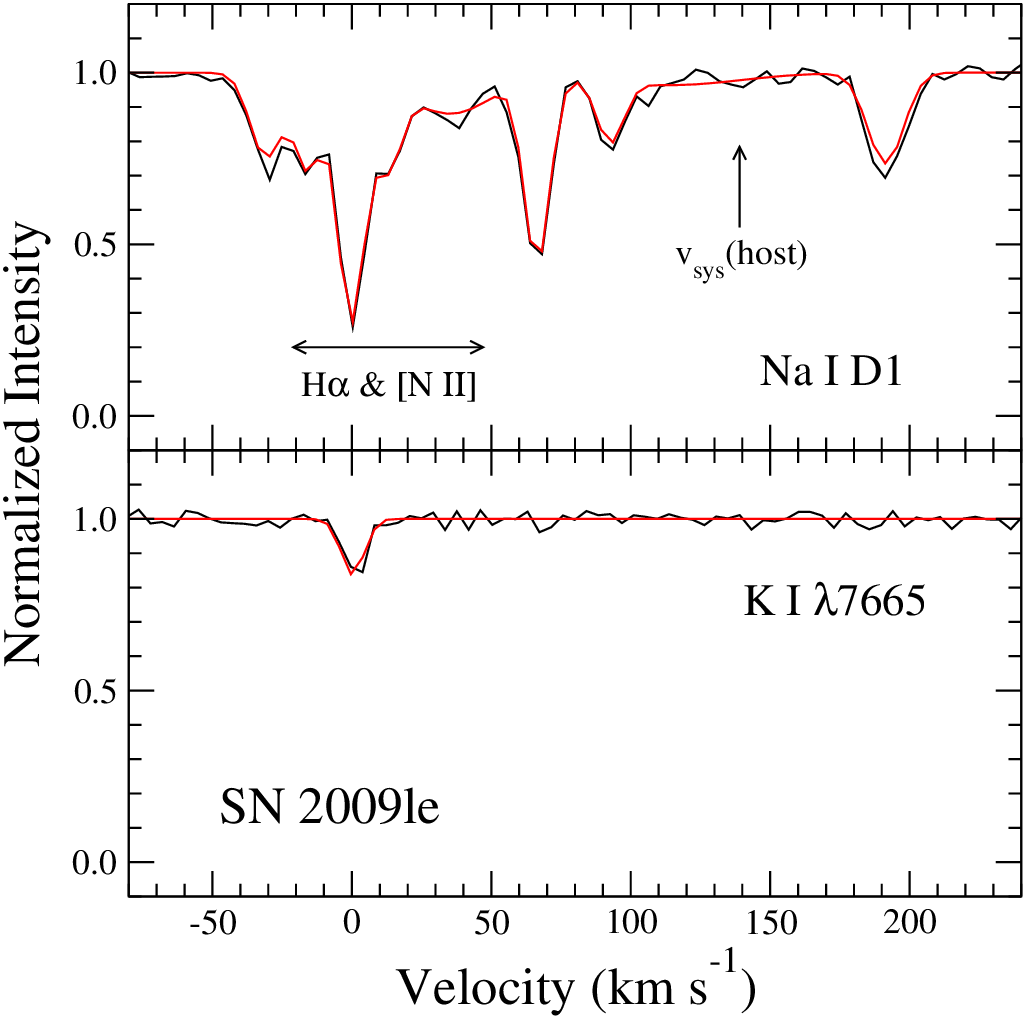}
\caption{Host \ion{Na}{1}~D1 and \ion{K}{1} $\lambda$7665 absorption 
in SN~2009le.  The observations correspond to the black line and the 
best-fitting profile VPFIT model is shown in red.  Zero velocity has
been arbitrarily set to correspond to the single component observed for of the 
\ion{K}{1} $\lambda$7665 line. The FWHM velocity range covered by H$\alpha$ 
and [N~II] emission in the extracted spectrum of the SN is indicated.  
The systematic velocity of the host galaxy is also shown.}\label{fig:fig12}
\end{figure}

 SN~2009le is a ``Redshifted'' SN~Ia with a total host  \ion{Na}{1} 
 column density more than 2-$\sigma$ above the
 Galactic $N_{Na~I}$ vs. $A_V$ relation.  The 
 lower-right plot in Figure~\ref{fig:fig5} shows that it is similar to SN~2009ig
 in having a large host \ion{Na}{1} column density, but relatively weak 
 5780~\AA\ DIB feature.  At $\log{(N_{Na~I} / N_{K~I})} \sim 2.1$, the
 $N_{Na~I} / N_{K~I}$ ratio is somewhat high, although not
 extraordinarily so (see \S\ref{sec:ki}).  Figure~\ref{fig:fig12} shows the observed
 \ion{Na}{1}~D1 and \ion{K}{1} $\lambda$7665 profiles
 of SN~2009le, along with the VPFIT model.  
 The host \ion{Na}{1} absorption is quite unusual in covering 
 $\sim250$~\kms\ in velocity space, more than any other SN
 in our sample.  Weak emission lines of H$\alpha$, H$\beta$, 
 [N~II]~$\lambda\lambda$6548,6584, and 
 [S~II]~$\lambda\lambda$6717,6731 are observed along
 the slit in the spectrum
 of the SN.  The [N~II]~$\lambda6584 / {\rm H}\alpha$ and
 [S~II]$/ {\rm H}\alpha$ flux ratios are both
 $\sim$0.7 and the [O~III]~$\lambda$5007 line is not detected,
 suggesting that the ionized gas is reasonably metal-rich \citep{pettini04}.
 The FWHM velocity range of the H$\alpha$ and [N~II] emission
 in the extracted spectrum of the SN is indicated in 
 Figure~\ref{fig:fig12}, as is the heliocentric systemic velocity 
 of 5,334~\kms\ of the host galaxy ESO~478-G006
 \citep{springob05}.
 
 The \ion{Na}{1} absorption extending from $+80$ to $+200$~\kms\
 in SN~2009le may be due to high velocity clouds in the host
 galaxy since $N_{Na~I}/N_{Ca~II} < 2$ for most of these 
 components \citep{siluk74,vallerga93}, whereas the
 stronger absorption between $-50$ and $+80$~\kms\ has
 $N_{Na~I}/N_{Ca~II} \sim 5$, typical of 
 cool dense gas in the disk of the Milky Way.  We find
 $\log{(N_{Na~I} / N_{K~I})} = 1.7$ for the zero velocity 
 component, consistent with the Milky Way ratio.  If we remove
 the high-velocity components from consideration, the 
 host \ion{Na}{1} column
 density decreases by only $\sim$0.1~dex, and so the strength
 of the D-line absorption in this SN remains strong, although
 not exceptionally so,
 for the amount of dust extinction implied by both the colors and 
 the equivalent width of the 5780~\AA\ DIB. 
 
  \subsection{The $N_{Na~I} / N_{K~I}$ Ratio}
 \label{sec:ki}

\begin{figure}
\epsscale{1.}
\plotone{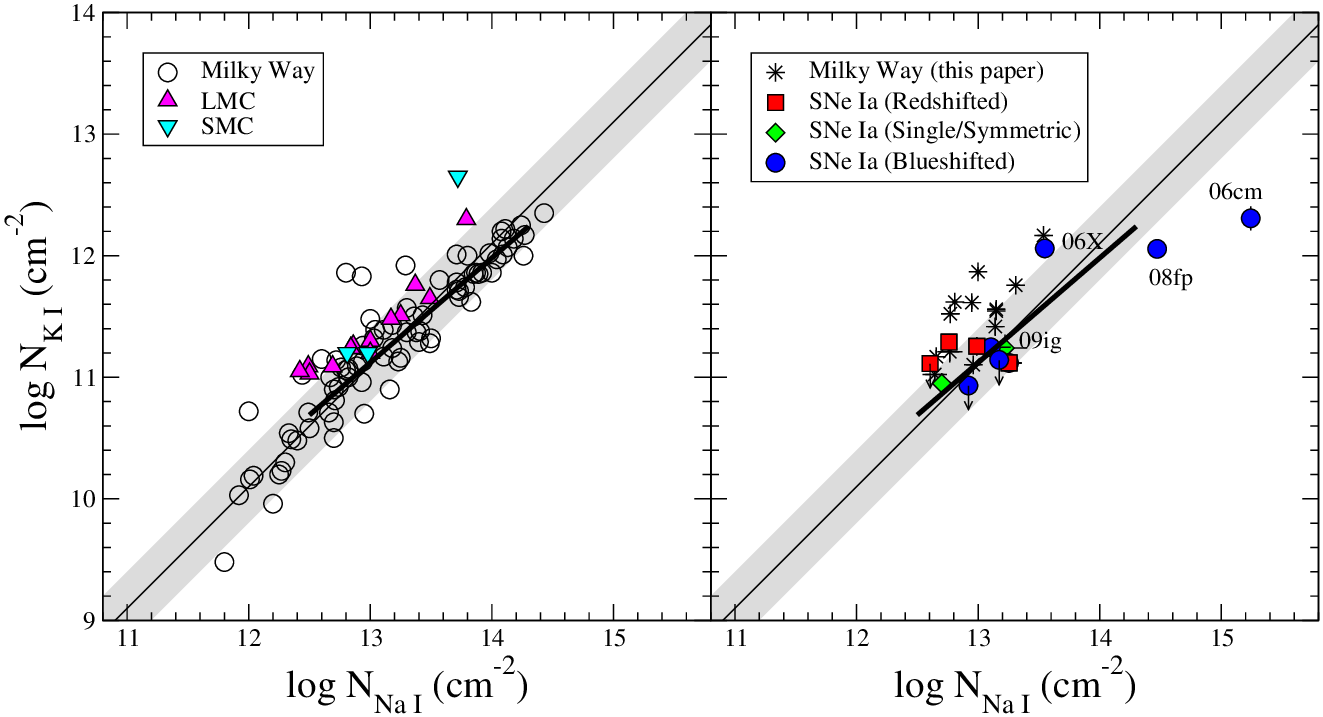}
\caption{Column densities of \ion{Na}{1} and \ion{K}{1} plotted against each other.
(Left) Measurements for the Milky Way \citep{welty01,kemp02}, LMC \citep{cox06,welty06},
and SMC \citep{welty06} are compared.  The diagonal line shows the average
value of $\log{(N_{Na~I} / N_{K~I})} = 1.9$ and rms dispersion of 0.3~dex obtained for
the Milky Way measurements.  The thick black line is a representation of the
$\log{N_{K~I}}$ versus $\log{N_{Na~I}}$ trend shown in 
Figure~2 of \citet{kemp02}. (Right) Plot of column densities of \ion{Na}{1} and \ion{K}{1}
for our Milky Way and SNe Ia host absorption samples.  The fit and rms dispersion
of the Milky Way stellar sample from the left half of the figure is duplicated for
comparison.  Also repeated is the $\log{N_{K~I}}$ vs. 
$\log{N_{Na~I}}$ trend from \citet{kemp02}.}\label{fig:fig13}
\end{figure}

 In the Milky Way ISM, the Na~I and K~I column densities are tightly 
 correlated as would be expected based on ionization potentials and 
 condensation temperatures \citep{welty01}.  This is illustrated in the 
 left half of  Figure~\ref{fig:fig13} which shows the measurements of 
 \citet{welty01} and  \citet{kemp02}.  A fit to these data gives 
 $\log{(N_{Na~I} / N_{K~I})} = 1.9$, with an rms dispersion of 0.3~dex. 
 %\citet{welty01} argue that this value is consistent with solar abundances
 %if K is depleted by $\sim$0.1 dex with respect to Na. 
 \citet{kemp02} argued from their observations that the value of the 
 $N_{Na~I} / N_{K~I}$ ratio actually increases slightly with column
 density as indicated by the thick black line in Figure~\ref{fig:fig13}.
 This is a small effect that is also clearly consistent with the data.
  
 Included in Figure~\ref{fig:fig13} are \ion{Na}{1} and \ion{K}{1} column
 density measurements in the LMC \citep{cox06,welty06} and SMC
 \citep{welty06}, which give a slightly lower value of $\log{(N_{Na~I} / N_{K~I})} = 1.6$.  
 This difference may be due to saturation effects as  mentioned in
  \S\ref{sec:coldensities} since the LMC and SMC values of $N_{Na~I}$
  were measured using the D~lines, whereas the Galactic
  measurements of \citet{welty01} and \citet{kemp02} were derived 
  from the UV $\lambda\lambda$3302, 3303 doublet for 
  $\log{N_{Na~I}} > 12.5$~\cmsq.
 
 Our observations of the Milky Way sample in Table~\ref{tab:tab1}
 are plotted in the right half of Figure~\ref{fig:fig13}, and give a 
 weighted mean of $\log{(N_{Na~I} / N_{K~I})} = 1.4$ with an rms 
 of 0.3.  This lower value is also likely due to saturation effects in the 
 D~lines. The weighted mean for the SN~Ia host absorption sample is 
 $\log{(N_{Na~I} / N_{K~I})} = 1.7$ with an rms of 0.6.  Excluding the 
 three SNe with anomalously large \ion{Na}{1} column densities --- 
 2006cm, 2008fp, and 2009ig --- lowers this to 1.6 with 
 an rms of 0.3.  The $\log{(N_{Na~I} / N_{K~I})}$ values for SNe~2006cm 
 and 2008fp of 2.9 and 2.4, respectively, are significantly
 greater than the mean our spectra give for the Milky Way. 
 Interestingly, however, the measurements for SNe~2009ig and 2009le
 are fully consistent with the Galactic $N_{Na~I} / N_{K~I}$ ratio.
 
 \citet{kemp02} argue on theoretical grounds that, for a cosmic abundance 
 ratio of $N_{Na}/N_{K} \approx 15$, standard values of the photoionization 
 rate and recombination coefficients, and electron densities between 
 0.001--1~${\rm cm}^{-3}$, the $N_{Na~I} / N_{K~I}$ ratio should be in 
 the range of $\approx$ 60--80.  Although at low column densities a harder 
 radiation field could lower the ratio close to the cosmic abundance 
 value of 15, at the highest column densities (i.e., as the electron density
 approaches zero), the $N_{Na~I} / N_{K~I}$ ratio should asymptotically
 approach a value of $\approx 80$, or $\log{(N_{Na~I} / N_{K~I})} \approx 1.9$.  
 Observations in the Milky Way show exactly this effect (see Figure~5 
 of \citet{kemp02}), although a few sight-lines reach values as large as 
 $\log{(N_{Na~I} / N_{K~I})} \sim 2.1$.
 
Thus, the much higher values of $\log{(N_{Na~I} / N_{K~I})} = 2.4$--2.9 
observed for the host absorption in SNe~2006cm and 2008fp imply 
either an enhancement of  \ion{Na}{1} or a depletion of \ion{K}{1}.  The fact 
that both the DIBs and \ion{K}{1} column densities for these two objects 
are more consistent with the value of $A_V$ derived from the SN colors 
than are the \ion{Na}{1} column densities would seem to suggest that 
an enhanced abundance of \ion{Na}{1} is the more likely explanation.  
%However, it 
%should be noted that the observations of $N_{Na~I} / N_{K~I}$ ratios in the 
%Milky Way do not extend significantly beyond $\log{N_{Na~I}} = 14$~\cmsq, 
%and if the slope of the \citet{kemp02} relation plotted in Figure~\ref{fig:fig11} 
%extends to such large column densities, the $N_{Na~I} / N_{K~I}$ ratio for
%SN~2008fp may not be quite so peculiar.  
More observations of the K~I doublet in SNe~Ia are needed to determine if 
the large $N_{Na~I} / N_{K~I}$ ratios observed for SNe~2006cm and 
2008fp are a common characteristic of SNe~Ia with anomalously strong
\ion{Na}{1}~D lines, or whether most, like SNe~2009ig and 2009le, show
normal $N_{Na~I} / N_{K~I}$ ratios. 
%In either case, however, the consistency 
%of the strength of the 5780~\AA\ DIB with the SN colors implies that the gas 
%producing the strong Na absorption is characterized by a low dust-to-gas ratio.

\subsection{Possible Origins of the Strong Na~I~D Absorption}
 
 Although one-fourth of the SNe~Ia in our sample displayed 
 anomalously large host \ion{Na}{1} column densities in comparison 
 with dust reddening deduced from their colors, the observations
 provide conflicting clues as to the origin of this phenomenon.  
 The fact that all such SNe have ``Blueshifted'' D-line profiles as 
 classified by \citet{sternberg11} is highly suggestive that an 
 outflowing CSM is responsible for the strong \ion{Na}{1} absorption.  
 The existence of events such as SN~2009ig that were essentially
 unreddened, as indicated by both their colors and the weakness
 of the DIB 5780~\AA\ absorption, yet showed strong host 
 \ion{Na}{1}~D lines is also difficult to understand in the ISM scenario.
 Nevertheless, as discussed in \S~\ref{sec:indiv}, the evidence
 seems to clearly favor an ISM origin for most, if not all, of
 the strong D-line absorption 
 observed in SN~2008fp, and, as detailed 
 in \S~\ref{sec:relation}, unusually strong 
 \ion{Na}{1}~D lines have not been observed in the few SNe Ia studied 
 to date for which there is independent evidence for the existence of CSM.
 
 If all SNe~Ia have the same progenitor systems, and if
 the anomalously large host \ion{Na}{1} column densities are due
 to CSM, then we would expect to see strong D-line absorption
 in at least some of the SNe~Ia that occur in elliptical galaxies
 (i.e., where there is no significant ISM).  Unfortunately, 
 only one of the SNe in our sample (2007on) occurred in
 a host galaxy that is classified as an elliptical.  \citet{sternberg11}
 obtained high-dispersion spectra of an additional two SNe~Ia  
 (2006ct and 2006eu) in elliptical galaxies.  In none
 of these three cases was \ion{Na}{1} absorption detected.
 It is important to continue to obtain echelle spectra of more
 SNe~Ia in nearby ellipticals to build up a 
 statistically-significant sample.  If it is found that  SNe~Ia in 
 elliptical hosts never show strong \ion{Na}{1}~D lines, that 
 would in turn imply that either the large \ion{Na}{1} column
 densities seen in some SNe~Ia are not due to CSM, or 
 the progenitors of SNe Ia in elliptical galaxies are fundamentally 
 different from (or, are a subset of) those in later-type galaxies.
 Some support for the latter hypothesis is found in the observation that 
 early-type galaxies tend to produce a factor of $\sim$20 less 
 SNe~Ia (normalized to stellar mass) than do late-type galaxies 
 \citep{mannucci05}, and that these SNe are, on average, faster-declining 
 and less-luminous than those observed in spirals 
 \citep{hamuy95,hamuy96}.   
 %Thus, we are led 
 %to conclude that both the ISM and CSM scenarios may be
 %required to explain the anomalously-strong host \ion{Na}{1} 
 %absorption.  In the remainder of this section, we discuss
 %possible ways for producing an enhanced Na abundance
 %in CSM associated with candidate progenitor systems 
 %of SNe~Ia.
 
 Novae have long been considered to be possible progenitors of some SNe~Ia, and 
there is evidence, both observational and theoretical, that highly enhanced Na is 
produced in the thermonuclear runaway that powers novae outbursts.   
In a review of abundance determinations for novae, \citet{gehrz98} identified 
several novae whose Na abundances were derived from emission-line 
analysis to be more than an order of magnitude enhanced over solar values.
%The highly enhanced (relative to Fe) heavy element abundances of many 
%novae have received the attention of theoretical modeling of the outburst, 
%and \citet{starrfield93} have demonstrated that hot CNO proton-capture 
%burning on massive ONeMg WDs produces large amounts of Al and Na, 
%such that enhancements of two orders of magnitude are expected for such 
%novae.  
It is therefore possible in the SD scenario that very high Na abundances 
in SNe Ia might be understood in terms of their having been produced in prior 
nova outbursts in progenitors whose eventual collapse formed the supernovae.
 
Giant stars in globular clusters show large Na  abundance 
 enhancements that are not observed in field giants 
 \citep[e.g., see][and references therein]{sneden04}.  The Na
 and O abundances are anti-correlated, consistent with proton-capture 
 fusion processes during CNO burning.  These abundance variations 
 most likely result from pollution by an earlier generation of more massive stars. 
 Similar star-to-star abundance variations in C, N, O and Na have been 
 observed in stars at or near the main-sequence turn-off 
 \citep{gratton01,ramirez03,james04,cohen05a,cohen05b},
 in support of
 the self-pollution scenario as first proposed by \citet{cottrell81}.
 Perhaps, then, the ``Blueshifted'' profiles and anomalously large
 \ion{Na}{1} column densities (with respect to \ion{K}{1}) observed in SNe~Ia
 such as 2006cm and 2008fp 
 point to a SD progenitor system
 with a massive asymptotic giant branch (AGB) star as the donor star.
 
The stellar yields of Na from intermediate-mass AGB stars of 
solar metallicity show only relatively minor enhancements of 
0.2--0.4 dex \citep{karakas10,karakas12}. This is down from the 
results published in \citet{karakas07b}, which predict yields of 
Na that are up to a factor of $\approx 8$ higher (for a 6$M_{\odot}$,
$Z = 0.02$ model). The cause of the variation is the use of revised
experimental reaction rates for the $^{23}$Na $+ p$ reactions
\citep[see discussion in][]{karakas10}. \citet{karakas12} 
varied the mass-loss rate and reaction rates, key uncertainties in
AGB nucleosynthesis calculations of intermediate-mass stars. They 
find that the only way to obtain Na enrichments of $\approx 1$ dex 
is to use the old rates for the $^{23}$Na $+ p$ reactions combined 
with a lower mass-loss rate in stars of mass 5--7$M_{\odot}$ at
$Z=0.02$. In summary, fine-tuning of model parameters is required 
in order to produce substantial Na enrichment ($\approx 1$~dex) 
in the wind.

There is another way to obtain substantial enrichments in Na.
Most of the Na produced during the AGB phase remains in
the core of the star. Sodium is produced by H-shell burning
but it is not destroyed by $\alpha$-captures during helium burning.
For this reason, a reservoir of Na builds up in the outer layers of
the H-exhausted core and only a small fraction of this is mixed to the surface
where it can be lost by stellar winds. By the beginning of the AGB
phase, the amount of Na is $\gtrsim 7$ times the initial value in 
the outer $\lesssim 0.01M_{\odot}$ layers of the core. 
This is a robust prediction that does not depend on reaction rates or
initial stellar mass, and is found for a range of masses from
1.9-7$M_{\odot}$ at $Z=0.02$.
By the end of the AGB phase, this increases to 8 to 30 
times the initial Na, depending on the details of AGB evolution. 
{\em Higher} values are in fact found in the lowest mass AGB 
star cores of $\approx 1.9-3M_{\odot}$. This is because of the formation of a 
$^{13}$C pocket, a necessary ingredient for the production of 
$s$-process elements \citep[see e.g.,][]{mowlavi99b,cristallo09,karakas10}.
The Na produced in this manner will remain in the white dwarf (WD) 
after the AGB phase ends.
This means that less fine tuning is required to produce substantial
Na enrichment in SNe~Ia with DD progenitors, assuming some of the
material from the outer layers of the disrupted WD can form CSM
\citep{raskin13,shen13}. If the WD merger involves a 
He WD plus a C-O WD, then stellar models also predict
Na enhancements of a factor of 7 spread over a larger
$\approx 0.1M_{\odot}$ of the helium core\footnote[28]{From a 1.9$M_{\odot}$ 
stellar model of $Z=0.02$ which produces a 0.45$M_{\odot}$ helium 
core at the tip of the RGB.}.

\subsection{Location of the Dust and the Nature of the Low Values of $R_V$}

 Our finding that the strength of the DIB at 5780~\AA\ 
 correlates well with the dust extinction derived from
 the SN~Ia colors provides an important clue to the location of the dust. 
 DIBs are weak or absent in the CSM of nearly all mass-losing stars,
 including post-AGB stars \citep{lebertre93,luna08}.  Most likely this
 means that either the molecular carriers responsible for the DIBs are 
 not present in the CSM, or that the excitation conditions are not 
 what is required to produce observable absorption.  If the latter
 were true, it is conceivable that a SN~Ia exploding 
 within a dusty CSM might produce the conditions to make DIBs visible.  
 However, with the exception of SN~1986G which may have 
 been located behind the dust lane of NGC~5128, all
 of the significantly reddened SNe~Ia in our sample exploded
 in the dusty disks of their host galaxies.  Thus, the simplest and most
 obvious interpretation of the correlation of EW(5780)~\AA\ with $A_V$ 
 is that most of the dust extinction 
 is interstellar in origin, and {\em is not} produced in CSM associated with 
 the SN progenitor system.  

The very low values of $R_V$ that seem to characterize
the dust reddening in many SNe~Ia are unusual in the Milky
Way, but not completely without precedent.  A well-studied
example is the B3~V star, HD~210121 for which photometry
and spectropolarimetry give $R_V \sim 2.0$ \citep{larson96,fitzpatrick07}.  
The dust reddening of this star is produced by a translucent molecular
cloud at high Galactic latitude characterized by an enhanced relative 
abundance of small grains and a dust-to-gas ratio $\sim$30\% smaller
than the standard Galactic value \citep{larson00}. Similarly low 
values of $R_V$ have been observed in a few other high Galactic 
latitude clouds \citep{larson05} and toward the Galactic bulge 
\citep{udalski03,nataf13}.  In their study of anomalous extinction
sight-lines, \citet{mazzei11} list two stars with $R_V = 0.60 \pm 0.18$
(HD~1337) and $1.10 \pm 0.18$ (HD~137569).  
Nevertheless, such extreme values
are rare in the solar neighborhood.  
In their study of 328 Galactic extinction curves of O and B stars
at a median distance of $\sim1.3$~kpc, \citet{fitzpatrick07} 
found that less than 3\% had 
$R_V < 2.5$. 

For comparison, four of the five SNe~Ia in our sample with
$E(B-V) > 0.4$ had  $R_V < 2.5$.  The average value of $R_V$ for these 
four SNe is $\sim$1.4.  Such unusually  low values led  
\citet{wang05} and \citet{goobar08} to suggest that $R_V$
for these reddened SNe~Ia is modified by multiple scattering
of photons in a dusty CSM which effectively steepens the
extinction law.  However, this idea appears to be at odds with our
finding that the dust extinction is largely produced in the ISM.
A possible exception is the heavily-reddened SN~2006X, 
for which our measurement of the upper limit on the strength of 
the 5780~\AA\ DIB lies significantly below the EW(5780) vs. $A_V$
 relation in the Milky Way (see Figure~\ref{fig:fig5}).  As mentioned
 in \S\ref{sec:relation}, SN~2006X 
 showed variable \ion{Na}{1}~D absorption, and thus almost 
 surely had CSM.  Evidence in favor of the dust reddening
 being produced in the CSM of this SN was reported by
 \citet{folatelli10}, who found that the reddening was  
 better matched by a \citet{goobar08}
 model than a normal reddening law.
 
Alternatively, the weak DIB absorption in SN~2006X may be
a consequence of conditions in the ISM.
A recent survey of sight-lines in the 
Scorpius OB2 association by \citet{vos11} showed that the correlation of 
 the 5780~\AA\ DIB with reddening is different for  a diffuse environment 
 exposed to a strong ultraviolet (UV) radiation field with low molecular content, 
 as opposed to a denser cloud protected from the impinging interstellar 
 UV radiation.  The relations plotted in Figure~\ref{fig:fig5} correspond 
 more to the ``UV exposed'' trend, whereas the observations indicate
 that SN~2006X exploded behind a dense molecular cloud \citep{cox08}
 which may have probed a ``UV protected'' environment 
 characterized by a shallower dependence of EW(5780) on $A_V$
 \citep{vos11}.  This may explain why SN~2006X falls below the
 general trend seen in Figure~\ref{fig:fig5}.

An independent test of the values of $R_V$ 
derived from the SN colors is provided by spectropolarimetric
measurements, since the wavelength of maximum polarization
is well-correlated with $R_V$ \citep{sarkowsky75,whittet78,clayton88}.  
Using this technique, \citet{hough87} found $R_V = 2.4 \pm 0.23$ for
SN~1986G, which compares quite well with our value of 
$2.57^{+0.23}_{-0.21}$ derived from the optical and NIR
light curves.  Spectropolarimetry of SN~2001el by \citet{wang03} 
gave $R_V = 2.88 \pm 0.15$, which is in fair agreement with 
our value of $2.25^{+0.46}_{-0.36}$.
More recently, \citet{patat09} obtained 
spectropolarimetry of SN~2006X, for which the SN colors
imply $R_V = 1.31^{+0.08}_{-0.10}$.  A strongly-polarized continuum 
was observed, peaking at $\lesssim 3500$~\AA, consistent with
$R_V \lesssim 2$.  Moreover, the angle of the interstellar polarization
was found to be tangential to the dust lane associated with the spiral 
arm close to the explosion site, in keeping with observations that
dust grains in the ISM of disk galaxies are aligned along the 
spiral arm pattern \citep[e.g.,][]{scarrot87}.  These results imply that the 
majority of the dust reddening of SN~2006X was produced in the ISM.

We conclude, therefore, that the low values of $R_V$ observed 
for these reddened SNe~Ia reflect the dust properties of the ISM 
in the neighborhood of the SN progenitor.   
The fact that we find $R_V < 2$ for such a large fraction of reddened 
events is clearly telling us something important about both the 
environment and nature of the
SN progenitors, many of which exploded in the dusty, inner regions of 
spiral galaxies.  In turn, SN observations appear to provide a
powerful new means of studying the  
extreme dust properties found in at least some spirals.

\section{CONCLUSIONS}
\label{sec:conclusions}

The observations in this paper lead us to two important conclusions
about SNe~Ia.  The first of these is that
the dust responsible for the observed reddening of SNe~Ia 
appears to be predominantly 
located in the ISM of the host galaxies and not in CSM associated with the 
progenitor system.  This conclusion is based on both the correlation
found between the strength of the DIB at 5780~\AA\ and the visual 
extinction, $A_V$, derived from the SN colors, and the fact that
DIBs are a characteristic of the ISM in our Galaxy, and not the CSM of 
mass losing stars.  A direct implication of
this finding is that the peculiarly-low values of $R_V$ derived for the
most reddened SNe~Ia are not generally due to multiple scattering
of photons in a dusty CSM as suggested by \citet{wang05} and \cite{goobar08}, 
but rather are characteristic of the properties of the dust in 
the ISM where the SNe exploded.
Of course, we cannot rule out the presence of some dust in the CSM,
but the evidence is consistent with the majority of reddening being produced in
the host galaxy ISM.

The second major conclusion of this paper is that approximately one quarter of all SNe~Ia
show anomalously large host \ion{Na}{1} column densities in comparison with
the amount of dust that reddens their spectra.  This result is based on
comparison with the correlation between the \ion{Na}{1} column density 
and $A_V$ observed in the Milky Way, but similar correlations are also
observed for the Magellanic Clouds where the ISM is of significantly 
lower metallicity.  Remarkably, all of the 
SNe with unusually strong D~lines have ``Blueshifted'' profiles in the 
classification scheme of \citet{sternberg11}.   It is tempting, therefore, to ascribe the
anomalous strength of the \ion{Na}{1} to outflowing CSM, especially since strong
D~lines are observed in several SNe~Ia that did not suffer significant
dust reddening.  We have identified ways of producing enhanced Na 
abundances in the CSM in both the SD and 
DD progenitor models.  Nevertheless, unusually strong \ion{Na}{1}~D 
lines have not been observed in the few SNe~Ia studied to date for which
there is independent evidence for the existence of CSM, casting some
doubt on the CSM interpretation.  
%And for one of the more heavily 
%reddened objects, SN~2008fp, the observations seem to clearly favor an ISM 
%origin. 
%Whatever the source of the strong \ion{Na}{1}, the gas 
%producing it is almost certainly characterized by an 
%unusually low dust-to-gas ratio.  

Our observations show that the column density and/or equivalent width of
the \ion{Na}{1}~D lines are, in general, unreliable indicators of
the dust extinction suffered by SNe~Ia.  The exception to this
statement is that weak or undetectable \ion{Na}{1} absorption appears
to be consistent with little or no extinction.  We find that a better 
predictor of individual SN~Ia extinction is the equivalent width of the 
DIB at 5780~\AA, although this method requires moderate
resolution ($\lambda / \Delta\lambda \gtrsim 4,000$), high signal-to-noise
ratio spectra, and yields host $A_V$ values with a precision of only $\pm$50\%.
  
Obviously there are many details of the CSM versus ISM puzzle left to resolve.  Nevertheless, 
high-dispersion spectroscopy has revealed a number of critical clues to 
the nature of SN~Ia progenitors, including the discovery of variable \ion{Na}{1}
absorption, the excess of blueshifted absorbers, the connection between
blueshifted absorption and ejecta velocities, and now the correlation of
blueshifted absorption with Na-rich gas and the finding that
the extinction affecting the SN light curves arises in the host galaxy ISM.  
Larger samples of such observations combined with optical and 
near-infrared light curves seem likely to shed further light on the physical
mechanism responsible for these apparently related phenomena.

%% If you wish to include an acknowledgments section in your paper,
%% separate it off from the body of the text using the \acknowledgments
%% command.

%% Included in this acknowledgments section are examples of the
%% AASTeX hypertext markup commands. Use \url without the optional [HREF]
%% argument when you want to print the url directly in the text. Otherwise,
%% use either \url or \anchor, with the HREF as the first argument and the
%% text to be printed in the second.

\acknowledgments

The work of the CSP has been supported by the National Science Foundation 
under grants  ASTÐ0306969, ASTÐ0607438, and ASTÐ1008343.  
M.M.P. gratefully acknowledges the Aspen Center for Physics and NSF 
Grant 1066293 for hospitality during the conception of this work.  M.M.P. also
thanks Brandon Lawton, Andy McWilliam, and Sebasti\'{a}n L\'{o}pez for helpful discussions, and 
the Australian Astronomical Observatory and the ARC Centre of Excellence 
for All-sky Astrophysics (CAASTRO) for hosting and supporting a 
3-month research leave during which this paper was completed.
A.I.K. is grateful for support from the Australian Research Council for a
Future Fellowship (FT110100475).
A.G. was supported by the EU/FP7 via an
ERC grant, a Minerva ARCHES prize, and the Kimmel Award for innovative Investigation.
M.S. acknowledges generous support provided by the Danish Agency for Science 
and Technology and Innovation realized through a Sapere Aude Level 2 grant.
M.~H. and G.~P. are grateful for  support from Millennium Center for Supernova Science 
(P10-064-F), with input from Fondo de Innovaci\'on para la Competitividad, 
del Ministerio de Economia, Fomento y Turismo de Chile.
Computing resources used for this work were made possible by a grant from the 
Ahmanson Foundation.  This research has made use of the NASA/IPAC 
Extragalactic Database (NED) which is operated by the Jet Propulsion Laboratory, 
California Institute of Technology, under contract with the National Aeronautics and 
Space Administration.

%% To help institutions obtain information on the effectiveness of their
%% telescopes, the AAS Journals has created a group of keywords for telescope
%% facilities. A common set of keywords will make these types of searches
%% significantly easier and more accurate. In addition, they will also be
%% useful in linking papers together which utilize the same telescopes
%% within the framework of the National Virtual Observatory.
%% See the AASTeX Web site at http://www.journals.uchicago.edu/AAS/AASTeX
%% for information on obtaining the facility keywords.

%% After the acknowledgments section, use the following syntax and the
%% \facility{} macro to list the keywords of facilities used in the research
%% for the paper.  Each keyword will be checked against the master list during
%% copy editing.  Individual instruments or configurations can be provided 
%% in parentheses, after the keyword, but they will not be verified.

{\it Facilities:} \facility{Magellan (MIKE)}, \facility{KECK (HIRES)}, \facility{HET (HRS)}.

%% The reference list follows the main body and any appendices.
%% Use LaTeX's thebibliography environment to mark up your reference list.
%% Note \begin{thebibliography} is followed by an empty set of
%% curly braces.  If you forget this, LaTeX will generate the error
%% "Perhaps a missing \item?".
%%
%% thebibliography produces citations in the text using \bibitem-\cite
%% cross-referencing. Each reference is preceded by a
%% \bibitem command that defines in curly braces the KEY that corresponds
%% to the KEY in the \cite commands (see the first section above).
%% Make sure that you provide a unique KEY for every \bibitem or else the
%% paper will not LaTeX. The square brackets should contain
%% the citation text that LaTeX will insert in
%% place of the \cite commands.

%% We have used macros to produce journal name abbreviations.
%% AASTeX provides a number of these for the more frequently-cited journals.
%% See the Author Guide for a list of them.

%% Note that the style of the \bibitem labels (in []) is slightly
%% different from previous examples.  The natbib system solves a host
%% of citation expression problems, but it is necessary to clearly
%% delimit the year from the author name used in the citation.
%% See the natbib documentation for more details and options.

\clearpage

%% Use the figure environment and \plotone or \plottwo to include
%% figures and captions in your electronic submission.
%% To embed the sample graphics in
%% the file, uncomment the \plotone, \plottwo, and
%% \includegraphics commands
%%
%% If you need a layout that cannot be achieved with \plotone or
%% \plottwo, you can invoke the graphicx package directly with the
%% \includegraphics command or use \plotfiddle. For more information,
%% please see the tutorial on "Using Electronic Art with AASTeX" in the
%% documentation section at the AASTeX Web site,
%% http://www.journals.uchicago.edu/AAS/AASTeX.
%%
%% The examples below also include sample markup for submission of
%% supplemental electronic materials. As always, be sure to check
%% the instructions to authors for the journal you are submitting to
%% for specific submissions guidelines as they vary from
%% journal to journal.

%% This example uses \plotone to include an EPS file scaled to
%% 80% of its natural size with \epsscale. Its caption
%% has been written to indicate that additional figure parts will be
%% available in the electronic journal.

\clearpage

\clearpage

\clearpage

\clearpage

\clearpage

\clearpage

\clearpage

\clearpage

\clearpage

\clearpage

\clearpage

\clearpage

\clearpage

%% Here we use \plottwo to present two versions of the same figure,
%% one in black and white for print the other in RGB color
%% for online presentation. Note that the caption indicates
%% that a color version of the figure will be available online.
%%

% \begin{figure}
% \plottwo{f2.eps}{f2_color.eps}
% \caption{A panel taken from Figure 2 of \citet{rudnick03}. 
% See the electronic edition of the Journal for a color version 
% of this figure.\label{fig2}}
% \end{figure}

%% If you are not including electonic art with your submission, you may
%% mark up your captions using the \figcaption command. See the
%% User Guide for details.
%%
%% No more than seven \figcaption commands are allowed per page,
%% so if you have more than seven captions, insert a \clearpage
%% after every seventh one.

%% Tables should be submitted one per page, so put a \clearpage before
%% each one.

%% Two options are available to the author for producing tables:  the
%% deluxetable environment provided by the AASTeX package or the LaTeX
%% table environment.  Use of deluxetable is preferred.
%%

\clearpage
\input{tab1.tex}

\clearpage
\input{tab2.tex}

\clearpage
\input{tab3.tex}

\clearpage
\input{tab4.tex}

\clearpage

%% Tables may also be prepared as separate files. See the accompanying
%% sample file table.tex for an example of an external table file.
%% To include an external file in your main document, use the \input
%% command. Uncomment the line below to include table.tex in this
%% sample file. (Note that you will need to comment out the \documentclass,
%% \begin{document}, and \end{document} commands from table.tex if you want
%% to include it in this document.)

%% \input{table}

%% The following command ends your manuscript. LaTeX will ignore any text
%% that appears after it.

\end{document}

%% file: tab1.tex
\begin{deluxetable}{lcccc}
%\tabletypesize{\footnotesize} 
\tabletypesize{\scriptsize} 
\tablecolumns{5} 
\tablewidth{0pt} 
%\rotate 
\tablecaption{Milky Way \ion{Na}{1} and \ion{K}{1} Column Density Measurements\label{tab:tab1}} 
\tablehead{ 
 \colhead{} & \colhead{$A_V$} & \colhead{$\log{N_{Na~I}}$)} & \colhead{$\log{N_{K~I}}$} \\
 \colhead{Object} & \colhead{mag} & \colhead{cm$^{-2}$} & \colhead{cm$^{-2}$} & Reference \\
 \colhead{(1)} & \colhead{(2)} & \colhead{(3)} & \colhead{(4)} & \colhead{(5)} }
\startdata
2003gd       & $0.19 \pm 0.03$ & $12.775 \pm 0.034$ & \nodata            & 1 \\
2006be       & $0.08 \pm 0.01$ & $12.068 \pm 0.038$ & \nodata            & 2 \\
2006ca       & $0.64 \pm 0.10$ & $13.181 \pm 0.052$ & \nodata            & 2 \\
2006eu       & $0.52 \pm 0.08$ & $12.914 \pm 0.039$ & \nodata            & 2 \\
2007af       & $0.11 \pm 0.02$ & $11.751 \pm 0.106$ & \nodata            & 2 \\
2007hj       & $0.26 \pm 0.04$ & $12.859 \pm 0.102$ & \nodata            & 2 \\
2007kk       & $0.64 \pm 0.10$ & $12.801 \pm 0.122$ & \nodata            & 2 \\
2007le       & $0.09 \pm 0.02$ & $11.924 \pm 0.055$ & \nodata            & 2 \\
2007on       & $0.03 \pm 0.01$ & $11.178 \pm 0.077$ & \nodata            & 2 \\
2007sr       & $0.13 \pm 0.02$ & $11.734 \pm 0.018$ & \nodata            & 2 \\
2008C        & $0.23 \pm 0.04$ & $12.777 \pm 0.467$ & \nodata            & 2 \\
2008fp       & $0.54 \pm 0.09$ & $13.141 \pm 0.061$ & $11.417 \pm 0.070$ & 2 \\
2008ge       & $0.04 \pm 0.01$ & $11.307 \pm 0.045$ & \nodata            & 2 \\
2008hv       & $0.09 \pm 0.01$ & $12.276 \pm 0.016$ & \nodata            & 2 \\
2008ia       & $0.62 \pm 0.10$ & $13.149 \pm 0.010$ & $11.545 \pm 0.070$ & 2 \\
2009ds       & $0.11 \pm 0.02$ & $12.489 \pm 0.020$ & \nodata            & 2 \\
2009ev       & $0.28 \pm 0.05$ & $12.737 \pm 0.040$ & \nodata            & 2 \\
2009iw       & $0.24 \pm 0.04$ & $12.543 \pm 0.021$ & \nodata            & 2 \\
2009le       & $0.05 \pm 0.01$ & $11.793 \pm 0.011$ & \nodata            & 2 \\
2009mz       & $0.08 \pm 0.01$ & $11.972 \pm 0.030$ & \nodata            & 2 \\
2009nr       & $0.07 \pm 0.01$ & $11.926 \pm 0.031$ & \nodata            & 2 \\
2010A        & $0.08 \pm 0.01$ & $11.431 \pm 0.032$ & \nodata            & 2 \\
2010ev       & $0.29 \pm 0.05$ & $12.564 \pm 0.028$ & \nodata            & 2 \\
2010jl       & $0.07 \pm 0.01$ & $11.791 \pm 0.119$ & \nodata            & 1 \\
2010ko       & $0.39 \pm 0.06$ & $12.767 \pm 0.103$ & $11.210 \pm 0.087$ & 1 \\
2011K        & $0.27 \pm 0.04$ & $12.500 \pm 0.016$ & \nodata            & 1 \\
2011di       & $0.29 \pm 0.05$ & $12.472 \pm 0.025$ & \nodata            & 1 \\
2011dn       & $0.49 \pm 0.08$ & $12.948 \pm 0.011$ & $11.614 \pm 0.052$ & 1 \\
2011dq       & $0.31 \pm 0.05$ & $12.808 \pm 0.022$ & $11.619 \pm 0.024$ & 1 \\
2011dy       & $0.19 \pm 0.03$ & $12.960 \pm 0.056$ & $11.103 \pm 0.041$ & 1 \\
2011ek       & $0.97 \pm 0.15$ & $12.999 \pm 0.024$ & $11.867 \pm 0.038$ & 1 \\
2011fj       & $0.47 \pm 0.07$ & $13.043 \pm 0.047$ & \nodata            & 1 \\
2012cg       & $0.07 \pm 0.01$ & $11.218 \pm 0.057$ & \nodata            & 1 \\
PTF11iqb       & $0.09 \pm 0.02$ & $12.004 \pm 0.007$ & \nodata            & 1 \\
3C273          & $0.06 \pm 0.01$ & $12.073 \pm 0.014$ & \nodata            & 1 \\
IC4329A        & $0.16 \pm 0.02$ & $11.999 \pm 0.137$ & \nodata            & 1 \\
Mk509          & $0.16 \pm 0.02$ & $12.142 \pm 0.041$ & \nodata            & 1 \\
NGC1068        & $0.09 \pm 0.02$ & $12.618 \pm 0.156$ & \nodata            & 1 \\
NGC2110        & $1.03 \pm 0.16$ & $13.149 \pm 0.071$ & $11.559 \pm 0.049$ & 1 \\
NGC3783        & $0.33 \pm 0.05$ & $12.656 \pm 0.013$ & $11.169 \pm 0.055$ & 1 \\
PDS456         & $1.42 \pm 0.23$ & $13.309 \pm 0.037$ & $11.757 \pm 0.022$ & 1 \\
Fairall51      & $0.30 \pm 0.05$ & $12.388 \pm 0.036$ & \nodata            & 1 \\
IRAS06213+0020 & $1.77 \pm 0.28$ & $13.538 \pm 0.049$ & $12.167 \pm 0.018$ & 1 \\
IRAS08311-2459 & $0.29 \pm 0.05$ & $12.668 \pm 0.040$ & \nodata            & 1 \\
IRAS09149-6206 & $0.50 \pm 0.08$ & $12.643 \pm 0.100$ & $11.023 \pm 0.075$ & 1 \\
IRAS11353-4854 & $0.53 \pm 0.08$ & $12.771 \pm 0.075$ & $11.522 \pm 0.027$ & 1 \\
\enddata 
\tablecomments{Columns: (1) Object name; (2) Milky Way dust extinction \citep{schlafly11}; 
(3) Logarithm of the total neutral sodium column density; (4) Logarithm of the total neutral potassium 
column density; (5) High-dispersion spectroscopy reference [1 = unpublished MIKE spectrum;
2 = \citet{sternberg11}].
}

\end{deluxetable}

%% file: tab2.tex
\begin{deluxetable}{lllccll} 
\tabletypesize{\scriptsize}
\tablecolumns{7} 
\tablewidth{0pt} 
%\rotate 
\tablecaption{SNe~Ia Host Galaxy Information and Light-Curve Measurements\label{tab:tab2}} 
\tablehead{ 
 \colhead{} & \colhead{} & \colhead{} & \multicolumn{2}{c}{Photometry References} & \colhead{} & \colhead{$A_V$} \\
  \cline{4-5}
 \colhead{SN} & \colhead{Host Galaxy} & \colhead{Morphology} & \colhead{Optical} & \colhead{Near-IR} & \colhead{$R_V$} & \colhead{mag} \\
 \colhead{(1)} & \colhead{(2)} & \colhead{(3)} & \colhead{(4)} & \colhead{(5)} & \colhead{(6)} & \colhead{(7)} } 
\startdata
1986G           & NGC 5128     & S0 pec               &  1 & 16      & $2.57~_{-0.21}^{+0.23}$ & $2.03~_{-0.13}^{+0.09}$ \\
1994D           & NGC 4526     & SAB(s)0\^{}0?(s)     &  2 & 17      & $2.24~_{-0.73}^{+0.62}$ & $< 0.08$ \\
2001cp          & UGC 10738    & Sbc                  &  3 & \nodata & $2.28~_{-0.80}^{+0.62}$ & $< 0.17$ \\
2001da          & NGC 7780     & Sab                  &  3 & \nodata & $1.82~_{-0.53}^{+0.76}$ & $0.49~_{-0.14}^{+0.19}$ \\
2001el          & NGC 1448     & SAcd? edge-on        &  4 &  4      & $2.25~_{-0.36}^{+0.46}$ & $0.62~_{-0.08}^{+0.08}$ \\
2002bo          & NGC 3190     & SA(s)a pec edge-on   &  5 &  5      & $1.22~_{-0.21}^{+0.26}$ & $0.62~_{-0.10}^{+0.09}$ \\
2002cr          & NGC 5468     & SAB(rs)cd            &  6 & \nodata & $2.16~_{-0.68}^{+0.66}$ & $0.21~_{-0.07}^{+0.09}$ \\
2002ha          & NGC 6962     & SAB(r)ab             &  6 & \nodata & $2.45~_{-0.77}^{+0.52}$ & $< 0.08$ \\
2002jg          & NGC 7253B    & S?                   &  3 & \nodata & $2.46~_{-0.63}^{+0.57}$ & $< 0.17$ \\
2006X           & NGC 4321     & SAB(s)bc             &  7 &  7      & $1.31~_{-0.10}^{+0.08}$ & $1.88~_{-0.13}^{+0.09}$ \\
2006cm          & UGC 11723    & Sb edge-on           &  4 & \nodata & $1.95~_{-0.33}^{+0.46}$ & $2.15~_{-0.35}^{+0.40}$ \\
2007af          & NGC 5584     & SAB(rs)cd            &  8 &  8      & $2.11~_{-0.48}^{+0.55}$ & $0.31~_{-0.06}^{+0.06}$ \\
2007fb          & UGC 12859    & Sbc                  &  9 & \nodata & $2.17~_{-0.60}^{+0.53}$ & $< 0.09$ \\
2007fs          & ESO 601-G005 & Sb?                  &  9 & \nodata & $2.24~_{-0.78}^{+0.64}$ & $0.12~_{-0.05}^{+0.08}$ \\
2007hj          & NGC 7461     & SB0                  & 10 & 10      & $1.96~_{-0.55}^{+0.59}$ & $< 0.13$ \\
2007kk          & UGC 2828     & SB(rs)bc             &  9 & \nodata & $2.13~_{-0.64}^{+0.55}$ & $< 0.04$ \\
2007le          & NGC 7721     & SA(s)c               &    8 &  8      & $1.46~_{-0.24}^{+0.32}$ & $0.54~_{-0.08}^{+0.08}$ \\
2007on          & NGC 1404     & E1                   &  8 &  8      & $1.93~_{-0.61}^{+0.80}$ & $< 0.03$  \\
2007sr          & NGC 4038     & SB(s)m pec           & 11 & 11      & $1.74~_{-0.64}^{+0.50}$ & $0.23~_{-0.07}^{+0.06}$ \\
2008C           & UGC 3611     & S0/a                 &  8 &  8      & $2.42~_{-0.72}^{+0.56}$ & $0.39~_{-0.11}^{+0.16}$ \\
2008ec          & NGC 7469     & (R')SAB(rs)a         & 10 & \nodata & $2.21~_{-0.73}^{+0.53}$ & $0.53~_{-0.16}^{+0.13}$ \\
2008fp          & ESO 428-G014 & SAB0\^{}0(r) pec     &  8 &  8      & $1.20~_{-0.14}^{+0.26}$ & $0.71~_{-0.08}^{+0.10}$ \\
2008hv          & NGC 2765     & S0                   &  8 &  8      & $2.23~_{-0.68}^{+0.63}$ & $< 0.08$ \\
2008ia          & ESO 125-G006 & S0                   &  8 &  8      & $2.35~_{-0.57}^{+0.62}$ & $0.11~_{-0.03}^{+0.05}$ \\
SNF20080514-002 & UGC 8472     & S0                   &  3 & \nodata & $2.03~_{-0.58}^{+0.66}$ & $< 0.02$ \\
2009ds          & NGC 3905     & SB(rs)c              & 10 & 10      & $2.20~_{-0.69}^{+0.55}$ & $0.16~_{-0.05}^{+0.08}$ \\
2009ig          & NGC 1015     & SB(r)a?              & 12 & \nodata & $2.29~_{-0.62}^{+0.68}$ & $< 0.05$ \\
2009le          & ESO 478-G006 & Sbc                  & 10 & 10      & $2.48~_{-0.83}^{+0.57}$ & $0.24~_{-0.08}^{+0.09}$ \\
2010A           & UGC 2019     & S?                   &  9 & \nodata & $2.24~_{-0.74}^{+0.75}$ & $0.39~_{-0.11}^{+0.14}$ \\
2010ev          & NGC 3244     & SA(rs)cd             & 13 & \nodata & $1.54~_{-0.59}^{+0.57}$ & $0.50~_{-0.19}^{+0.17}$ \\
2011fe          & NGC 5457     & SAB(rs)cd            & 14 & 18      & $1.63~_{-0.53}^{+0.60}$ & $0.15~_{-0.06}^{+0.19}$ \\
2012cg          & NGC 4424     & SB(s)a?              & 15 & 15      & $2.7 \pm 0.5$ & $0.67 \pm 0.12$ \\
\enddata 
\tablecomments{Columns: (1) SN name; (2) Host galaxy name; 
(3) Host galaxy morphology from the NASA/IPAC Extragalactic Database (NED); 
(4) Optical photometry reference; (5) Near-infrared photometry reference;
(6) Host dust extinction $R_V$ value; (7) Host dust extinction $A_V$ value.
}
\tablerefs{
(1) \citet{phillips87}; (2) CTIO~0.9~m unpublished; (3) \citet{ganes10}; 
(4) \citet{hicken09a}; (5) \citet{krisciunas04}; (6) \citet{krisciunas03}; 
(7) \citet{contreras10}; (8) \citet{stritzinger11}; (9) \citet{hicken12}; 
(10) CSP, unpublished; (11) \citet{schweizer08}; (12) \citet{foley12a};
(13) \citet{gutierrez11}; (14) \citet{richmond12}; 
%(22) \citet{silverman12};
(15) CfA Supernova Group, unpublished (http://www.cfa.harvard.edu/supernova/sn12cg.html); 
(16) \citet{frogel87}; (17) \citet{richmond95}; (18) \citet{matheson12}.
}

\end{deluxetable}

%% file: tab3.tex
\begin{deluxetable}{lccclll} 
\tabletypesize{\scriptsize}
\tablecolumns{7} 
\tablewidth{0pt} 
%\rotate 
\tablecaption{SNe~Ia Host Absorption-Line Measurements\label{tab:tab3}} 
\tablehead{ 
 \colhead{} & \colhead{Sternberg} & \colhead{} & \colhead{} & \colhead{$\log{N_{Na~I}}$} & \colhead{$\log{N_{K~I}}$} & \colhead{EW(5780)} \\
 \colhead{SN} & \colhead{Type} & \colhead{Reference} & \colhead{$\lambda / \Delta\lambda$} & \colhead{cm$^{-2}$} & 
  \colhead{cm$^{-2}$} & \colhead{m\AA} \\
 \colhead{(1)} & \colhead{(2)} & \colhead{(3)} & \colhead{(4)} & \colhead{(5)} &
  \colhead{(6)} & \colhead{(7)}} 
\startdata
1986G           & B       & 1 & 20,000 & 13.740~$\pm~0.050$ & \nodata            &  335~$\pm~5$ \\
1994D           & S       & 2 & 38,000 & 11.800~$\pm~0.020$ & \nodata            & \nodata      \\
2001cp          & R       & 3 & 30,000 & 12.970~$\pm~0.470$ & \nodata            & \nodata      \\
2001da          & R       & 3 & 30,000 & 12.512~$\pm~0.073$ & \nodata            & \nodata      \\
2001el          & R       & 4 & 50,000 & 12.76~$\pm~0.03$   & 11.29~$\pm~0.03$   &  189~$\pm~3$ \\
%2001ep          & B       & 3 & 30,000 & 12.876~$\pm~0.214$ & \nodata           & \nodata      \\
2002bo          & B       & 3 & 30,000 & 14.406~$\pm~0.862$ & \nodata            &   37~$\pm~7$ \\
2002cr          & S       & 3 & 30,000 & 12.329~$\pm~0.088$ & \nodata            & $ < 47 $     \\
2002ha          & B       & 3 & 30,000 & 12.886~$\pm~0.125$ & \nodata            & \nodata      \\
2002jg          & B       & 3 & 30,000 & 13.246~$\pm~0.021$ & \nodata            & \nodata      \\
2006X           & B       & 5 & 48,000 & 13.779~$\pm~0.041$ & 11.995~$\pm~0.035$ & $ < 72 $     \\
2006cm          & B       & 5 & 48,000 & 15.242~$\pm~0.069$ & 12.307~$\pm~0.099$ & 441~$\pm~20$ \\
2007af          & R       & 5 & 48,000 & 12.044~$\pm~0.142$ & \nodata            & $ < 37 $     \\
2007fb          & B       & 5 & 48,000 & 12.844~$\pm~0.024$ & \nodata            & \nodata      \\
2007fs          & S       & 5 & 48,000 & 12.583~$\pm~0.029$ & \nodata            & $ < 31 $     \\
2007hj          & \nodata & 5 & 52,000 & $ < 12.05 $        & \nodata            & \nodata      \\
2007kk          & B       & 5 & 52,000 & 12.595~$\pm~0.171$ & \nodata            & \nodata      \\
2007le          & B       & 5 & 52,000 & 13.281~$\pm~0.012$ & $ < 11.14 $        &  139~$\pm~4$ \\
2007on          & \nodata & 5 & 48,000 & $ < 11.32 $        & \nodata            & \nodata      \\
2007sr          & S       & 5 & 25,000 & 13.220~$\pm~0.182$ & 11.241~$\pm~0.109$ &   71~$\pm~4$ \\
2008C           & B       & 5 & 48,000 & 12.720~$\pm~0.184$ & \nodata            & $ < 43 $     \\
2008ec          & B       & 5 & 54,000 & 13.055~$\pm~0.044$ & \nodata            &   77~$\pm~7$ \\
2008fp          & B       & 5 & 35,000 & 14.472~$\pm~0.036$ & 12.056~$\pm~0.033$ &  86~$\pm~11$ \\
2008hv          & \nodata & 5 & 35,000 & $ < 11.18 $        & \nodata            & $ < 25 $     \\
2008ia          & \nodata & 5 & 35,500 & $ < 11.39 $        & \nodata            & \nodata      \\
SNF20080514-002 & \nodata & 5 & 52,000 & $ < 11.65 $        & \nodata            & \nodata      \\
2009ds          & B       & 5 & 35,500 & 12.920~$\pm~0.058$ & $ < 10.93 $        & \nodata      \\
2009ig          & B       & 5 & 34,500 & 13.105~$\pm~0.161$ & 11.247~$\pm~0.034$ & $ < 18 $     \\
2009le          & R       & 5 & 35,500 & 13.254~$\pm~0.103$ & 11.118~$\pm~0.073$ & $ < 25 $     \\
2010A           & R       & 5 & 35,500 & 12.605~$\pm~0.021$ & $ < 11.11 $        &  99~$\pm~10$ \\
2010ev          & S       & 5 & 35,500 & 12.701~$\pm~0.029$ & 10.949~$\pm~0.085$ & 227~$\pm~10$ \\
2011fe          & S       & 6 & 82,000 & 11.495~$\pm~0.024$ & \nodata            & \nodata      \\
2012cg          & R       & 7 & 47,000 & 12.989~$\pm~0.035$ & 11.255~$\pm~0.075$ &   84~$\pm~5$ \\
\enddata 
\tablecomments{Columns: (1) SN name; (2) Classification of Na~I~D profile as per \citet{sternberg11} (B = Blueshifted, 
R = Redshifted, S = Single/Symmetric); (3) High-dispersion spectroscopy reference; (4) Spectral resolution;
(5) Logarithm of the total neutral sodium column density; (6) Logarithm of the total neutral potassium column density;
(7) Equivalent width of DIB feature at 5780~\AA.
}
\tablerefs{
(1) \citet{dodorico89}; (2) \citet{ho95}; (3) \citet{foley12b}; (4) \citet{sollerman05}; 
(5) \citet{sternberg11}; (6) \citet{patat13}; (7) \citet{raskutti13}. 
}

\end{deluxetable}

%% file: tab4.tex
\begin{deluxetable}{llcc}
\tabletypesize{\footnotesize} 
%\tabletypesize{\scriptsize} 
\tablecolumns{4} 
\tablewidth{0pt} 
%\rotate 
%\tablecaption{$\chi^2$ per Degree of Freedom with Respect to the Galactic $\log{N_{Na~I}}$ vs. $A_V$ Relation\label{tab:tab4}} 
\tablecaption{Goodness of Fit, $\chi^2_\nu$, to the Galactic $\log{N_{Na~I}}$ vs. $A_V$ Relation\label{tab:tab4}} 
\tablehead{ 
 \colhead{} & \colhead{}       & \colhead{}          & \colhead{Excluding} \\
 \colhead{} & \colhead{Sample} & \colhead{All} & \colhead{Upper Limits}
}
\startdata
Total Column Densities:   & \citet{sembach93} + Milky Way (this paper) & 135         & \nodata \\
%                         & \citet{sembach93}                          & 190         & \nodata \\
%                         & Milky Way (this paper)                     &  81         & \nodata \\
                          & LMC/SMC \citet{cox06,welty06}              & 201         & \nodata \\
\\
                          & Single/Symmetric + Redshifted              & $\geq  114$ &  130    \\
                          & Blueshifted                                & $\geq  434$ &  301    \\
\\
Most Prominent Component: & Blueshifted                                & $\geq  233$ &   60    \\
\\
Blueshifted Components:   & Blueshifted                                & $\geq  179$ &  233    \\
\enddata 
\tablecomments{The value of the $\chi^2$ per degree of freedom, 
$\chi^2_\nu$, is calculated with respect to the fit to the combined sample
of the \citet{sembach93} data and the Milky Way measurements given
in this paper (equation~\ref{eq:mwnai}).}

\end{deluxetable}

%% file: ms.bbl
\begin{thebibliography}{}

\bibitem[Astier et al.(2006)]{astier06}
Astier, P., Guy, J., Regnault, N., et al. 2006, \aap, 447, 31

\bibitem[Bernstein et al.(2003)]{bernstein03}
Bernstein, R., et al. 2003, Proc. SPIE, 4841, 1694

\bibitem[Blondin et al.(2009)]{blondin09}
Blondin, S., Prieto, J.~L., Patat, F., et al. 2009, \apj, 693, 207

\bibitem[Bloom et al.(2012)]{bloom12}
Bloom, J.~S., Kasen, D., Shen, K.~J. et al. 2012, \apjl, 744, L17

\bibitem[Bouret et al.(2003)]{bouret03}
Bouret, J.-C., Lanz, T., Hillier, D.~J., et al. 2003, \apj, 595, 1182

\bibitem[Burns et al.(2011)]{Burns12}
Burns, C., et al. 2011, \aj, 141, 19

\bibitem[Burns et al.(2013)]{burns13}
Burns, C., et al. 2013, in preparation

\bibitem[Cardelli et al.(1989)]{cardelli89}
Cardelli, J.~A., Clayton, G.~C., \& Mathis, J.~S. 1989, \apj, 345, 245

%\bibitem[Chugai(2008)]{chugai08}
%Chugai, N.~N. 2008, Astron. Lett., 34, 389

\bibitem[Clayton \& Mathis(1988)]{clayton88}
Clayton, G.~C., \& Mathis, J.~S. 1988, \apj, 327, 911

\bibitem[Cohen et al.(2005)]{cohen05b}
Cohen, J.~G., Briley, M.~M., \& Stetson, P.~B. 2005, \aj, 130, 1177

\bibitem[Cohen \& Kirby(2012)]{cohen12}
Cohen, J.~G., \& Kirby, E.~N. 2012, \apj, 760, 86

\bibitem[Cohen \& Mel{\'e}ndez(2005)]{cohen05a}
Cohen, J.~G. \& Mel{\'e}ndez, J. 2005, \aj, 129, 303

\bibitem[Conley et al.(2007)]{conley07}
Conley, A., Carlberg, R.~G., Guy, J., et al. 2007, \apj, 664, L13

\bibitem[Contreras et al.(2010)]{contreras10}
Contreras, C., Hamuy, M., Phillips, M.~M., et al. 2010, \aj, 139,  519

\bibitem[Cordiner et al.(2011)]{cordiner11}
Cordiner, M.~A., Cox, N.~L.~J., Evans, C.~J., et al. 2011, \apj, 726, 39

\bibitem[Cottrell \& Da Costa(1981)]{cottrell81}
Cottrell, P.~L. \& Da Costa, G.~S. 1981, \apjl, 245, L79

\bibitem[Cox et al.(2008)]{cox08a}
Cox, N. L. J., \& Cordiner, M. A. 2008, in IAU Symp. 251, Organic Matter in
Space, ed. S. Kwok (Cambridge: Cambridge Univ. Press), 237

\bibitem[Cox et al.(2006)]{cox06}
Cox,~N.~L.~J., Cordiner, M.~A., Cami, J., et al. 2006, \aap, 447, 991

\bibitem[Cox et al.(2007)]{cox07}
Cox,~N.~L.~J., Cordiner, M.~A., Ehrenfreund, P., et al. 2007, \aap, 470, 941

\bibitem[Cox \& Patat(2008)]{cox08}
Cox, N.~L.~J., \& Patat, F. 2008, \aap, 485, L9

\bibitem[Cox \& Patat(2013)]{cox13}
Cox, N.~L.~J., \& Patat, F. 2013, \aap, submitted

\bibitem[Cristallo et~al.(2009)]{cristallo09}
Cristallo, S., Straniero, O., Gallino, R., et al. 2009, \apj, 696, 797

%\bibitem[Crotts \& Yourdon(2008)]{crotts08}
%Crotts,~A.~P.~S., \& Yourdon, D. 2008, 689, 1186

\bibitem[Dilday et al.(2012)]{dilday12}
Dilday, B., Howell, D.~A., Cenko, S.~B., et al. 2012, Science, 337, 942

\bibitem[D'Odorico et al.(1989)]{dodorico89}
D'Odorico, S., di Serego Alighieri, S., Pettini, M., et al. 1989, \aap, 215,  21

\bibitem[Dufour(1984)]{dufour84}
Dufour, R.~J. 1984, IAU Symp. 108, 353

\bibitem[Fitzpatrick \& Massa(2007)]{fitzpatrick07}
Fitzpatrick, E.~L., \& Massa, D. 2007, \apj, 663, 320

\bibitem[Folatelli et al.(2010)]{folatelli10}
Folatelli, G., Phillips, M.~M., Burns, C.~R., et al. 2010, \aj, 139, 120

\bibitem[Foley et al.(2012a)]{foley12a}
Foley, R., Challis, P.~J., Filippenko, A.~V., et al. 2012a, \apj, 744, 28

\bibitem[Foley \& Kasen(2011)]{foley_kasen11}
Foley, R.~J., \& Kasen, D. 2011, \apj, 729, 55

\bibitem[Foley et al.(2012b)]{foley12b}
Foley, R., Simon, J.~D., Burns, C.~R., et al. 2012b, \apj, 752, 101

\bibitem[F\"{o}rster et al.(2012)]{forster12}
F\"{o}rster, F., Gonz‡lez-Gait‡n, S., Anderson, J., et al. 2012, \apjl, 754, L21

\bibitem[Frogel et al.(1987)]{frogel87}
Frogel, J.~A., Gregory, B., Kawara, K., et al. 1987, \apjl, 315, L129

\bibitem[Friedman et al.(2011)]{friedman11}
Friedman, S.~D., York, D. G., McCall, B. J., et al. 2011, \apj, 727, 33

\bibitem[Galazutdinov, Lo Curto, \& Kre\l owski(2008)]{galazutdinov08}
Galazutdinov, G.~A., Lo Curto, G., \& Kre\l owski, J. 2008, \mnras, 386, 2003

\bibitem[Galazutdinov et al.(2000)]{galazutdinov00}
Galazutdinov, G.~A., Musaev, F.~A., Kre\l owski, J., Walker, G.~A. 2000, \pasp, 112, 648 

\bibitem[Ganeshalingam et al.(2010)]{ganes10}
Ganeshalingam, M., Li, W., Filippenko, A.~V., et al. 2010, \apjs, 190, 418

\bibitem[Gehrz et al.(1998)]{gehrz98}
Gehrz, R.~D., Truran, J.~W., Williams, R.~E., \& Starrfield, S. 1998, \pasp, 110, 3

%\bibitem[Garnavich et al.(2013)]{garnavich13}
%Garnevich, P.~M., et al. 2013, American Astronomical Society, AAS Meeting \#221, \#209.04

\bibitem[Goobar(2008)]{goobar08}
Goobar, A. 2008, \apjl, 686, L103

\bibitem[Gordon et al.(2003)]{gordon03}
Gordon, K.~D., Clayton, G.~C., Misselt, K.~A., et al. 2003, \apj, 594, 279

\bibitem[Gratton et~al.(2001)]{gratton01}
Gratton, R.~G., Bonifacio, P., Bragaglia, A., et al. 2001, \aap, 369, 87

\bibitem[Guti\'errez et al.(2011)]{gutierrez11}
Guti\'errez, C., Folatelli, G., Pignata, G., et~al. 2011, Bulet\'in Asociaci\'on Argentina de Astronom\'ia, 54, 109

\bibitem[Hamuy et al.(1995)]{hamuy95}
Hamuy, M., Phillips, M.~M., Maza, J., et al. 1995, \aj, 109, 1

\bibitem[Hamuy et al.(1996)]{hamuy96}
Hamuy, M., Phillips, M.~M., Suntzeff, N.~B., et al. 1996, \aj, 112, 2391

\bibitem[Hamuy et al.(2006)]{hamuy06}
Hamuy, M., Folatelli, G., Morrell, N.~I., et al. 2006, \pasp, 118, 2

\bibitem[Hamuy et al.(2003)]{hamuy03}
Hamuy, M., Phillips, M.~M., Suntzeff, N.~B., et al. 2003, Nature, 424, 651

\bibitem[Heckman \& Lehnert(2000)]{heckman00}
Heckman, T.~M., \& Lehnert, M.~D. 2000, \apj, 537, 690

\bibitem[Herbig(1993)]{herbig93}
Herbig, G.~H. 1993, \apj, 407, 142

\bibitem[Hicken et al.(2009a)]{hicken09a}
Hicken, M., Challis, P.; Jha, S., et al. 2009a, \apj, 700, 331

\bibitem[Hicken et al.(2012)]{hicken12}
Hicken, M., Challis, P., Kirshner, R.~P., et al. 2012, \apjs, 200, 12

\bibitem[Ho \& Filippenko(1995)]{ho95}
Ho, L.~C., \& Filippenko, A.~V. 1995, \apj, 444, 165 [Erratum: 1996, \apj, 463, 818]

\bibitem[Hobbs(1974)]{hobbs74}
Hobbs, L.~M. 1974, \apj, 191, 381

\bibitem[Hobbs et al.(2008)]{hobbs08}
Hobbs, L.~M., York, D.~G., Snow, T.~P., et al. 2008, \apj, 680, 1256

\bibitem[Hobbs et al.(2009)]{hobbs09}
Hobbs, L.~M., York, D.~G., Thorburn, J.~A., et al. 2009, \apj, 705, 32

\bibitem[Horesh et al.(2012)]{horesh12}
Horesh, A., Kulkarni, S.~R., Fox, D.~B., et al. 2012, \apj, 746, 21

\bibitem[Hough et al.(1987)]{hough87}
Hough, J.~H., Bailey, J.~A., Rouse, M.~F., \& Whittet, D.~C.~B. 1987, \mnras, 227, P1

\bibitem[Howell(2011)]{howell11}
Howell, D. A. 2011, NatCo, 2, 350

\bibitem[James et~al.(2004)]{james04}
James, G., Fran\c{c}ois, P., Bonifacio, P., et al. 2004, \aap, 414, 1071

\bibitem[Junkkarinen et al.(2004)]{junkkarinen04}
Junkkarinen, V.~T., Cohen, R.~D., Beaver, E.~A., et al. 2004, \apj, 614, 658

\bibitem[Karakas(2010)]{karakas10}
Karakas, A.~I. 2010, \mnras, 403, 1413

\bibitem[Karakas et al.(2012)]{karakas12}
Karakas, A.~I., Garc{\'i}a-Hern{\'a}ndez, D.~A., \& Lugaro, M. 2012, \apj, 751, 8

\bibitem[Karakas \& Lattanzio(2007)]{karakas07b}
Karakas, A.~I. \& Lattanzio, J.~C. 2007, PASA, 24, 103

\bibitem[Kemp et al.(2002)]{kemp02}
Kemp, S.~N., Bates, B., Beckman, J.~E., et al. 2002, \mnras, 333, 561

\bibitem[Kessler et al.(2009)]{kessler09}
Kessler, R., Becker, A.~C., Cinabro, D., et al. 2009, \apj, 185, 32

\bibitem[Krisciunas et al.(2003)]{krisciunas03}
Krisciunas, K., Suntzeff, N.~B., Candia, P., et~al. 2003, \aj, 125, 166

\bibitem[Krisciunas et al.(2004)]{krisciunas04}
Krisciunas, K., Suntzeff, N.~B., Phillips, M.~M., et~al. 2004, \aj, 128, 3034 [Erratum: 2005, \aj, 130, 350]

\bibitem[Larson \& Whittet(2005)]{larson05}
Larson, K.~A., \& Whittet, D.~C.~B. 2005, \apj, 623, 897

\bibitem[Larson, Whittet, \& Hough(1996)]{larson96}
Larson, K.~A., Whittet, D.~C.~B., \& Hough, J.~H. 1996, \apj, 472, 755

\bibitem[Larson et al.(2000)]{larson00}
Larson, K.~A., Wolff, M.~J., Roberge, W.~G., et al. 2000, \apj, 532, 1021

\bibitem[Le Bertre \& Lequeux(1993)]{lebertre93}
Le Bertre, T., \& Lequeux, J. 1993, \aap, 274, 909

\bibitem[Luna et al.(2008)]{luna08}
Luna, R., Cox, N.~L.~J., Satorre, M.~A., et al. 2008, \aap, 480, 133

\bibitem[Maeda et al.(2010)]{maeda10}
Maeda K., Benetti, S., Stritzinger, M., et al., 2010, Nature, 466, 82

\bibitem[Maeda et al.(2011)]{maeda11}
Maeda K., Leloudas, G., Taubenberger, S., et al., 2011, \mnras, 413, 3075

\bibitem[Maguire et al.(2013)]{maguire13}
Maguire, K., Sullivan, M., Patat, F., et al. 2013, arXiv:1308.3899

%\bibitem[Mandel et al.(2011)]{mandel11}
%Mandel, K.~S., Narayan, G., \& Kirshner, K.~P. 2011, \apj, 731, 120

\bibitem[Mannucci et al.(2005)]{mannucci05}
Mannucci, F., Della Valle, M., Panagia, N., et al. 2005, \aap, 433, 807

\bibitem[Maoz \& Mannucci(2012)]{maoz12}
Maoz, D., \& Mannucci, F. 2012, PASA, 29, 447

\bibitem[Matheson et al.(2012)]{matheson12}
Matheson, T., Joyce, R.~R., Allen, L. E., et~al. 2012, \apj, 754, 19

\bibitem[Mazzei \& Barbaro(2011)]{mazzei11}
Mazzei, P., \& Barbaro, G. 2011, \aap, 527, 34

\bibitem[Merrill \& Wilson(1938)]{merrill38}
Merrill, P.~W., \& Wilson, O.~C. 1938, \apj, 87, 9

\bibitem[Moore \& Bildsten(2012)]{moore12}
Moore, K., \& Bildsten, L. 2012, \apj, 761, 182

\bibitem[Mowlavi(1999)]{mowlavi99b}
Mowlavi, N. 1999, \aap, 350, 73

\bibitem[Munari \& Zwitter(1997)]{munari97}
Munari, U., \& Zwitter, T. 1997, \aap, 318, 269

%\bibitem[Murdin(1972)]{murdin72}
%Murdin, P. 1972, \mnras, 157, 461

\bibitem[Nataf et al.(2013)]{nataf13}
Nataf, D.~M., Gould, A., Fouqu\'{e}, P., et al. 2013, \apj, 769, 88

\bibitem[Patat(2005)]{patat05}
Patat, F. 2005, \mnras, 357, 1161

\bibitem[Patat(2013)]{patat13a}
Patat,~F. 2013, in Binary Paths to Type Ia Supernovae Explosions, 
Proceedings of the International Astronomical Union, IAU Symposium, 
Vol. 281, 291

\bibitem[Patat et al.(2009)]{patat09}
Patat,~F., Baade, D., H\"{o}flich, P., et~al. 2009, \aap, 508, 229

\bibitem[Patat et al.(2007)]{patat07}
Patat,~F., Chandra, P., Chevalier, R., et~al. 2007, Science, 317, 924

\bibitem[Patat et al.(2013)]{patat13}
Patat,~F., Cordiner, M.~A., Cox, N.~L.~J., et~al. 2013, \aap, 549, 62

\bibitem[Peimbert \& Torres-Peimbert(1976)]{peimbert76}
Peimbert, M., \& Torres-Peimbert, S. \apj, 203, 581

\bibitem[Pettini \& Pagel(2004)]{pettini04}
Pettini, M., \& Pagel, B.~E.~J. 2004, \mnras, 348, L59

\bibitem[Phillips(1993)]{phillips93}
Phillips, M. M. 1993, \apjl, 413, L105 

\bibitem[Phillips(2012)]{phillips12}
Phillips, M. M. 2012, PASA, 29, 434

\bibitem[Phillips et al.(1987)]{phillips87}
Phillips, M. M., Phillips, A.~C., Heathcote, S.~R., et al. 1987, \pasp, 99, 592

\bibitem[Poznanski et al.(2011)]{poznanski11}
Poznanski, D., Ganeshalingam M., Silverman J.~M., \& Filippenko, A.~V. 2011, \mnras, 415, L81

\bibitem[Poznanski et al.(2012)]{poznanski12}
Poznanski, D., Prochaska, J.~X., \& Bloom, J.~S. 2012, \mnras, 426, 1465

\bibitem[Ram{\'i}rez \& Cohen(2003)]{ramirez03}
Ram{\'i}rez, S.~V. \& Cohen, J.~G. 2003, \aj, 125, 224

\bibitem[Raskin \& Kasen(2013)]{raskin13}
Raskin, C., \& Kasen, D. 2013, arXiv:1304.4957

\bibitem[Raskutti et al.(2013)]{raskutti13}
Raskutti, S., et~al. 2013, in preparation

\bibitem[Richmond et al.(1995)]{richmond95}
Richmond, M.~W., Treffers, R.~R., Filippenko, A.~V., et~al. 1995, \aj, 109, 2121

\bibitem[Richmond \& Smith(2012)]{richmond12}
Richmond, M.~W., \& Smith, H.~A. 2012, arXiv:1203.4013

\bibitem[Rolleston, Trundle, \& Dufton(2002)]{rolleston02}
Rolleston, W.~R.~J., Trundle, C., \& Dufton, P.~L. 2002, \aap, 396, 53

\bibitem[Sarkowsky et al.(1975)]{sarkowsky75}
Sarkowsky, K., Mathewson, D.~S., \& Ford, V.~L. 1975, \apj, 196, 261

\bibitem[Sarre(2006)]{sarre06}
Sarre, P. J. 2006, JMoSp, 238, 1

\bibitem[Scarrot et al.(1987)]{scarrot87}
Scarrot, S.~M., Ward-Thompson, D., \& Warren-Smith, R.~F. 1987, \mnras, 224, 299 

\bibitem[Schlafly \& Finkbeiner(2011)]{schlafly11}
Schlafly, E.~F., \& Finkbeiner, D.~P. 2011, \apj, 737, 103

\bibitem[Schlegel et al.(1998)]{schlegel98} 
Schlegel,~D.~J., Finkbeiner,~D.~P., \& Davis,~M. 1998, \apj, 500, 525\

\bibitem[Schweizer et al.(2008)]{schweizer08}
Schweizer, F., Burns, C.~R., Madore, B.~F., et~al. 2008, \aj, 136, 1482

\bibitem[Scolnic et al.(2013)]{scolnic13}
Scolnic, D.~M., Riess, A.~G., Foley, R.~J., et al. 2013, arXiv:1306.4050

\bibitem[Sembach et al.(1993)]{sembach93}
Sembach, K.~R., Danks, A.~C., \& Savage, B.~D. 1993, \aaps, 100, 107

\bibitem[Shen et al.(2013)]{shen13}
Shen, K.~J., Guillochon, J., \& Foley, R.~J. 2013, arXiv:1302.2916

\bibitem[Silverman et al.(2012)]{silverman12}
Silverman, J.~A., Ganeshalingam, M., Cenko, S.~B., et~al. 2012, \apjl, 756, L7

\bibitem[Siluk \& Silk(1974)]{siluk74}
Siluk, R.~S., \& Silk, J. 1974, \apj, 192, 51

\bibitem[Simon et al.(2009)]{simon09}
Simon,~J.~D., Gal-Yam, A., Gnat, O., et~al. 2009, \apj, 702, 1157

\bibitem[Sneden et al.(2004)]{sneden04}
Sneden, C., Ivans, I.~I., \& Fulbright, J.~P. 2004, in
Carnegie Observatories Astrophysics Series, Vol. 4: 
Origin and Evolution of the Elements, ed. A.~McWilliam and M.~Rauch 
(Cambridge: Cambridge Univ. Press), 170

\bibitem[Sollerman et al.(2005)]{sollerman05}
Sollerman, J., Cox, N., Mattila, S., et al. 2005, \aap, 429, 559

\bibitem[Sparks, Carollo, \& Macchetto(1997)]{sparks97}
Sparks, W.~B., Carollo, C.~M., \& Macchetto, F. 1997, \apj, 486, 253

\bibitem[Springob et al.(2005)]{springob05}
Springob, C.~M., Haynes, M.~P.,, Giovanelli, R., \& Kent, B.~R. 2005, \apjs, 160, 149

\bibitem[Starrfield et al.(1993)]{starrfield93}
Starrfield, S., Truran, J.~W., Politano, M., et al. 1993, Phys. Rept., 227, 223

\bibitem[Sternberg et al.(2011)]{sternberg11}
Sternberg,~A., Gal-Yam, A., Simon, J.~D., et al. 2011, Science, 333, 856

\bibitem[Stritzinger et al.(2010)]{stritzinger10}
Stritzinger,~M., Burns, C.~R., Phillips, M.~M., et al. 2010, \aj, 140, 2036

\bibitem[Stritzinger et al.(2011)]{stritzinger11}
Stritzinger,~M., Phillips, M.~M., Boldt, L.~N., et al. 2011, \aj, 142, 156

%\bibitem[Takeda et al.(2009)]{takeda09}
%Takeda, Y, et al. 2009, \pasj, 61, 563

\bibitem[Tripp(1998)]{tripp98}
Tripp, R. 1998, \aap, 331, 815

\bibitem[Tuairisg et al.(2000)]{tuairisg00}
Tuairisg, S.~\'{O}., Cami, J., Foing, B.~H., Sonnentrucker, P., Ehrenfreund, P.  2000, \aaps, 142, 225

\bibitem[Turatto et al.(2003)]{turatto03}
Turatto, M., Benetti, S., \& Capellaro, E. 2003, in
From Twilight to Highlight: The Physics of Supernovae,
ed. W.~Hillebrandt and B.~Leibundgut (Berlin: Springer-Verlag), 200

\bibitem[Udalski(2003)]{udalski03}
Udalski, A. 2003, \apj, 590, 284

\bibitem[Vallerga et al.(1993)]{vallerga93}
Vallerga, J.~V., Vedder, P.~W., Craig, N., \& Welsh, B.~Y. 1993, \apj, 411, 729 

\bibitem[Ventura \& D'Antona(2008)]{ventura08}
Ventura, P., \& DÕAntona, F. 2008, \aap, 479, 805

\bibitem[Vos et al.(2011)]{vos11}
Vos, D.~A.~I., Cox, N.~L.~J., Kaper, L., Spaans, M., \& Ehrenfreund, P. 2011, \aap, 533, 129

\bibitem[Wang et al.(2003)]{wang03}
Wang, L., Baade, D., H\"{o}flich, P., et al. 2003, \apj, 591, 1110

\bibitem[Wang(2005)]{wang05}
Wang,~L. 2005, \apjl, 635, L33

%\bibitem[Wang et al.(2009)]{wang09}
%Wang,~X., et al. 2009, \apjl, 699, L139

\bibitem[Wang et al.(2013)]{wang13}
Wang,~X., Wang, L., Filippenko, A.~V., Zhang, T., \& Zhao, X. 2013, Science, 340, 170

\bibitem[Welsh et al.(2010)]{welsh10}
Welsh, B.~Y., Lallement, R., Vergely, J.-L., \& Raimond, S. 2010, \aap, 510, 54

\bibitem[Welty \& Hobbs(2001)]{welty01}
Welty, D.~E., \& Hobbs, L.~M. 2001, \apjs, 133, 345

\bibitem[Welty et al.(2006)]{welty06}
Welty, D.~E., Federman, S.~R., Gredel, R., Thorburn, J.~A., \& Lambert, D.~L. 2006, \apjs, 165, 138

\bibitem[Whittet \& van Breda(1978)]{whittet78}
Whittet, D.~C.~B., \& van Breda, I.~G. 1987, \aap, 66, 57

\bibitem[Yaron \& Gal-Yam(2012)]{yaron12}
Yaron, O., \& Gal-Yam, A. 2012, \pasp, 124, 668

\bibitem[York et al.(2006)]{york06}
York, B.~A., Ellison, S.~L., Lawton, B., et al. 2006, \apjl, 647 L29

\bibitem[Yuan \& Liu(2012)]{yuan12}
Yuan, H.~B., \& Liu, X.~W. 2012, \mnras, 425, 1763

\end{thebibliography}
